\documentclass{SciPost}

\usepackage{hyperref}
\hypersetup{
    colorlinks,
    linkcolor={red!50!black},
    citecolor={blue!50!black},
    urlcolor={blue!80!black}
}

\usepackage[bitstream-charter]{mathdesign}
\usepackage{subcaption}
\usepackage{graphicx}
\usepackage{caption}
\usepackage{xspace}
\usepackage{physics}
\usepackage[lined,ruled]{algorithm2e}
\usepackage{amsmath}
\usepackage{mathtools}
\usepackage{fusioncategories}
\usepackage{xfrac}
\usepackage{lscape}

\DeclareSymbolFont{usualmathcal}{OMS}{cmsy}{m}{n}
\DeclareSymbolFontAlphabet{\mathcal}{usualmathcal}

\fancypagestyle{SPstyle}{
\fancyhf{}
\lhead{\colorbox{scipostblue}{\bf \color{white} ~SciPost Physics Codebases }}
\rhead{{\bf \color{scipostdeepblue} ~Submission }}

\fancyfoot[C]{\textbf{\thepage}}
}

\newcommand{\eg}{{e.g.}}
\newcommand{\ie}{{i.e.}}

\newcommand{\library}[1]{\textsf{#1}\xspace}
\newcommand{\programminglanguage}[1]{\textsc{#1}\xspace}
\newcommand{\code}[1]{\verb|#1|}

\newcommand{\Julia}{\programminglanguage{Julia}}
\newcommand{\Python}{\programminglanguage{Python}}

\newcommand{\Matlab}{\programminglanguage{Matlab}}
\newcommand{\Cpp}{\programminglanguage{C++}}

\newcommand{\TensorKit}{\library{TensorKit.jl}}
\newcommand{\TensorOperations}{\library{TensorOperations.jl}}
\newcommand{\Stridedjl}{\library{{Strided.jl}}}
\newcommand{\MPSKit}{\library{MPSKit.jl}}
\newcommand{\PEPSKit}{\library{PEPSKit.jl}}
\newcommand{\ITensor}{\library{ITensor}}
\newcommand{\ITensorsjl}{\library{ITensors.jl}}
\newcommand{\TeNPy}{\library{TeNPy}}
\newcommand{\YASTN}{\library{YASTN}}
\newcommand{\pepstorch}{\library{peps-torch}}
\newcommand{\Cytnx}{\library{Cytnx}}

\newcommand{\Uniten}{\library{Uni10}}
\newcommand{\quimb}{\library{quimb}}
\newcommand{\SyTen}{\library{SyTen}}
\newcommand{\QSpace}{\library{QSpace}}
\newcommand{\TNRKit}{\library{TNRKit.jl}}
\newcommand{\FiniteMPS}{\library{FiniteMPS.jl}}

\newcommand{\PyTorch}{\library{PyTorch}}
\newcommand{\JAX}{\library{JAX}}
\newcommand{\BLAS}{\library{BLAS}}
\newcommand{\LAPACK}{\library{LAPACK}}
\newcommand{\CUDA}{\library{CUDA}}
\newcommand{\cuTENSOR}{\library{cuTENSOR}}
\newcommand{\TBLIS}{\library{TBLIS}}
\newcommand{\HPTT}{\library{HPTT}}

\newcommand{\reals}{\ensuremath{\mathbb{R}}}
\newcommand{\complexs}{\ensuremath{\mathbb{C}}}

\newcommand{\U}[1]{\ensuremath{U_{#1}}}
\newcommand{\SU}[1]{\ensuremath{SU_{#1}}}
\newcommand{\SO}[1]{\ensuremath{SO_{#1}}}
\newcommand{\Z}[1]{\ensuremath{\mathbb{Z}_{#1}}}
\newcommand{\Sym}[1]{\ensuremath{\mathcal{S}_{#1}}}

\newcommand{\gradedbra}[3][]{\ensuremath{\bra{#2}^{\left(#3\right)}_{#1}}}
\newcommand{\gradedket}[3][]{\ensuremath{\ket{#2}^{\left(#3\right)}_{#1}}}
\newcommand{\gradedtensor}[3]{\ensuremath{#1_{#2}^{\left(#3\right)}}}
\newcommand{\gradedspace}[2]{\ensuremath{{#1}^{\left(#2\right)}}}

\NewSymbol{B}{ true }{ true }{ true }{ true }
\NewSymbol[text=\tilde{X}]{tX}{ true }{ true }{ }{ true }
\NewSymbol[text=F^\dagger]{Fd}{ true }{ true }{ true }{ true }

\renewcommand{\XSymbol}[3]{\ensuremath{\XMatrix{#1}{#2;#3}}}
\renewcommand{\tXSymbol}[3]{\ensuremath{\tXMatrix{\tilde{#1}}{#2;#3}}}

\newcommand{\group}[1]{\ensuremath{\mathsf{#1}}}
\newcommand{\category}[1]{\ensuremath{\mathbf{#1}}}
\newcommand{\field}[1]{\ensuremath{\mathbb{#1}}}

\newcommand{\Ob}{\ensuremath{\mathrm{Ob}}}
\newcommand{\Hom}{\ensuremath{\mathrm{Hom}}}
\newcommand{\End}{\ensuremath{\mathrm{End}}}
\newcommand{\Rep}[1]{\ensuremath{\category{Rep}^{\group{#1}}}}
\newcommand{\Irr}{\ensuremath{\mathrm{Irr}}}
\newcommand{\id}[1]{\ensuremath{\mathbb{1}_{#1}}}
\renewcommand{\tr}{\ensuremath{\mathrm{tr}}}
\renewcommand{\ev}[1]{\ensuremath{\epsilon_{#1}}} 
\newcommand{\coev}[1]{\ensuremath{\eta_{#1}}}

\newcommand{\diagram}[2]{\;\vcenter{\hbox{\includegraphics[scale=0.45,page=#2]{./figures/#1.pdf}}}\;}

\newcommand{\fusiontreemanipulations}[1]{\;\vcenter{\hbox{\includegraphics[scale=0.5,page=#1]{./figures/fusiontreemanipulations.pdf}}}\;}

\newtheorem{definition}{Definition}
\newtheorem{lemma}{Lemma}
\newtheorem{theorem}{Theorem}

\newcommand{\algref}[1]{Algorithm~(\ref{#1})}
\renewcommand{\eqref}[1]{Equation~(\ref{#1})}

\begin{document}

\pagestyle{SPstyle}

\begin{center}{\Large \textbf{\color{scipostdeepblue}{
TensorKit.jl: A Julia package for large-scale tensor computations, with a hint of category theory.\\
}}}\end{center}

\begin{center}\textbf{
Lukas Devos\textsuperscript{1,2$\star$} and
Jutho Haegeman\textsuperscript{2$\dagger$}
}\end{center}

\begin{center}
{\bf 1} Center for Computational Quantum Physics, Flatiron Institute, New York, New York 10010, USA
\\
{\bf 2} Department of Physics and Astronomy, Ghent University, Krijgslaan 281, 9000 Gent, Belgium
\\[\baselineskip]
$\star$ \href{mailto:ldevos98@gmail.com}{\small ldevos98@gmail.com}\,,\quad
$\dagger$ \href{mailto:Jutho.Haegeman@ugent.be}{\small Jutho.Haegeman@ugent.be}
\end{center}

\section*{\color{scipostdeepblue}{Abstract}}
\boldmath\textbf{%
\TensorKit is a Julia-based software package for tensor computations, especially focusing on tensors with internal symmetries. This paper introduces the design philosophy, core functionalities, and distinctive features, including how to handle abelian, non-abelian, and anyonic symmetries through the ``TensorMap'' type. We highlight the software's flexibility, performance, and its capability to extend to new tensor types and symmetries, illustrating its practical applications through select case studies.
}

\vspace{\baselineskip}

\noindent\textcolor{white!90!black}{%
\fbox{\parbox{0.975\linewidth}{%
\textcolor{white!40!black}{\begin{tabular}{lr}%
  \begin{minipage}{0.6\textwidth}%
    {\small Copyright attribution to authors. \newline
    This work is a submission to SciPost Physics Codebases. \newline
    License information to appear upon publication. \newline
    Publication information to appear upon publication.}
  \end{minipage} & \begin{minipage}{0.4\textwidth}
    {\small Received Date \newline Accepted Date \newline Published Date}%
  \end{minipage}
\end{tabular}}
}}
}


\vspace{10pt}
\noindent\rule{\textwidth}{1pt}
\tableofcontents
\noindent\rule{\textwidth}{1pt}
\vspace{10pt}

\section{Introduction}
\label{sec:intro}

\TensorKit is an open-source software package, developed using the Julia programming language, that provides a set of datatypes and algorithms to represent and manipulate tensors, with a particular focus on tensors that ``exhibit internal symmetries'', i.e.\ that act covariantly with respect to a symmetry action on the tensor indices. 
This package aims to offer both an implementation of and an interface for the different algorithmic components typically used in tensor network methods --in particular tensor contraction and tensor decompositions-- that is sufficiently generic to be used in combination with a wide range of symmetries as they appear in the context of quantum many-body systems.
The structure induced by those symmetries is leveraged to significantly enhance performance in terms of both memory usage and computational efficiency.


The design philosophy behind \TensorKit is to provide an interface that is as symmetry-agnostic as possible, thereby enabling the users to interact with complex tensor operations, while the required symmetry management is taken care of behind the scenes.
At the same time, \TensorKit aims to support a rich variety of symmetries, including those associated with both abelian and non-abelian groups, as well as the larger class of ``categorical'' or ``non-invertible'' symmetries that arises in modern theoretical physics. This paper explores the structure and manipulation of symmetric tensors from the ground up, touching upon various technical aspects and implementation details. The complexity is gradually increased by discussing first abelian, then non-abelian and finally categorical symmetries. Along the way, the motivation for some of the design decisions and interface choices should also become clear.



In the remainder of this section, we first put the functionality of this package into its natural context, namely that of tensor network algorithms, primarily in the domain of condensed matter physics and quantum chemistry. Secondly, we compare the functionality of this package to some of the main alternatives, again focusing on those that are primarily targeting applications in quantum many-body physics.

\subsection{Context And Background}

Since the advent of Steve White's Density Matrix Renormalization Group (DMRG) algorithm \cite{white1992}, and especially over the past two decades, tensor network methods have been established as a powerful and essential set of techniques \cite{Verstraete01032008,montangero2018introduction,RevModPhys.93.045003,annurev:/content/journals/10.1146/annurev-conmatphys-040721-022705,xiang2023density} to complement more traditional techniques of computational physics, such as exact diagonalization, Hartree-Fock and other mean-field treatments, and various flavors and incarnations of Monte Carlo sampling.
Initially developed to target the lowest energy states of (quasi-) one-dimensional quantum spin chain Hamiltonians, the scope of tensor network methods has dramatically expanded.

Today, a comprehensive suite of algorithms exists for approximating and studying a wide array of quantum phenomena including thermal states, low- and high-energy excited states, dynamics of closed and open systems, and non-equilibrium steady states. These methods are being applied beyond simple one-dimensional lattices to systems that may be gapped or critical, possess higher-dimensional lattice geometries, or lack a clear locality structure due to long-range interactions. Even tensor network-inspired ans\"atze for systems in the continuum have been devised. As a consequence, the tensor network toolbox is now being used for a variety of quantum many-body problems that arise in condensed matter, cold atom physics and quantum chemistry, as well as in high-energy physics and even quantum gravity. Additionally, tensor network techniques have proven instrumental in studying partition functions of models in classical statistical mechanics.

Similar decompositions of high-order tensors into networks of partially contracted low-order tensors have been conceived in the numerical analysis and linear algebra community, for instance, in solving high-dimensional partial differential equations.
Moreover, tensor networks share clear parallels with certain architectures in machine learning, such as specific formulations of neural networks.

Despite this great variety in methods and application domains, there are some fundamental operations at the level of individual tensors that all of these algorithms have in common.
By far the most basic and crucial operation that appears in all of these techniques is the contraction of two tensors into a third one or a scalar.
The most operationally efficient method for contracting multiple tensors --forming a ``network'' of tensors-- is to contract pairs sequentially.
Although this approach involves creating intermediate tensors, which may seem less memory efficient, it compensates by avoiding the higher computational costs associated with contracting all tensors simultaneously.
However, determining the most efficient sequence for tensor contraction remains a challenging problem that has attracted significant attention. 
A second primitive building block of tensor-based algorithms is the factorization of a tensor into lower-rank tensors.
These are often directly related or generalized from the well known matrix factorizations from linear algebra (eigenvalue and singular value decompositions).

More recently, driven by advances in the machine learning community, the integration of automatic differentiation (AD) has facilitated rapid developments in gradient-based optimization strategies, now considered state-of-the-art for exploring tensor network geometries beyond the traditional matrix product state (a.k.a.\ tensor train) format.

So far, we have referred to tensors under the assumption that the reader is familiar with this concept.
In its simplest form, a tensor in computational contexts is akin to a multi-dimensional array.
This means it extends the idea of a matrix (which you can think of as a two-dimensional array) into an object indexed by multiple integers, each representing a different dimension with its own range of values.
In scientific applications, this indexing operation typically returns a numerical value.
A deeper mathematical interpretation would consider each index of a tensor as linked to a vector space and the tensor itself as part of the combined tensor product of these spaces.
However, various fields may offer slightly different but nearly equivalent definitions of what constitutes a tensor.
Tensors find utility in numerous other domains not yet discussed, such as classical mechanics, elasticity theory, general relativity, and various methods in nuclear physics and quantum chemistry.
We will revisit the issue of defining a tensor to accommodate the applications we have in mind in section \ref{sec:tensor}.

An important consideration in the practical use of tensors is their inherent structure, which impacts how they are best stored and manipulated.
In many-body physics, for instance, tensors are often required to remain invariant under the combined action of a symmetry group on the vector spaces associated with their indices.
This requirement imposes complex structural constraints on the tensors.
For abelian symmetry groups, this leads to a block sparsity in the tensor, where certain hyper-rectangular regions of the tensor entries are entirely zero.
Exploiting this structure in both storage and computations can significantly reduce the required resources.
For non-abelian groups, the structure becomes more intricate to exploit, but the efficiency gains can be even larger.
Additionally, the notion of symmetry can be further extended to include fermionic and ``anyonic'' generalizations. The latter are also referred to as non-invertible symmetries and require mathematics that goes beyond the theory of groups and representations, namely into the realm of category theory.
The manipulation of tensors with these types of symmetries arises for example in the context of topological quantum computation and error correction \cite{PhysRevA.95.022309,PhysRevX.12.021012}.
Various case studies and examples where symmetries were incorporated in tensor network studies have appeared in the literature, including the non-abelian \cite{singh2012,singh2013,weichselbaum2012,weichselbaum2020,schmoll2020,bruognolo2021} and anyonic \cite{pfeifer2010,singh2014,ayeni2016} case.
However, an all-encompassing framework with an open-source implementation has been lacking for a long time.

In other tensor applications, very different forms of structure can also be relevant.
True (unstructured) sparsity, characterized by randomly distributed zeros, is rarely observed in many-body physics applications, but might appear in the context of random walks or stochastic processes.
Symmetries associated with the permutation of indices are often encountered in fields like classical physics, or in the description of quantum systems of identical particles (fermions or bosons) in first quantization.
While such index permutation structure has an interesting interplay with the action of the general linear group on the vector spaces associated with the indices via the Schur-Weyl duality, we here limit our scope and do not include such types of sparsity or structure into our discussion.

\subsection{Related Work And Distinguishing Features}

Given the widespread use of tensor contractions and factorizations in various application domains, we do not attempt to provide a full overview of available software packages here, and instead refer to the excellent Ref.~\citeonline{psarrasLandscapeSoftwareTensor2022a} for this. Even the DMRG algorithm specifically has seen such widespread use that by now an extensive list of open-source packages exists that implement some specific flavor of it \cite{sehlstedt2025softwarelandscapedensitymatrix}.
Here, we will focus on those packages with a similar scope as \TensorKit, i.e.\ that have quantum many-body physics as primary application, support some form of symmetry structure and are sufficiently general to be used for building general tensor network algorithms beyond DMRG and matrix product states.
Recent years have seen the rise of such packages, alongside an evolution of more publications now being combined with associated code and scripts (or data) to reproduce the results.

Among the oldest and most established open-source packages is \ITensor, which started as a \Cpp library that includes an efficient implementation of the DMRG algorithm and a convenient interface for specifying Hamiltonians.
It has seen widespread use for ground state calculations in many publications.
Since 2019, the \ITensor team started the development of a \Julia version known as \ITensorsjl, which is now the main and recommend package \cite{10.21468/SciPostPhysCodeb.4}.
It is important to note that \ITensorsjl is not a \Julia interface to the corresponding \Cpp library, but a completely rewritten package using the native \Julia language, exactly as with the \TensorKit package that is described in this manuscript.
There are several libraries that are built using the popular \Python programming language, or that have a \Python interface around a compiled core (often in \Cpp).
Notable examples here are \TeNPy \cite{10.21468/SciPostPhysLectNotes.5}, \quimb \cite{gray2018quimb}, the combination of \YASTN and \pepstorch \cite{10.21468/SciPostPhysCodeb.52}, and the combination of \Cytnx and \Uniten \cite{kao2015uni10,10.21468/SciPostPhysCodeb.53}.
Along similar lines, \QSpace \cite{10.21468/SciPostPhysCodeb.40} is a \Cpp library embedded via the MEX interface into \Matlab, whereas \SyTen \cite{hubig2015syten} is predominantly a \Cpp library, with Python bindings that are work in progress.
\SyTen is also the only closed-source package in our overview. 
Without a doubt, even within the very specific domain of tensor networks for quantum many-body systems, additional packages and projects exist that have escaped our attention.

Regarding the choice of programming language, it should be noted that the benefits from exploiting the symmetry structure is only guaranteed if the associated overhead from the symmetry management is low, which requires a sufficiently performant language.
The aforementioned \Python and \Matlab packages solve this by having a compiled core written using \Cpp, which is an illustration of the infamous two-language problem.
It is exactly this problem that the open-source and scientifically oriented programming language \Julia promises to solve.
By being a just-in-time compiled language\footnote{Sometimes referred to as just-ahead-of-time compiled.}, it combines the ease of use of a dynamic scripting language with the performance characteristics of a statically compiled language. 
This is definitely one of the main motivations why \Julia was chosen for this project. 

In terms of functionality, all of the aforementioned libraries support at least the contraction of tensor networks made up from dense (\ie\ unstructured or without symmetries) tensors, evaluated using the computer's main central processing unit (CPU).
Pairwise tensor contractions are almost always evaluated by first permuting the data, such that the contraction becomes equivalent to a regular matrix multiplication\footnote{In computer science, permuting of tensor indices is often referred to as a generalization of matrix transposition, and this approach for contracting tensors is known as Transpose-Transpose-GEMM-Transpose (TTGT), where GEMM is the name of the general matrix multiplication kernel in \BLAS.}, which is then handled by a dedicated and optimized implementation of the Basic Linear Algebra Subprograms (\BLAS) library.
This part of the computation, which is often the most time consuming one, can then benefit from multi-threading capabilities provided by the \BLAS library.
Similarly, multi-threading support for tensor decompositions is provided at the level of the corresponding matrix decomposition via the specific implementation of the Linear Algebra Package (\LAPACK) that is being used.
In order to further enhance the performance of these algorithms, many packages also focus their attention towards supporting various dedicated hardware setups such as graphical processing units (GPUs), or multi-node computing setups commonly encountered in the realm of high performance computing (HPC).

Aside from \quimb, all of the aforementioned libraries have support for tensors with abelian symmetries.
However, so far only \QSpace provides support for general non-abelian symmetries, and none of the open-source packages provide generalizations to categorical symmetries. From the publicly available documentation, we gather that \SyTen offers specific support for $\SU{2}$, $\SU{2} \times \U{1}$, $\SU{2} \times \U{1} \times \Z{k}$, as well as a single instance of $\SU{3}$ in the context of MPS (at most rank-$3$ tensors), whereas higher order tensors seem to be restricted to at most $\SU{2}$ or $\SU{2} \times \U{1}$ symmetries.

In terms of algorithms, most of the aforementioned packages come with support for at least finite-size MPS using the DMRG algorithm.
\TeNPy seems most complete in terms of MPS algorithms and support for physical models and Hamiltonians, including both finite- and infinite-size DMRG to target ground states, as well as three different time-evolution algorithms for studying dynamics using MPS.
\YASTN additionally provides a framework to work with projected entangled-pair states (PEPS) using the \library{peps-torch} package, and there are some tensor-based renormalization group schemes that are provided, typically via examples.
As such these are mid- to high-level packages, providing the basic functionality to construct higher-level tensor network algorithms and including some of them, while resorting to lower-level libraries (such as \BLAS and \LAPACK, \CUDA, \PyTorch or \JAX) for mapping the elementary operations at the level of vectors, matrices or multidimensional arrays to dedicated processing units.

The Julia package \TensorKit presented in this manuscript is similarly a mid-level package, that implements operations on tensors while supporting a tensor structure that is determined by general abelian, non-abelian or categorical symmetries (which includes e.g.\ fermionic statistics).
It is designed to support various backends for more low-level matrix and array operations (matrix multiplication, linear algebra factorizations, array transposition) on the CPU and GPU, and strives to offer a complete set of pullback rules to support (reverse-mode) automatic differentiation (currently via the \library{ChainRules.jl} ecosystem, as supported by the \library{Zygote.jl} AD engine \cite{innes2018}).
The development of \TensorKit started around the end of 2017, and forms the basis for a number of packages that provide higher-level tensor network algorithms.
These include in particular \MPSKit, \PEPSKit and \TNRKit \cite{MPSKit2025,Brehmer2025,TNRKit2025}, the development of which was started in the Quantum Group of Ghent University (but with several external contributors since then), as well as \FiniteMPS \cite{Li2025}.
In the following sections, we discuss how we arrive from mathematical constructions to actual implementations of symmetries, vector spaces and tensors, and we conclude with some tensor contraction benchmarks.
\section{Fundamentals}
\label{sec:fundamentals}

Before diving into the complexities of symmetric tensors, it is beneficial to establish some general terminology, notations and conventions, as well as to define the tensor operations considered in this manuscript.
Essentially, we here define the public Application Programming Interface (API) of \TensorKit, and illustrate the implementations of these operations for ``plain'' tensors in the absence of any symmetry structure.
As will become clear in the following sections, the chosen interface is motivated by its compatibility with symmetries, but is otherwise agnostic to it.
The additional management that takes place (behind the scenes) to exploit and preserve the structure of tensors with symmetries is explained in the following sections. 


Concretely, this section provides the definition of a tensor for our purposes, a brief overview of the graphical notation prevalent in tensor network literature, and introduces the basic building blocks for tensor network algorithms, including index manipulations, contractions, and decompositions.
Readers already familiar with these concepts may choose to skip this section and proceed to the next, returning to reference specific examples as needed.

\subsection{What Is A Tensor?}
\label{sec:tensor}
Tensors are often introduced through abstract definitions, yet fundamentally they just extend the well-understood concepts of vectors and matrices.
This section aims to explore how tensors generalize these concepts, distilling properties that allow us to apply familiar linear algebra techniques.
Specifically, we focus on understanding both the vector-like and matrix-like properties of tensors, as the ability to transition between these interpretations forms the cornerstone of many subsequent operations.

To establish a suitable definition of tensors, consider the following.
\begin{definition}[Tensors as vectors]
    A rank-$N$ tensor $t$ is a vector in the tensor product space of $N$ vector spaces $V_1, \ldots, V_N$:
    \begin{equation}
        t \in V_1 \otimes \cdots \otimes V_N.
    \end{equation}
\end{definition}
In particular, since the tensor product of vector spaces is itself a vector space, this implies that a tensor \emph{is} a vector.
Consequently, tensors from the same tensor product space can be added together, or multiplied by scalars in the underlying field.
For simplicity, we will work with vector spaces over the complex numbers $\complexs$ throughout this manuscript, but note that \TensorKit also supports working over $\reals$, with real numbers.

Importantly, if a basis $\{\ket{i}, i=1,\ldots, \dim(V)\}$ is defined for each vector space $V$, a canonical method exists to construct a basis for their tensor product space.
This process essentially involves enumerating all possible combinations of the tensor products of these basis vectors.
\begin{equation}
\label{eq:basisexpansion}
    t \in V_1 \otimes \cdots \otimes V_N \implies t \equiv t_{i_1\ldots i_N} \ket{i_1} \cdots \ket{i_N}.
\end{equation}
Here, and in the following, we will assume the Einstein summation convention and implicitly sum over repeated labels.
This formulation also immediately yields a practical storage scheme.
Specifically, the coefficients of a tensor can be organized linearly, aligning with \Julia's $1$-based column-major storage scheme: 
\begin{equation}
\label{eq:linearindex}
    t_{i_1\ldots i_N} = t_I \quad \text{where} \quad I = 1 + \sum_{j=1}^N \left((i_j - 1)\prod_{k<j}\dim(V_k)\right).
\end{equation}

In addition to interpreting a tensor as a vector, it is also useful to view it as a linear map or matrix.
This perspective facilitates the definition of operations like compositions and decompositions.
To that end, we introduce the notion of a \emph{tensor map}.
\begin{definition}[Tensors as matrices]
    A rank-$(N_1,N_2)$ tensor map $t$ is a linear map from the tensor product of $N_2$ vector spaces $W_1,\ldots,W_{N_2}$ to the tensor product of $N_1$ vector spaces $V_1,\ldots,V_{N_1}$, where $W_1 \otimes \cdots \otimes W_{N_2}$ is called the \emph{domain} and $V_1 \otimes \cdots \otimes V_{N_1}$ is called the \emph{codomain} of the tensor map:
    \begin{equation}
        t \colon W_1 \otimes \cdots \otimes W_{N_2} \rightarrow V_1 \otimes \cdots \otimes V_{N_1}
    \end{equation}
\end{definition}
A \emph{tensor} of rank $N$ is then just a special case of an $(N, 0)$ tensor map.
This can be made more explicit by the identification of $\complexs \rightarrow V_1 \otimes \cdots \otimes V_N$ with $V_1 \otimes \cdots \otimes V_N$.

The basis of each individual vector space allows us to construct an explicit representation of tensor maps.
Specifically, the action of the linear map $t$ on the basis vectors of the domain is given by
\begin{equation}
    \ket{j_1}\cdots\ket{j_{N_2}} \mapsto t_{i_1 \ldots i_{N_1} ; j_1 \ldots j_{N_2}} \ket{i_1}\cdots\ket{i_{N_1}},
\end{equation}
where we use subscripts to index into the tensor elements.
This definition fully specifies the action of the tensor map, as it can be extended by linearity.
For clarity, our notation will make the distinction between domain and codomain indices explicit by separating them with a semicolon.

Furthermore, using \eqref{eq:linearindex} to define the rows $I$ and columns $J$ now produces a straightforward matrix representation.
Using this format, the composition of tensor maps is completely equivalent to matrix multiplication.
Thus, storing a tensor map in this form enables the use of high-performance linear algebra libraries.

Such operators are often represented using bra-ket notation:
\begin{equation}
\label{eq:homtobraket}
    t \equiv  t_{i_1 \ldots i_{N_1} ; j_1 \ldots j_{N_2}} \ket{i_1}\cdots\ket{i_{N_1}}\bra{j_{N_2}}\cdots\bra{j_1}.
\end{equation}
where, for a vector space $W$ with basis $\{\ket{j},j=1,\ldots,\dim(W)\}$, the set of dual vectors $\{\bra{j},j=1,\ldots,\dim(W)\}$ represent the canonical basis for the dual space $W^\ast$.
However, in the most general case, some care is required for implementing this equivalence between tensor maps $W_1 \otimes \cdots \otimes W_{N_2} \rightarrow V_1 \otimes \cdots \otimes V_{N_1}$ and tensors in $V_1 \otimes \cdots \otimes V_{N_1} \otimes W^*_{N_2} \otimes \cdots \otimes W^*_1$.
This is not necessarily a completely trivial operation, and it is elaborated upon in the following sections.

To manage the complexity of multiple indices and subscripts, the Penrose graphical notation or tensor network notation has been developed.
Here, we introduce the foundational rules of this notation.
As we progress through this chapter, we will elaborate on these rules and detail additional meanings, uses of the diagrams, and permitted manipulations.
Each tensor or tensor map is represented as a vertex in a graph, typically highlighted with a colored shape, and lines corresponding to the component vector spaces are added.

The flow from the domain to the codomain, and thus the direction of morphism composition $f \circ g$ (colloquially known as the flow of ``time'') could be chosen left to right (like the arrow in $f \colon V \rightarrow W$), right to left (like the composition order $f \circ g$ or the matrix product), bottom to top (quantum field theory convention) or top to bottom (quantum circuit convention).
Throughout this paper, we will stick to the latter convention.
Furthermore, the direction of the arrows on the lines, which only becomes relevant once we introduce duals, is also subject to convention and is here chosen to follow the arrow in $f \colon V \rightarrow W$, \ie pointing downward along the flow of time.
Finally, whenever we require a linear order of the indices, we will use the convention first to enumerate the indices of the codomain, followed by the indices of the domain.

\begin{figure}[ht]
\centering
\begin{subfigure}[c]{0.4\textwidth}
\centering
\includegraphics[scale=0.50,page=1]{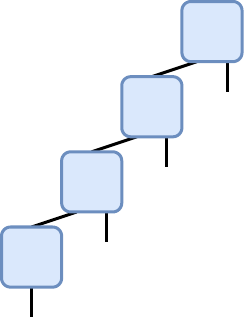}
\caption{}
\end{subfigure}
\begin{subfigure}[c]{0.4\textwidth}
\centering
\includegraphics[scale=0.4,page=2]{./figures/networks.pdf}
\caption{}
\end{subfigure}
\caption{Examples of common networks for quantum state ans\"atze. Panel (a) represents a Matrix Product State (MPS), while panel (b) is an example of a Multi-scale Entanglement Renormalization Ansatz (MERA).}
\label{fig:networks}
\end{figure}

This graphical representation is illustrated by a $(1, 0)$ tensor map (vector) $v$, a $(1, 1)$ tensor map (matrix) $A$ and a $(3, 2)$ tensor map $t$.
\begin{align}
    \diagram{networks}{3} = v \colon \complexs \rightarrow V,\qquad
    \diagram{networks}{4} = A \colon W \rightarrow V \\
    \diagram{networks}{5} = t \colon W_1 \otimes W_2 \rightarrow V_1 \otimes V_2 \otimes V_3 
\end{align}
Connections between lines in these diagrams indicate the composition of tensor maps, giving rise to the complex structures known as tensor networks.
For example, Fig.~\ref{fig:networks} includes some common networks in many-body physics.

\subsection{Index Manipulations}
\label{sec:indexmanipulations}

In discussing tensor operations, it is crucial to clarify that the equivalences between $W \rightarrow V$ and $V \otimes W^*$, as well as between $V \otimes W$ and $W \otimes V$, should not be viewed as mere identifications.
More accurately, these relationships are governed by isomorphisms that map between these interpretations.
Typically, these isomorphisms are acted with implicitly, especially when the vector spaces involved have no additional structure, effectively reducing the mapping to trivial identification.

However, in our explanation, we choose to make these isomorphisms explicit.
By precisely defining how these isomorphisms function and how to implement them, we establish both a theoretical and operational foundation for all the methods implemented in \TensorKit.
For completeness, we also link back to the more well-known cases that do not involve symmetries and show how our discussion provides a natural generalization of these cases.

\subsubsection{Grouping And Splitting Indices}

We begin by considering the ability to group and split the indices of a tensor map.
Given two vector spaces $V$ and $W$, we have seen that $V \otimes W$ itself forms a vector space.
Consequently, any isomorphism $V \otimes W \rightarrow Z$ with $Z \simeq V \otimes W$ establishes an equivalence between a tensor with two separate indices for $V$ and $W$, and a tensor with a single composite index in $Z$.

Merging indices involves the composition of a tensor with an isomorphism.
This process is reversible through the application of the inverse isomorphism.
The choice of isomorphism is arbitrary, and as a result there is no unique method for grouping.
Accordingly, splitting the indices requires knowledge of the prior operations applied to the tensor map, since the isomorphism used for splitting must be the inverse of the one used during grouping.

Although the choice of isomorphism is arbitrary, specific choices can offer computational advantages and can often be implemented at reduced or no additional cost compared to a generic isomorphism.
For efficient computation of inverses, it is typically advantageous to opt for unitary isomorphisms.
This consideration becomes crucial when discussing orthogonal tensor decompositions in \ref{sec:adjoints} and \ref{sec:decompositions}, where unitarity is a prerequisite.
For example, \eqref{eq:linearindex} can be seen as implementing such an isomorphism, where the matrix representation corresponds to the identity matrix.

In \TensorKit, we adhere to the convention of storing all tensor maps in their matrix representation.
This implies that every tensor map $t \colon W_1 \otimes \cdots \otimes W_{N_2} \rightarrow V_1 \otimes \cdots \otimes V_{N_1}$ is implicitly structured as the composition of a unitary isomorphism grouping the domain indices, a matrix representation of the data, and an inverse isomorphism splitting the codomain indices.
We often denote the matrix representation with $\tilde{t}$, but whenever it is clear from the context we also drop the tilde and simply write $t$.
\begin{equation}
\label{eq:tm_storage}
    t = G_V^\dagger \circ \tilde{t} \circ G_W \iff
    \diagram{indexmanipulations}{23} = \diagram{indexmanipulations}{24}\ .
\end{equation}

This results in the following canonical grouping isomorphism for non-symmetric tensors, which is equivalent to a reshape operation in the context of multi-dimensional arrays:
\begin{equation}
\label{eq:reshapemap}
    G_{I; i_1\ldots i_N} = \begin{cases}
        1 & I = 1 + \sum_{j=1}^N \left((i_j - 1)\prod_{k<j}\dim(V_k)\right) \\
        0 & \text{otherwise}
    \end{cases}
\end{equation}

Standardizing the isomorphisms used in our storage scheme eliminates the need to store them explicitly.
Instead, we maintain awareness of their presence and ensure that subsequent index manipulations are adjusted to account for them.
In particular, we only need to store the matrix representation $\tilde{t}$ and the data that characterizes the domain and codomain.

\subsubsection{Transpositions}
\label{sec:transpose}

In this subsection, we formalize the transition from $\complexs \rightarrow V \otimes W^*$ to $W \rightarrow V$.
Firstly, it is useful to recognize that the field $\complexs$ can be considered a trivial vector space over itself, with the property that $\complexs \otimes V = V \otimes \complexs = V$ for any other vector space $V$.
In diagrams, such vector spaces will be represented by dotted lines, which can be created and removed arbitrarily.

We then utilize the concept of duality in vector spaces. For a given vector space $W$, the dual space $W^*$ is given by the space of linear maps $W \to \complexs$. We denote elements of $W^*$ with a bra $\bra{\omega} \in W^*$, and the action of $\bra{\omega}$ on an vector $\ket{w} \in W$ as $\braket{\omega}{w} \in \complexs$.\footnote{It is important to note that $\braket{\omega}{w}$ does not denote an inner product, which we have not yet introduced at this point, and which we will denote as $\langle w_1,w_2\rangle$ for two elements $\ket{w_1}, \ket{w_2} \in W$.}

Because the action of a dual vector on a vector is a bilinear operation, we can reinterpret it as the linear map $\ev{W} \colon W^* \otimes W \rightarrow \complexs : \bra{\omega} \otimes \ket{w} \mapsto \braket{\omega}{w}$, which is known as the evaluation map and graphically depicted as:
\begin{equation}
    \ev{W} = \diagram{indexmanipulations}{60} = \diagram{indexmanipulations}{61}
\end{equation}

This enables the transition of an index from the codomain to the domain of a tensor map, or a \emph{right bend}:
\begin{equation}
    \diagram{indexmanipulations}{1} \mapsto \diagram{indexmanipulations}{2} = \diagram{indexmanipulations}{3}.
\end{equation}
To clarify, arrows can be added to indicate whether lines correspond to vector spaces (pointing downward) or their duals (pointing upward).
In the following, we opt to drop these arrows as much as possible to reduce visual clutter, only adding them whenever a distinction is necessary.

For finite-dimensional vector spaces, for any given basis $\{\ket{w_i}, i=1,\ldots,\dim(W)\}$ of $W$, there is a canonical dual basis $\{\bra{\omega_i}, i=1,\ldots,\dim(W^*)=\dim(W)\}$ of $W^*$, with the property that $\braket{\omega_i}{w_j} = \delta_{ij}$.
For the computational basis $\{\ket{i}, i=1,\ldots,\dim(W)\}$ with respect to which tensor data is stored, the elements of the canonical dual basis are simply denoted as $\bra{i}$ for $i=1,\ldots,\dim(W)\}$, as we had already used above in \eqref{eq:homtobraket}. We can now construct the special element $\coev{W} = \sum_{i} \ket{i} \otimes \bra{i}$ of $W \otimes W^*$, henceforth interpreted as a map $\coev{W} \colon \complexs \rightarrow W \otimes W^*$ and referred to as the coevaluation map,
\begin{equation}
    \coev{W} = \diagram{indexmanipulations}{62} = \diagram{indexmanipulations}{63}.
\end{equation}
that can easily be verified to be independent of the chosen basis.
Using this map, the line bending can be reversed
\begin{equation}
    \diagram{indexmanipulations}{1} = \diagram{indexmanipulations}{4}.
\end{equation}
Futhermore, $\coev{W}$ is required for the definition of the traditional \emph{matrix transpose}:
\begin{equation}
    A = \diagram{indexmanipulations}{11} \mapsto A^T = \diagram{indexmanipulations}{13} = \diagram{indexmanipulations}{12}.
\end{equation}
Indeed, the transpose of $A: W \to V$ is a map $A^T:V^* \to W^*$ and can be written in formulas as $A^T = (\ev{V} \otimes \id{W^*}) \circ (\id{V^*} \otimes A \otimes \id{W^*}) \circ (\id{V^*} \otimes \coev{W})$.
The complexity of even this simple operator already illustrates the usefulness of the graphical notation.

In general, there are four different maps $\ev{W}, \coev{W}, \ev{V^*}, \coev{V^*}$, where we make use of the relation $(V^*)^* \equiv V$ that holds for arbitrary finite-dimensional vector spaces. In particular, $\ev{V^*}:V \otimes V^\ast \to \complexs$ can be used to implement the right bend of a normal space $V$ from the codomain to the domain, or to left bend a dual space $V^*$ in the codomain of a tensor map.

For higher-rank tensor maps, multiple bending operations can be chained together to reconfigure the partitioning of indices, enabling the transition between different configurations of the $N_1 + N_2 = N_1^\prime + N_2^\prime$ indices.
\begin{equation}
    \diagram{indexmanipulations}{5} \mapsto \diagram{indexmanipulations}{7} = \diagram{indexmanipulations}{6}
\end{equation}
The composition of several (co)evaluation maps in this way naturally leads to the canonical choice $(V_1 \otimes \ldots \otimes V_N)^* \equiv V_N^* \otimes \ldots \otimes V_1^*$, which explains the reversal of the order of domain indices in the bras in \eqref{eq:homtobraket}.

The interaction between the storage scheme of \eqref{eq:tm_storage} and these tensor operations then consists of first bending the relevant lines, followed by a projection step.
This projection step ensures that the final result adheres to the chosen storage scheme.
As the isomorphisms of our storage scheme are unitary isomorphisms, this can be achieved efficiently by inverting them, as demonstrated below:
\begin{align}
    \diagram{indexmanipulations}{24} \mapsto \diagram{indexmanipulations}{25} &= \diagram{indexmanipulations}{26}
    &\implies
    \diagram{indexmanipulations}{27} &= \diagram{indexmanipulations}{28}
\end{align}
For completeness, we also add the corresponding coefficient notation, translating the diagrams back into equations.
In particular, the bending of a tensor map $A$ with grouping (splitting) morphisms $G_A$ and $G^\dagger_A$ into a tensor map $C$ with grouping (splitting) morphisms $G_C$ and $G^\dagger_C$ results in:
\begin{align}
\label{eq:bend_coeffs}
C_{I;J} = (G_C)_{I; i_1 i_2} \ev{i_3 j_4} (G_A^\dagger)_{i_1 i_2 i_3; K} A_{K; L} (G_A)_{L; j_1 j_2 j_3} (G_C^\dagger)_{j_1 j_2 j_3 j_4; J}
\end{align}

In cases where no additional structure is present in the vector spaces, the bending maps $\ev{}$ and $\coev{}$ and the grouping and splitting maps can be chosen as trivial.
As a result, \eqref{eq:bend_coeffs} simplifies, resulting in an expression equivalent to an index permutation of a multi-dimensional array:
\begin{equation}
    C_{I;J} \equiv C_{i_1 i_2; j_1 j_2 j_3 j_4} = A_{i_1 i_2 j_4; j_1 j_2 j_3} \equiv A_{I^\prime; J^\prime}
\end{equation}
This effectively becomes a reshape operation of the domain and codomain indices, followed by a transposition and another reshape.
Because of the linear memory layout of a computer, a reshape operation does not have to move any data, and is thus virtually free.
The remaining index permutation can be efficiently executed using high-performance array transposition libraries such as \Stridedjl, \HPTT and \TBLIS \cite{githubGitHubJuthoStridedjl,hptt2017,matthewsHighPerformanceTensorContraction2018}.

Using these bending operations, we can thus define a generalized tensor transpose that allows arbitrary cyclic permutations of the indices\footnote{More precisely, the general transposition of a tensor map $t_{i_1\ldots i_{N_1}; j_1 \ldots j_{N_2}}$ corresponds to a cyclic permutation of the indices ordered as $(i_1,\ldots,i_{N_1},j_{N_2},\ldots,j_1)$, followed by a novel bipartition of the resulting set into $N_1'$ codomain indices and $N_2'$ (reversely ordered) domain indices.
In the convention of Algorithm~\ref{alg:transpose}, this yields the requirement that $(p_1,\ldots, p_{N_1'}, q_{N_2'}, \ldots, q_1)$ is a cyclic permutation of $(1, 2, \ldots, N_1, N_1+N_2, N_1+N_2 -1, \ldots, N_1 + 1)$.}.
This operation forms the first algorithmic primitive for index manipulation, which is outlined for non-symmetric tensors in \algref{alg:transpose}.
Here, we used \verb|transpose| to denote the index permutation of the multi-dimensional arrays, dispatched to high-performance libraries.

It is important to note that our definition of a transpose operation does not fully specify how the lines are bent.
For example, in the definition of a transpose, it is also possible to bend domain indices to the left and codomain indices to the right.
In theory, this introduces the requirement to distinguish between left and right transposes.
However, for the purposes of \TensorKit, we will only consider scenarios where these operations yield the same result, with a more detailed exploration of these complications reserved for Section \ref{sec:category}.

\begin{algorithm}[ht]
\caption{Tensor Transpose}\label{alg:transpose}
\SetKwInOut{Input}{Inputs}\SetKwInOut{Output}{Output}
\SetKwFunction{transpose}{transpose}
\DontPrintSemicolon

\Input{A tensor map $A \colon W_1 \otimes \cdots \otimes W_{N_2} \rightarrow V_1 \otimes \cdots \otimes V_{N_1}$ and two tuples of integers $(p, q)$ with $|p| = N_1'$, $|q|=N_2'$ and $N_1'+N_2' = N_1 + N_2$}
\Output{A transposed tensor map $C$}
\BlankLine
\For{$i \leftarrow 1$ to $N_1'$}{
    $V^\prime_i \leftarrow$ $p_i^\text{th}$ vector space of $(V_1 \ldots, W_1^* \ldots)$\;
}
\For{$j \leftarrow 1$ to $N_2'$}{
    $W^\prime_j \leftarrow$ $q_j^\text{th}$ vector space of $(V_1^* \ldots, W_1 \ldots)$\;
}
Assign $C \colon W^\prime_1 \otimes \cdots \otimes W^\prime_{N_2'} \rightarrow V^\prime_1 \otimes \cdots \otimes V^\prime_{N_1'}$\;
$C \leftarrow$ \transpose{$A$, $(p\ldots, q\ldots)$}\;
\end{algorithm}

\subsubsection{Permutations}

To accommodate scenarios that require arbitrary permutations of the input and output spaces, we require an isomorphism $\tau \colon V \otimes W \rightarrow W \otimes V$.
This isomorphism enables the interchange of vector spaces and is depicted in diagrams as two crossing lines:
\begin{equation}
    \diagram{indexmanipulations}{14} \mapsto \diagram{indexmanipulations}{16} = \diagram{indexmanipulations}{15}.
\end{equation}

The permutation group $\Sym{N}$ is generated by the set of elementary swaps of adjacent indices: $\tau^{(1, 2)}, \tau^{(2, 3)}, \ldots, \tau^{(N-1, N)}$.
This implies that any permutation can be constructed from a series of these elementary swaps.\footnote{A practical algorithm to find these is found through a process similar to the bubble sort algorithm \cite{friendSortingElectronicComputer1956}.
Given a permutation $p$ of length $N$, this algorithm repeatedly iterates through that list, comparing and swapping adjacent entries.
This process is guaranteed to achieve the desired decomposition after at most $N$ sweeps through that list.}

Moreover, to allow for the permutation of indices between the domain and codomain, additional steps are necessary.
Using the transposition techniques of \ref{sec:transpose}, we first redirect all indices to the domain or codomain.
The permutation is then applied via a sequence of adjacent swaps.
Finally, the tensor map is reorganized back to the desired partition of indices.
Unlike the tensor transpose, this method enables an arbitrary reordering of the indices.
For example, we could have
\begin{equation}
    \diagram{indexmanipulations}{29} \mapsto \diagram{indexmanipulations}{31} = \diagram{indexmanipulations}{30}
\end{equation}

To further clarify and consider the interplay with our storage scheme, consider the following diagrammatic operation and its translation into index notation:
\begin{align}
    \diagram{indexmanipulations}{32} \mapsto \diagram{indexmanipulations}{34} &= \diagram{indexmanipulations}{33}
    & \implies 
    \diagram{indexmanipulations}{27} &= \diagram{indexmanipulations}{35}
\end{align}
\begin{equation}
    \label{eq:permute_coeffs}
    C_{I;J} = (G_C)_{I; i_1 i_2 i_3} 
        \ev{j_2^\prime j_2} 
        \tau_{i_3 j_2^\prime; k_2^\prime l_2^\prime}
        \tau_{i_2 k_2^\prime ; k_2 l_3^\prime}  
        (G_A^\dagger)_{i_1 k_2; K} A_{K; L} 
        (G_A)_{L; j_1 l_2 l_3}
        \ev{l_3 l_3^\prime}
        \ev{l_2 l_2^\prime} 
        (G_C^\dagger)_{j_1 j_2; J}
\end{equation}

In scenarios where no additional structure is imposed on the vector spaces, the isomorphism $\tau$ simply switches the positions of the elements of the tensor product, expressed as $\tau (v \otimes w) = w \otimes v$, or $\tau_{ij;kl} = \delta_{il}\delta_{jk}$.
Thus, \eqref{eq:permute_coeffs} again simplifies to a sequence of reshapes and a permutation of a multi-dimensional array.
\begin{equation}
    C_{I;J} \equiv C_{i_1 i_2 i_3; j_1 j_2} = A_{i_1 j_2; j_1 i_3 i_2} \equiv A_{I^\prime; J^\prime}
\end{equation}
This method constitutes the second algorithmic primitive for index manipulations: a generalized tensor permute, as outlined in \algref{alg:permute}.

The decomposition of a permutation into a series of adjacent swaps is not unique.
To maintain consistency, our framework assumes that the swapping operator $\tau$ is coherent, \ie\ that the result is independent of the decomposition path chosen.
In particular, the coherence also implies that for cyclic permutations, the result is independent of whether it is handled via the transpose operation (only applying evaluation and coevaluation maps) or via the permute operation (which might additionally apply a sequence of swap operations). 
More detailed considerations of these consistency requirements will be postponed to Section \ref{sec:category}.

\begin{algorithm}[ht]
\caption{Tensor Permute}\label{alg:permute}
\SetKwInOut{Input}{Inputs}\SetKwInOut{Output}{Output}
\SetKwFunction{permute}{permute}
\DontPrintSemicolon

\Input{A tensor map $A \colon W_1 \otimes \cdots \otimes W_{N_2} \rightarrow V_1 \otimes \cdots \otimes V_{N_1}$ and two tuples of integers $(p, q)$ with $|p| = N_1'$, $|q|=N_2'$ and $N_1'+N_2' = N_1 + N_2$}
\Output{A permuted tensor map $C$}
\BlankLine
\For{$i \leftarrow 1$ to $N_1'$}{
    $V^\prime_i \leftarrow$ $p_i^\text{th}$ vector space of $(V_1 \ldots, W_1^* \ldots)$\;
}
\For{$j \leftarrow 1$ to $N_2'$}{
    $W^\prime_j \leftarrow$ $q_j^\text{th}$ vector space of $(V_1^* \ldots, W_1 \ldots)$\;
}
Assign $C \colon W^\prime_1 \otimes \cdots \otimes W^\prime_{N_2'} \rightarrow V^\prime_1 \otimes \cdots \otimes V^\prime_{N_1'}$\;
$C \leftarrow$ \permute{$A$, $(p\ldots, q\ldots)$}\;
\end{algorithm}

\subsubsection{Traces}

Traces and partial traces represent specialized forms of index manipulations where indices within the same tensor map are connected.
This operation reduces the tensor's dimensionality by summing over specified index pairs, similar to taking the sum of diagonal elements in the context of matrices.

For a linear map $A \colon V \rightarrow V$, the trace operation yields a scalar representing the sum of diagonal elements.
This scalar is represented by a loop connecting the output and input of $A$:
\begin{equation}
\tr(A) = \sum_i A_{ii} = \diagram{indexmanipulations}{49}
\end{equation}
\begin{equation}
\tr(A) = \coev{;IK} A_{I;J} \ev{JK;}
\end{equation}
Note that we could have equally well connected the lines on the left side, as these operations are assumed equal, as discussed in section \ref{sec:transpose}.

To generalize to partial traces where an arbitrary subset of indices is contracted, we extend the operation to allow for the connection of arbitrary pairs of indices.
These traces are then resolved by first making these indices adjacent using the previously defined index manipulations and then pairing them up through the $\ev{}$ and $\coev{}$ maps.
Diagrammatically, this involves manipulating the tensor lines to be adjacent and then connecting them.
In our API of a partial trace of a tensor map, we also include a post-processing step to rearrange the remaining indices.\footnote{This addition makes it possible to support some libraries with specialized algorithms that allow for computing this in one step, which avoids the need of an intermediate tensor.}
This algorithmic primitive is outlined for non-symmetric tensors in \algref{alg:trace}.

As a concrete example, consider the following manipulations: 
\begin{equation}
    \diagram{indexmanipulations}{55} \mapsto \diagram{indexmanipulations}{51} = \diagram{indexmanipulations}{56}\ .
\end{equation}
As usual, the resulting operation for our storage scheme can be obtained using the following diagrammatic manipulations
\begin{align}
    \diagram{indexmanipulations}{57} \mapsto \diagram{indexmanipulations}{50} &= \diagram{indexmanipulations}{52}
    &\implies
    \diagram{indexmanipulations}{53} = \diagram{indexmanipulations}{54}
\end{align}
with corresponding index notation
\begin{equation}
\begin{aligned}
    C_{I;J} =& (G_C)_{I; k_4^\prime} \eta_{;k_1 k_2} \eta_{; k_3^\prime l_3^\prime} \tau_{k_4^\prime k_3^\prime; k_3 k_4} (G_A^\dagger)_{k_1 k_2 k_3 k_4; K} A_{K; L} \\
            & (G_A)_{L; l_1 l_2 l_3} \tau_{l_1 l_2; j_1 j_2} \eta_{l_3 l_36\prime;} (G_C^\dagger)_{j_1 j_2; J}\ .
\end{aligned}
\end{equation}

Again, whenever unstructured vector spaces are used, this definition coincides with the usual notion of partial traces, and the example given above reduces to:
\begin{equation}
    C_{I;J} \equiv C_{i_1; j_1 j_2} = A_{k k l i_1; j_2 j_1 l} .
\end{equation}

\begin{algorithm}[ht]

\caption{Tensor Trace}\label{alg:trace}
\SetKwInOut{Input}{Inputs}\SetKwInOut{Output}{Output}
\SetKwFunction{trace}{trace}
\SetKwFunction{permute}{permute}
\DontPrintSemicolon

\Input{A tensor map $A \colon W_1 \otimes \cdots \otimes W_{N_2} \rightarrow V_1 \otimes \cdots \otimes V_{N_1}$, two tuples of integers for tracing $(p^{\tau}, q^{\tau})$, and two tuples of integers for permuting $(p^{\pi}, q^{\pi})$}
\Output{A traced tensor map $C$}
\BlankLine
\For{$i \leftarrow 1$ to $|p^\pi|$}{
    $V^\prime_i \leftarrow$ $p^\pi_i {}^\text{th}$ vector space of $(V_1 \ldots, W_1^* \ldots)$\;
}
\For{$j \leftarrow 1$ to $|q^\pi|$}{
    $W^\prime_j \leftarrow$ $q^\pi_j {}^\text{th}$ vector space of $(V_1^* \ldots, W_1 \ldots)$\;
}
Assign $C \colon W^\prime_1 \otimes \cdots \otimes W^\prime_{|q^\pi|} \rightarrow V^\prime_1 \otimes \cdots \otimes V^\prime_{|p^\pi|}$\;
\BlankLine
\tcp{Adapt permutation tuples to account for traced indices}
$\tilde{p}^\pi \leftarrow$ $p^\pi$, shifted with indices from $p^\tau, q^\tau$ removed\;
$\tilde{q}^\pi \leftarrow$ $q^\pi$, shifted with indices from $p^\tau, q^\tau$ removed\;
$C \leftarrow$ \permute{\trace{$A$, $(p^\tau \ldots, q^\tau \ldots$)}, $(\tilde{p}^\pi \ldots, \tilde{q}^\pi \ldots)$}\;
\end{algorithm}

In summary, the ability to manipulate the indices of a tensor map is based on the existence of specific isomorphisms.
The interplay between vector spaces and their duals allows for the definition of a transposition map that bends lines to the left or right.
Finally, the existence of a swapping map permits the crossing of lines.

For the framework to remain consistent, certain compatibility conditions between these maps are required.
At this stage, we simply assume that these conditions are met, ensuring that sequences of manipulations that have the same final result share the same outcome.
A detailed discussion on the minimal set of necessary compatibility conditions is addressed in Chapter \ref{sec:category}.

\subsection{Contractions}

With the ability to manipulate the indices of tensor maps, we can define the essential primitives required for contracting networks of tensor maps.
We assert that to effectively evaluate a tensor network, only three basic operations are essential: pairwise contractions, outer products, and traces.
These operations can be combined in various ways to contract any tensor network configuration.

The functionality for these operations is encapsulated within the \TensorOperations library \cite{TensorOperations.jl}, which provides tools for generating code to evaluate tensor networks.
This library supports operations defined through the Einstein summation notation or the \library{NCON} convention \cite{pfeiferNCONTensorNetwork2014}, facilitating the most common approaches to specifying tensor networks.

The process of reducing a tensor network to these primitive operations is however not uniquely determined.
To ensure a consistent framework, it is crucial that the sequence of operations does not impact the final result, meaning all computation paths must yield the same output.
Again, this puts some constraints on the isomorphisms from Section \ref{sec:indexmanipulations}, which will be discussed in Section \ref{sec:category}.

Furthermore, it is important to note that even for a consistent framework, not all contraction paths are equal in terms of computational cost or memory requirements.
Identifying a contraction order that minimizes these costs, is known as the Tensor Network Contraction Order (TNCO) problem.
This combinatorial challenge has been classified as NP-hard in its most general form \cite{chi-chungOptimizingClassMultiDimensional1997,xuNPHardnessTensorNetwork2023}.
Despite this complexity, various strategies have been developed to tackle this problem, including efficient heuristics, graph-theoretic approaches, and even methods employing neural networks \cite{pfeiferFasterIdentificationOptimal2014,schindlerAlgorithmsTensorNetwork2020,meiromOptimizingTensorNetwork2022a}.

Currently, \TensorOperations supports the implementation of the exhaustive search algorithm described in \cite{pfeiferFasterIdentificationOptimal2014}, as well as a manual specification of order using the \library{NCON} syntax.
Nonetheless, the interface is sufficiently generic to allow easy integration of other and future algorithms designed to address the TNCO problem.

As tensor traces have already been discussed, we will now focus on the two remaining operations: pairwise contractions and outer products.
In particular, we will detail how both of these cases can be reduced to a simple composition of tensor maps, and how this operation can be efficiently implemented using high-performance linear algebra libraries.

\subsubsection{Compositions}

The composition of tensor maps $C = A \circ B$ is a special case of a pairwise contraction, where all domain indices of $A$ are contracted with all codomain indices of $B$.
In this scenario, the operation simplifies to regular matrix multiplication by making use of the unitarity of the splitting and grouping maps defined in \eqref{eq:tm_storage}.
The graphical representation of this operation makes this more clear:
\begin{equation}
    \diagram{indexmanipulations}{42} = \diagram{indexmanipulations}{43} \implies \diagram{indexmanipulations}{44} = \diagram{indexmanipulations}{45}
\end{equation}
Alternatively, the index notation for this operation is given by:
\begin{equation}
\label{eq:tm_composition}
\begin{aligned}
    C_{I;J} &= (G_C)_{I; i_1 i_2} (G_A^\dagger)_{i_1 i_2; I^\prime} A_{I^\prime ; K^\prime} (G_A)_{K^\prime; k_1 k_2 k_3} (G_B^\dagger)_{k_1 k_2 k_3; K} B_{K; J^\prime} (G_B)_{J^\prime; j_1 j_2} (G_C^\dagger)_{j_1 j_2; J} \\
    &= A_{I;K}B_{K;J}
\end{aligned}
\end{equation}

In particular, we find that the composition of tensor maps is equivalent to the matrix multiplication of their matrix representations.
This is one of the primary motivations for storing tensor maps in their matrix form, as it allows for the utilization of high-performance linear algebra libraries (typically \BLAS) to efficiently evaluate these multiplications.

\subsubsection{Pairwise Contractions}

To generalize this operation to arbitrary pairwise contractions, we incorporate pre-processing permutation steps for both input tensors $A$ and $B$ as well as a post-processing permutation step for $C$.
These extra steps are handled by the index manipulations of \ref{sec:indexmanipulations}.

Whenever no additional structure is present in the vector spaces, these pre- and post-processing steps reduce to simple transpositions of multi-dimensional arrays.
This operation is then analogous to the aforementioned Transpose-Transpose-GEMM-Transpose (TTGT), which can leverage specialized libraries for efficient execution.
This leads to the definition of the pairwise tensor contraction, which is another algorithmic primitive.
For non-symmetric tensors, this procedure is outlined in \algref{alg:contraction}.

There is a notable trade-off between two classes of algorithms that implement such schemes.
The first class, as described above, involves allocating intermediate objects that store the permuted tensor maps before contracting them.
Since this contraction then reduces to regular matrix multiplication, these methods enjoy the efficiency of the unrivaled BLAS routines for the theoretical bottleneck of the algorithm.
The second class aims to avoid these intermediate objects, as their allocation can have non-negligible impact on the runtime for certain regimes of tensor sizes.
This approach is comparatively newer but has shown promising results, with more and more algorithms emerging to rival the standard \BLAS implementations.
Additionally, hardware accelerators such as GPUs have also been shown to provide significant speedups for these types of operations.
Noteworthy libraries that implement these kinds of methods are \TBLIS, \library{TCCG} or \cuTENSOR \cite{matthewsHighPerformanceTensorContraction2018, springer2018design, springerCuTENSORHighPerformanceCUDA2019}.
For a more complete list of relevant software, we refer to \cite{psarrasLandscapeSoftwareTensor2022a}.

\TensorOperations accommodates some of these strategies through a backend selection mechanism, allowing users to dynamically choose between different implementations based on their specific requirements.
This feature is inherited and supported by \TensorKit, thus allowing for easy experimentation with new and emerging developments in this field.
In particular, \cuTENSOR is supported out-of-the-box and there is a wrapper library for \TBLIS called \library{TBLIS.jl}.

\begin{algorithm}[H]
\caption{Tensor Contraction}\label{alg:contraction}
\SetKwInOut{Input}{Inputs}\SetKwInOut{Output}{Output}
\SetKwFunction{permute}{permute}
\DontPrintSemicolon

\Input{A tensor map $A$ with permutation indices $(p^A, q^A)$, a tensor map $B$ with permutation indices $(p^B, q^B)$ and permutation indices for their resulting contraction $(p^{AB}, q^{AB})$}
\Output{The contracted tensor map $C$}
\BlankLine

$\tilde{A}$ $\leftarrow$ \permute{$A$, $(p^A, q^A)$}\;
$\tilde{B}$ $\leftarrow$ \permute{$B$, $(p^B, q^A)$}\;
$C$ $\leftarrow$ \permute{$\tilde{A} \circ \tilde{B}$, $(p^{AB}, q^{AB})$}
\end{algorithm}

\subsubsection{Outer Products}

One notable case occurs when contracting two tensor maps without pairing up any indices.
Because the underlying field $\complexs$ acts as a trivial vector space, we are allowed to insert an additional auxiliary (trivial) index to both tensor maps and perform the contraction as in the other cases.

The simplest example consists of considering the outer product of a vector and a covector, \ie\ $A \colon \complexs \rightarrow V$ and $B \colon W \rightarrow \complexs$.
In this case, the outer product of these tensor maps reduces to the outer product of their matrix representations, which are single-column and single-row matrices, respectively.
This scenario remains largely unchanged when tensor product spaces are involved, as illustrated by the graphical representation:
\begin{equation}
    \diagram{indexmanipulations}{46} = \diagram{indexmanipulations}{47} \implies \diagram{indexmanipulations}{44} = \diagram{indexmanipulations}{48}
\end{equation}

This operation can be generalized to include pre- and post-processing steps that manipulate the indices, therefore also allowing the tensor product of tensor maps.
Whenever the vector spaces are unstructured, this operation coincides with the usual definition of the Kronecker product.


\subsection{Orthogonality and Adjoints}
\label{sec:adjoints}

In most of physics, the relevant vector spaces come with the additional well-know structure of having an inner product, and thus a notion of orthogonality.

This will also be an essential property for many of the decompositions discussed in the next subsection.

The inner product on a vector space $V$ is denoted as $\langle \cdot, \cdot \rangle_V \colon V \otimes V \rightarrow \complexs$ and is, using the physics convention, linear in the second argument and antilinear in the first.
Given an inner product on the vector spaces $V$ and $W$, one can also introduce the adjoint of a linear map $A \colon W \rightarrow V$ as the linear map $A^\dagger \colon V \rightarrow W$ characterized by the property:
\begin{equation}
    \langle v, A w \rangle_V = \langle A^\dagger v, w \rangle_W, \qquad \forall v \in V, w \in W.
\end{equation}
This is an anti-linear involution, $(A^\dagger)^\dagger = A$ and $(\lambda A + \mu B)^\dagger = \bar{\lambda} A^\dagger + \bar{\mu} B^\dagger$, that satisfies $(A \circ B)^\dagger = B^\dagger \circ A^\dagger$.

The adjoint mapping exchanges the domain and codomain.
We can therefore graphically represent it as a reflection along the horizontal axis.
However, the arrows maintain their original direction, in this case downwards, as there is no line-bending involved here:
\begin{equation}
    \dagger\, \colon A = \diagram{indexmanipulations}{17} \mapsto \diagram{indexmanipulations}{18} = \diagram{indexmanipulations}{19} = A^\dagger
\end{equation}

It is typically assumed that an orthnormal basis has been chosen, such that the inner product takes the well-known Euclidean form
\begin{equation}
    \langle v, w \rangle \coloneqq \overline{v_i} w_i,
\end{equation}
where $\overline{v_i}$ denotes complex conjugation to avoid confusion with the dual spaces.
As a result, the matrix representation of the adjoint map $A^\dagger$ is then the Hermitian conjugate of the matrix representation of $A$, \ie
\begin{equation}
    A^\dagger_{i;j} = \overline{A_{j;i}}.
\end{equation}

Maps $A \colon W \rightarrow V$ that preserve the inner product are called \emph{isometric maps}, or simply isometries, and thus satisfy
\begin{equation}
    \langle A \circ w_1, A \circ w_2 \rangle = \langle v_1, v_2 \rangle, \qquad \forall w_1, w_2 \in W
\end{equation}
which leads to the condition
\begin{equation}
A^\dagger \circ A = \id{W}
\end{equation}
Furthermore, if $A^\dagger$ is also isometric, \ie~$A \circ A^\dagger = \id{V}$, these maps are called \emph{unitary}.

Unitarity is also useful to extend the adjoint to tensor maps $t \colon W_1 \otimes \cdots \otimes W_{N_2} \rightarrow V_1 \otimes \cdots \otimes V_{N_1}$.
Making use of the matrix representation of the map, \ie~$t = G_V^\dagger \circ \tilde{t} \circ G_W$, we can efficiently compute the adjoint of $t$ because the grouping and splitting isomorphisms are chosen to be unitary:
\begin{equation}
\label{eq:adjoint_tm}
    t^\dagger = \left(G_V^\dagger \circ \tilde{t} \circ G_W\right)^\dagger = G_W^\dagger \circ \tilde{t}^\dagger \circ G_V.
\end{equation}
Crucially, the right-hand side of \eqref{eq:adjoint_tm} is the same expression as the matrix representation of $t^\dagger \colon V_1 \otimes \cdots \otimes V_{N_1} \rightarrow W_1 \otimes \cdots \otimes W_{N_2}$.
In other words, the adjoint of an arbitrary tensor map is obtained by taking the Hermitian conjugate of the matrix representation, on the condition that the grouping and splitting isomorphisms are unitary.
Note also that this implies that unitary maps have unitary matrix representations.
This operation is graphically represented as
\begin{align}
    \dagger \colon \diagram{indexmanipulations}{36} \mapsto \diagram{indexmanipulations}{38} &= \diagram{indexmanipulations}{37}
    &\implies 
    \diagram{indexmanipulations}{39} &= \diagram{indexmanipulations}{40} = \diagram{indexmanipulations}{41}.
\end{align}

It is important to note that this operation modifies the linear order of the indices differently compared to the transpose operation.
In graphical terms, the adjoint operation reflects a diagram about the horizontal axis, while the transpose operation corresponds to a rotation of the diagram by 180 degrees.
Additionally, the adjoint exchanges codomain and domain, while the transpose uses the duality to permute the vector spaces.
As a result, both the order of the indices and the arrows of the resulting maps are transformed rather differently.
Indeed, for a tensor map $t \colon W_1 \otimes \cdots \otimes W_{N_2} \rightarrow V_1 \otimes \cdots \otimes V_{N_1}$, the adjoint is the tensor map 
$t^\dagger \colon V_1 \otimes \cdots \otimes V_{N_1} \rightarrow W_1 \otimes \cdots \otimes W_{N_2}$ whereas the natural generalization of the matrix transpose yields 
$t^T \colon V_{N_1}^* \otimes \cdots \otimes V_{1}^* \rightarrow W_{N_2}^* \otimes \cdots \otimes W_{1}^*$.
To aid in distinguishing these operations in the diagrammatic notation, it can often be helpful to introduce some asymmetry in the shape representing the tensor map:
\begin{equation}
    t = \diagram{indexmanipulations}{20},\quad t^\dagger = \diagram{indexmanipulations}{21}, \quad t^T = \diagram{indexmanipulations}{22}\ .
\end{equation}



\subsection{Decompositions}
\label{sec:decompositions}

The final class of tensor operations are tensor decompositions, which are complementary to tensor contractions.
Where a contraction takes a network of several tensors and produces a single tensor, a decomposition takes a single tensor and splits it into several parts.
These decompositions are crucial for creating orthogonal bases, finding low-rank approximations, solving linear equations, and computing eigenvalues and eigenvectors.
As such, they are an essential part of any tensor toolbox.
We focus on a few well-known decompositions, but the same techniques can be applied to develop other types of decompositions.

In particular, these techniques fundamentally depend on the interpretation of tensor maps as linear maps from their domain to codomain.
We assert that directly applying the decomposition to their matrix representation and combining it in a proper way with the grouping and splitting isomorphisms, yields a tensor decomposition with the expected properties.
For convenience, each of those decompositions can be further extended to include arbitrary index manipulations that reorganize the domain and codomain, but these are mere pre- and postprocessing steps.

\subsubsection{Eigenvalue Decomposition}

The eigen decomposition of a (square) matrix $A$ is a decomposition of the form
\begin{equation}
\label{eq:eigenvalue}
    A = V \Lambda V^{-1} \iff
    \diagram{factorizations}{3} = \diagram{factorizations}{4}
\end{equation}
where each column of $V$ is an eigenvector of $A$, and $\Lambda$ is a diagonal matrix of the corresponding eigenvalues\footnote{Note that not all square matrices admit an eigenvalue decomposition and one should in principle use the Jordan decomposition instead.
From a numerical perspective, however, the set of non-diagonalizable matrices has measure zero and the distinction between diagonalize and non-diagonalizable matrices cannot be made in the presence of finite numerical precision.
As such, in floating-point arithmetic, any matrix is assumed to be diagonalizable, but the resulting matrix of eigenvectors might have a vanishingly small determinant and thus linearly dependent columns of eigenvectors when a nontrivial Jordan structure is present.}.
For normal matrices, \ie\ matrices satisfying $A^\dagger A = A A^\dagger$, the eigenspaces are orthogonal and the matrix $V$ can be made unitary.


The eigenvalue decomposition extends to tensor maps with equal domain and codomain.
It can then be noted that if we take the eigenvalue decomposition of the matrix representation of the tensor map, and compose the matrix of eigenvectors with the splitting isomorphism, we obtain the proper generalization of the eigenvalue decomposition for tensor maps.
Hereto, we make use of the fact that the grouping and splitting isomorphisms are also inverses to each other.
For example, for a tensor map $A \colon W_1 \otimes W_2 \rightarrow W_1 \otimes W_2$, we obtain the eigenvectors $V \colon W \rightarrow W_1 \otimes W_2$ with $W \simeq W_1 \otimes W_2$ and the eigenvalues $\Lambda \colon W \rightarrow W$ by
\begin{equation}
\begin{aligned}
    \tilde{A} = \tilde{V} \circ \Lambda \circ \tilde{V}^{-1}
        \iff G_W^\dagger \circ \tilde{A} \circ G_W &= G_W^\dagger \circ \tilde{V} \circ \Lambda \circ \tilde{V}^{-1} \circ G_W
        \iff A = V \circ \Lambda \circ V^{-1} \\
    \diagram{factorizations}{3} = \diagram{factorizations}{4}
        \iff \diagram{factorizations}{1} &= \diagram{factorizations}{2}
        \iff \diagram{factorizations}{5} = \diagram{factorizations}{6}\ .
\end{aligned}
\end{equation}
In particular, this shows that this decomposition can be implemented directly on the matrix representation of the tensor map.
For a rank-$(N,N)$ tensor map $A$, the rank-$(N,1)$ tensor map $V$ inherits the splitting isomorphism of the tensor.
The domain of $V$, as well as the domain and codomain of the rank-$(1,1)$ tensor map $\Lambda$ is a single space $W$ that is isomorphic to the (co)domain $A$, but can itself not be given a tensor product structure and does therefore not come with additional grouping and splitting isomorphisms.
Indeed, it is only for a rank $(1,1)$ tensor map with no grouping and splitting isomorphisms that a diagonal matrix representation is meaningfully defined. 

\subsubsection{Singular Value Decomposition}

The (full) singular value decomposition (SVD) of a rectangular $(m \times n)$ matrix is a decomposition of the form
\begin{equation}
    \label{eq:svd}
    A = U \Sigma V^\dagger \iff \diagram{factorizations}{11} = \diagram{factorizations}{10}
\end{equation}
where $U$ and $V^\dagger$ are unitary matrices of size $(m \times m)$ and $(n \times n)$, respectively, and the $(m \times n)$ matrix $\Sigma$ only has nonzero entries on the diagonal---the singular values---which are non-negative and non-increasing along the diagonal.\footnote{
The definition of the SVD in terms of $V^\dagger$ is both a historical convention, as well as a result of how these are typically computed.
Linear algebra libraries often directly return the matrix representing $V^\dagger$ instead of $V$.
}
Here, the number of non-zero singular values is equal to the \emph{rank} of the matrix (not to be confused with the rank of a tensor map, indicating the number of spaces in the domain and codomain).
This decomposition is typically used to find low-rank approximations of a matrix, and it can be shown that the best rank-$k$ approximation is obtained through the SVD \cite{eckartApproximationOneMatrix1936}.
This decomposition can also be used to determine null spaces, or as a starting point for other decompositions, such as the polar decomposition.

As the grouping and splitting isomorphisms are chosen unitary in inner product spaces, we can once again extend the SVD to tensor maps by directly acting on the matrix representation. For a tensor map $A \colon W_1 \otimes W_2 \rightarrow Z_1 \otimes Z_2 \otimes Z_3 $ and with $W \simeq W_1 \otimes W_2$ and $Z \simeq Z_1 \otimes Z_2 \otimes Z_3$, we obtain unitary tensor maps $U \colon Z \rightarrow Z_1 \otimes Z_2 \otimes Z_3$ and $V^\dagger \colon W_1 \otimes W_2 \rightarrow W$, along with a rank-(1,1) tensor map $\Sigma \colon W \rightarrow Z$ with nonnegative real entries on the diagonal, via
\begin{equation}
\begin{aligned}
    \tilde{A} = \tilde{U} \circ \Sigma \circ \tilde{V}^\dagger
        \iff G_Z^\dagger \circ \tilde{A} \circ G_W &= G_Z^\dagger \circ \tilde{U} \circ \Sigma \circ \tilde{V}^\dagger \circ G_W
        \iff A = U \circ \Sigma \circ V^\dagger \\
    \diagram{factorizations}{11} = \diagram{factorizations}{10}
        \iff \diagram{factorizations}{15} &= \diagram{factorizations}{9}
        \iff \diagram{factorizations}{7} = \diagram{factorizations}{8}\ .
\end{aligned}
\end{equation}
Again, no tensor product structure exists on the new spaces $W$ and $Z$. For a rank $(N_1,N_2)$ tensor map $A$, $U$ [$V^\dagger$] is a rank $(N_1,1)$ [$(1,N_2)$] tensor map that inherits the splitting [grouping] isomorphism of $A$ in the codomain [domain], whereas $\Sigma$ is a rank $(1, 1)$ diagonal tensor map.

\subsubsection{QR Decomposition}

The QR decomposition is another decomposition that is commonly used for constructing orthonormal bases or in the context of solving (overconstrained) linear equations.
For an $(m \times n)$ matrix $A$, the QR decomposition takes the form
\begin{equation}
    A = QR \iff \diagram{factorizations}{11} = \diagram{factorizations}{14}
\end{equation}
where the $(m \times \min(m,n))$ matrix $Q$ is isometric and the $(\min(m,n) \times n)$ matrix $R$ is upper triangular.
In particular, the columns of $Q$ provide an orthonormal subspace for the image of $A$ (assuming that $A$ has full matrix rank).
Analogously, there exist an LQ decomposition $A = L Q$ where $L$ is lower triangular, and the rows of $Q$ provide an orthonormal basis for the image of $A^\dagger$.


Again, the extension to tensor maps follows from the matrix representations and the unitarity of the grouping and splitting isomorphisms.
For a tensor map $A \colon W_1 \otimes W_2 \rightarrow Z_1 \otimes Z_2 \otimes Z_3 $ and with $Z \simeq Z_1 \otimes Z_2 \otimes Z_3$, this results in an isometric tensor map $Q \colon Z \rightarrow Z_1 \otimes Z_2 \otimes Z_3$ and a tensor map $R \colon W_1 \otimes W_2 \rightarrow Z$ with $Z \simeq Z_1 \otimes Z_2 \otimes Z_3$:
\begin{equation}
\begin{aligned}
    \tilde{A} = \tilde{Q} \circ \tilde{R}
        \iff G_Z^\dagger \circ \tilde{A} \circ G_W &= G_Z^\dagger \circ \tilde{Q} \circ \tilde{R} \circ G_W
        \iff A = Q \circ R \\
    \diagram{factorizations}{11} = \diagram{factorizations}{14}
        \iff \diagram{factorizations}{15} &= \diagram{factorizations}{13}
        \iff \diagram{factorizations}{7} = \diagram{factorizations}{12}\ .
\end{aligned}
\end{equation}
For a $(N_1, N_2)$ tensor map $A$, $Q$ is an isometric $(N_1,1)$ tensor map that absorbs the splitting isomorphism of $A$ along its codomain, and $R$ is a $(1, N_2)$ tensor map that absorbs the grouping isomorphism of $A$. Since in this case there is a (possibly nontrivial) isomorphism between $R$ and its matrix representation $\tilde{R}$, one can wonder what the meaning is of the upper triangular nature of $\tilde{R}$. However, unlike the diagonal structure of the eigenvalues or singular values, there is nothing fundamental about the upper triangular nature of the second factor in the QR decomposition. This structure has a merely computational origin, and many other decompositions exist which split a $(m \times n)$ matrix $A$ into an isometric matrix containing an orthonormal basis for the image of $A$, and a remaining factor. For the typical case $m \geq n$, it is for example also possible to make the second factor lower triangular, known as the QL decomposition. Different decompositions simply correspond different choices for the orthonormal basis for the image of $A$ that is being constructed. Among these, the QR decomposition can be computed very efficiently. 

One notable alternative choice for the case $m \geq n$ is the polar decomposition, where the second factor is a positive semidefinite matrix, and can for example be constructed from the (thin) SVD decomposition as
$A = U \Sigma V^\dagger = (U V^\dagger) (V \Sigma V^\dagger)$. In the generalization to tensor maps, this is the only choice that can meaningfully introduce a tensor product structure on the new space that appears in the decomposition. In this case, the isometric factor has exactly the same domain and codomain as the original tensor, and the positive semidefinite factor has both codomain and domain equal to the original domain.
\section{Abelian Symmetries}
\label{sec:abeliansymmetries}

Having established our framework's notation and general operations, we now turn our attention to the role of symmetries.
Symmetries intuitively imply some kind of order, or structure.
Here, we try to formalize this intuition and clarify the precise nature of the additional structure symmetries introduce.

This chapter delves into the implications of symmetries on the operations defined in Section \ref{sec:fundamentals}.
For clarity and manageability, we initially focus on adapting our framework to accommodate abelian symmetry groups, deferring the discussion of non-abelian groups to Section \ref{sec:non-abelian}.

\subsection{What Is A Symmetry?}

Symmetry, in a broad sense, refers to the invariance or preservation of certain properties under a set of transformations.
These can range from geometric transformations, such as rotations and reflections, to more abstract transformations that affect internal degrees of freedom.

To articulate this concept more precisely, we define \emph{invariance} as the scenario where a function remains unchanged under the actions of a group.
Specifically, if we denote the action of a group element $g \in \group{G}$ on a vector $v \in V$ as $g \triangleright v$, we find that a linear map $f \colon V \rightarrow \complexs$ is invariant under this action if:
\begin{equation}
    f(g \triangleright v) = f(v), \quad \forall g \in \group{G} \text{ and } v \in V.
\end{equation}

Furthermore, \emph{equivariance} refers to a function that respects the group structure imposed on the vector space.
Formally, a map $f \colon W \rightarrow V$ is equivariant under the action of $\group{G}$ if:
\begin{equation}
    f(g \triangleright w) = g \triangleright f(w), \quad \forall g \in \group{G} \text{ and } w \in W.
\end{equation}
This intertwining property of equivariance is crucial because it guarantees that for a computation composed of several steps, the transformations applied at each step are coherent.
As a consequence, the final result is invariant under the group action.

More precisely, a group action on a vector space is known as a representation\footnote{We use the term \emph{representation} quite loosely, to denote both the map $\rho \colon \group{G} \rightarrow \End(V)$, as well as the more typical combination of $(\rho, V)$.
The exact meaning should be clear from the context.} $\rho \colon \group{G} \rightarrow \End(V)$, which maps group elements to (invertible) linear maps such that the group structure is respected, \ie~if $g = g_1 \circ g_2$, then $\rho(g) = \rho(g_1) \circ \rho(g_2)$.
This is important in the context of symmetric tensor networks, where the combination of the group action and equivariance facilitates an intuitive diagrammatic representation:
\begin{equation}
     g \triangleright \diagram{equivariance}{4} 
        = \diagram{equivariance}{1}
        = \diagram{equivariance}{2}
        = \diagram{equivariance}{3}
        = \diagram{equivariance}{4}\ .
\end{equation}

For compact groups $\group{G}$, all finite-dimensional representations can be chosen unitary, such that each element of the group $g$ acts through a unitary map $\rho(g)$.
Furthermore, any unitary representation $\rho_V$ on a vector space $V$ can be decomposed into a direct sum of irreducible representations (irreps) $\rho^{(a)}$, which we label by a \emph{charge} $a$.
This implies that the space $V$ admits a graded direct sum structure
\begin{equation}
    V \eqsim \bigoplus_a \left( \bigoplus_{i=1}^{\NSymbol{V}{a}} \gradedspace{V}{a}_i \right) \eqsim \bigoplus_a \complexs^{\NSymbol{V}{a}} \otimes \gradedspace{V}{a}\ ,
\end{equation}
where $\NSymbol{V}{a}$ is the number of times that irrep $a$ appears in the decomposition of $\rho_V$, henceforth called the degeneracy of irrep $a$, and $\gradedspace{V}{a}_i$ denotes the $i$th copy of the representation space $\gradedspace{V}{a}$ associated with irrep $a$.
In this decomposition, $\rho_V$ acts as
\begin{equation}
\label{eq:gradedrepresentation}
\rho_V(g) \eqsim \bigoplus_a \id{\NSymbol{V}{a}} \otimes \rho^{(a)}(g)\ .
\end{equation}

The most important property of irreps is the following \cite{serreLinearRepresentationsFinite2012}:
\begin{lemma}[Schur's lemma]
    Given two irreducible representations $\rho_V$ and $\rho_W$ and a linear map $A \colon W \rightarrow V$ between the respective representation spaces, then $\rho_V(g) \circ A = A \circ \rho_W(g)$ for all $g \in \group{G}$ implies one of the following:
    \begin{itemize}
        \item If $\rho_V$ is isomorphic to $\rho_W$, then $A = \lambda I$, where $\lambda$ is a scalar and $I$ is the (unitary) isomorphism between the irreducible representations.
        \item If $\rho_V$ is not isomorphic to $\rho_W$, then $A = 0$.
    \end{itemize}
\end{lemma}
This lemma implies that any equivariant linear map $A \colon W \rightarrow V$ between general representation spaces $V$ and $W$ is block-diagonalizable, with each block associated with an irreducible representation appearing in both the domain and codomain.
Specifically, this entails that with respect to the direct sum decomposition of $V$ and $W$, the matrix block
\begin{equation}
    \gradedtensor{A}{i;j}{a; b} \colon \gradedspace{W}{b}_j \rightarrow \gradedspace{V}{a}_i
\end{equation}
is non-zero only if $a \simeq b$, and in this case is a multiple of the isomorphism $\gradedspace{V}{b} \rightarrow \gradedspace{V}{a}$.

Henceforth, we assume a fixed (basis) choice for the irreps, such that two irreps $a$ and $b$ are either equal ($a=b$) or non-isomorphic for $a \neq b$, \ie\ the label $a$ will range over the set of equivalence classes of irreps and a fixed basis choice $\gradedspace{V}{a} = \mathop{span}\{\gradedket{m}{a},m=1,\ldots,d_a\}$, with $d_a$ the dimension of the representation.
We furthermore build every general representation space directly as $V=\bigoplus_a \complexs^{\NSymbol{V}{a}} \otimes \gradedspace{V}{a}$, as spanned by a basis
\begin{equation}
\label{eq:gradedbasis}
    \gradedket[m]{i}{a} \coloneqq \gradedket{i}{a} \otimes \gradedket{m}{a},  \qquad i \in \{1, \ldots, \NSymbol{V}{a}\}, \quad m \in \{1, \ldots, d_a\}\ .
\end{equation}
We will refer to the label $i$ as an \emph{outer} index, labeling the degeneracy space, whereas the index $m$ labeling the basis of the representation space is referred to as the \emph{inner} index.

The equivalence in \eqref{eq:gradedrepresentation} then becomes an equality, and an equivariant linear map $A: W\to V$ between $W=\bigoplus_a \complexs^{\NSymbol{W}{a}} \otimes \gradedspace{V}{a}$ and $V=\bigoplus_a \complexs^{\NSymbol{V}{a}} \otimes \gradedspace{V}{a}$ acquires the explicit block-diagonal form
\begin{equation}
\label{eq:gradedlinearmap}
    A = \bigoplus_a \gradedtensor{A}{}{a} \otimes \id{\gradedspace{V}{a}}
\end{equation}
with $\gradedtensor{A}{}{a}$ a matrix of size $\NSymbol{V}{a} \times \NSymbol{W}{a}$, also called the \emph{reduced matrix coefficients}.

The resulting block-diagonal structure of linear maps, ultimately dictated by Schur's lemma, is not merely of theoretical interest but has practical implications as well.
This structural property allows for the decomposition of linear algebra algorithms into simpler, smaller sub-tasks that are computationally more manageable.
For instance, consider the matrix multiplication of two $n \times n$ matrices, which generally requires $\mathcal{O}(n^3)$ operations.
If these matrices are block-diagonal, divided into $k$ equal blocks of size $\frac{n}{k} \times \frac{n}{k}$, the total computational cost effectively reduces to $\mathcal{O}(k \cdot \left(\frac{n}{k}\right)^3) = \mathcal{O}(\frac{n^3}{k^2})$, offering a significant reduction when $k$ is large.
Similar considerations apply to matrix decompositions.
Furthermore, this approach is amenable to multi-threaded implementations, as each block can be processed independently.

As we henceforth generalize this discussion to higher-rank tensor maps, we will be particularly interested in the decomposition of tensor product representations.
Starting from two irreps $a$ and $b$, the tensor product $\rho^{(a)} \otimes \rho^{(b)}$ acting on $\gradedspace{V}{a} \otimes \gradedspace{V}{b}$ is a representation that may no longer be irreducible.
This is denoted using the \emph{fusion rules}
\begin{equation}
\label{eq:fusionproduct}
    a \otimes b \mapsto \bigoplus_c \NSymbol{ab}{c} ~ c.
\end{equation}
which is a symbolic representation of
\begin{equation}
\gradedspace{V}{a} \otimes \gradedspace{V}{b} \eqsim \bigoplus_c \complexs^{\NSymbol{ab}{c}} \otimes \gradedspace{V}{c}
\end{equation}
at the level of the vector spaces, or
\begin{equation}
\rho^{(a)}(g) \otimes \rho^{(b)}(g) \eqsim \bigoplus_c \id{\NSymbol{ab}{c}} \otimes \rho^{(c)}(g)
\end{equation}
for the corresponding representations.
These equations hold up to a similarity transform, that will be elaborated upon in the next section.
For the remainder of this section, we restrict to abelian groups, where all irreps are one-dimensional.
Then the tensor product representation $a \otimes b$ is also one-dimensional, and thus irreducible, corresponding to a charge that we denote as $c = a b$, \ie\ $\NSymbol{ab}{c} = \delta_{c,ab}$.

\subsection{Index Manipulations}
This section explores how symmetries impose structural properties on the tensor maps in our framework, and how this influences the different index manipulations of these tensors when graded vector spaces are involved.

As mentioned before, we limit our focus to abelian groups for simplicity during the remainder of this section.
As all irreducible representations are one-dimensional, we can omit inner index labels [index $m$ in \eqref{eq:gradedbasis}], as well as the identity maps $\id{\gradedspace{V}{a}}$ in the context of \eqref{eq:gradedlinearmap}.
We adapt our index notation to account for this simplification, using $\gradedket{i}{a}$ to denote basis vectors, allowing for a tensor $t$ to be expanded as:
\begin{equation}
\label{eq:abeliantensor}
\begin{aligned}
    t &\in V_1 \otimes \cdots \otimes V_N \\
    t &\equiv t_{i_1\ldots i_N}^{(a_1\ldots a_N)} \gradedket{i_1}{a_1} \cdots \gradedket{i_N}{a_N} \ .
\end{aligned}
\end{equation}
For tensor maps over graded spaces, we define:
\begin{equation}
\label{eq:abeliantensormap}
\begin{aligned}
    t \colon W_1 \otimes \cdots \otimes W_{N_2} &\rightarrow V_1 \otimes \cdots \otimes V_{N_1} \\
    \gradedket{j_1}{b_1}\cdots\gradedket{j_{N_2}}{b_{N_2}} &\mapsto \gradedtensor{t}{i_1 \ldots i_{N_1} ; j_1 \ldots j_{N_2}}{a_1 \ldots a_{N_1} ; b_1 \ldots b_{N_2}} \gradedket{i_1}{a_1}\cdots\gradedket{i_{N_1}}{a_{N_1}}\ .
\end{aligned}
\end{equation}

We can now make use of Schur's lemma, from which we derive additional constraints on these coefficients. The uniqueness of the fusion product $a \otimes b$ for one-dimensional irreps makes this particularly simple and yields the following constraints on the coefficients of the tensors:
\begin{equation}
    \bigotimes_i a_i \not\simeq \bigotimes_j b_j \implies
        \gradedtensor{t}{i_1 \ldots i_{N_1} ; j_1 \ldots j_{N_2}}{a_1 \ldots a_{N_1} ; b_1 \ldots b_{N_2}} = 0\ .
\label{eq:abeliantensorzeros}
\end{equation}

To express a similar result for \eqref{eq:abeliantensor}, we note that each group has a trivial irreducible representation, denoted $I$, which acts on the associated one-dimensional space $\gradedspace{V}{I} \simeq \complexs$ by mapping all group elements to the identity, \ie\ the number 1.
Therefore, the formulation of $V_1 \otimes \cdots \otimes V_N$ as $\complexs \rightarrow V_1 \otimes \cdots \otimes V_N$ allows us to obtain a similar result:
\begin{equation}
    \bigotimes_i a_i \neq I \implies t_{i_1\ldots i_N}^{(a_1\ldots a_N)} = 0\ .
\end{equation}

This formulation defines each tensor map as comprising a set of allowable charge combinations, termed \emph{fusion channels}, each associated with a specific multi-dimensional array of parameters whose dimensions are defined by the outer degeneracies of the irreducible representations in the component spaces.

\subsubsection{Grouping And Splitting Indices}

The initial step in handling symmetric tensor maps involves generalizing the storage scheme introduced in \eqref{eq:tm_storage}.
In this scheme, for every tensor product spaces of the form $V_1 \otimes \cdots \otimes V_N$, an equivalent isomorphic space $Z \eqsim V_1 \otimes \cdots \otimes V_N$ and a corresponding (grouping) isomorphism $G: V_1 \otimes \cdots \otimes V_N \to Z$ was constructed.
Applying such isomorphisms and their inverses to both the codomain and the domain of a tensor map allows us to store the tensor in matrix form.
In the case of symmetries, this isomorphism must now respect the graded structure of the vector spaces.

As a result, we employ a grouping isomorphism that operates on the irrep labels by fusing them, while it acts as before on the remaining indices as in \eqref{eq:linearindex}.
For each block in the block-diagonal structure, labeled by the coupled charge $c$, we obtain the following explicit isomorphism:
\begin{equation}
\label{eq:abeliangrouping}
    \gradedtensor{G}{I; i_1\ldots i_N}{c; a_1 \ldots a_N} = \begin{cases}
        1 & I - \text{offset}_c(a_1, \ldots, a_N) = 1 + \sum_{j=1}^N \left((i_j - 1)\prod_{k<j}\dim(\gradedspace{V}{a_k}_k) \right) \\
        0 & \text{otherwise}
    \end{cases}
\end{equation}
The offset accounts for the position within each block structure, determined by looping over all fusion channels that contribute to that specific block.
This choice ensures that the resulting matrix representation is block-diagonal, where the blocks are now labeled by a single, coupled irrep.
In practice, this requires us to define an order to loop over all fusion channels, which is arbitrary but should be fixed.
Note that the resulting grouping isomorphism has the structure of a permutation matrix, and is thus definitely unitary.

Indeed, this procedure effectively means that the matrix representation of \eqref{eq:abeliantensormap} is constructed by sorting all coefficients with the same coupled charge $c = \bigotimes_{i=1}^{N_1} a_i = \bigotimes_{j=1}^{N_2} b_j$ into a matrix, and then combining these matrices in one block-diagonal matrix.
We will again use the notation $\gradedtensor{t}{}{c}$ to denote the block associated with the coupled charge $c$.

To illustrate this procedure, we consider a tensor that is invariant under the action of $\Z{2}$.
The group $\Z{2}$ has two irreps, the trivial and the sign irrep, which we will denote as $+$ and $-$.
The fusion rules of the $\Z{2}$ irreps are given by:
\begin{table}[h]
    \centering
    \begin{tabular}{c|cc}
        $\otimes$ & $+$ & $-$ \\
        \hline
        $+$ & $+$ & $-$ \\
        $-$ & $-$ & $+$
    \end{tabular}
\end{table}

For simplicity, we choose a graded space $V = \gradedspace{V}{+} \oplus \gradedspace{V}{-}$ without degeneracies, to provide a concrete example.
A tensor map $t \colon V \otimes V \rightarrow V \otimes V$, naively reshaped into a matrix, produces the following form:
\begin{equation}
    t = \begin{pmatrix}
        t^{(++;++)} & t^{(++;-+)} & t^{(++;+-)} & t^{(++;--)} \\
        t^{(-+;++)} & t^{(-+;-+)} & t^{(-+;+-)} & t^{(-+;--)} \\
        t^{(+-;++)} & t^{(+-;-+)} & t^{(+-;+-)} & t^{(+-;--)} \\
        t^{(--;++)} & t^{(--;-+)} & t^{(--;+-)} & t^{(--;--)}
    \end{pmatrix}
\end{equation}
For a $\Z{2}$-symmetric tensor map, zero components arise from \eqref{eq:abeliantensorzeros}, yielding
\begin{equation}
     t = \begin{pmatrix}
        t^{(++;++)} & \cdot & \cdot & t^{(++;--)} \\
        \cdot & t^{(-+;-+)} & t^{(-+;+-)} & \cdot \\
        \cdot & t^{(+-;-+)} & t^{(+-;+-)} & \cdot \\
        t_{(--;++)} & \cdot & \cdot & t^{(--;--)}
    \end{pmatrix}\ .
\end{equation}
The block-diagonal structure only emerges once we group the indices according to their coupled charges.
This grouping is exactly implemented by the unitary basis transformation $G$, leading to a final matrix structure:
\begin{equation}
\label{eq:abeliangroupexample}
    G t G^\dagger = \begin{pmatrix}
        t^{(++;++)} & t^{(++;--)} & \cdot & \cdot \\
        t^{(--;++)} & t^{(--;--)} & \cdot & \cdot \\
        \cdot & \cdot & t^{(-+;-+)} & t^{(-+;+-)} \\
        \cdot & \cdot & t^{(+-;-+)} & t^{(+-;+-)}
    \end{pmatrix} = t^{(+)} \oplus t^{(-)}\ .
\end{equation}
While this example excludes degeneracy considerations for simplicity, the approach is readily extendable to more complex scenarios.
The general results follow by replacing each single entry with a block of coefficients.

\subsubsection{Transpositions}
\label{subsec:abeliantransposition}

Transitioning to the integration of duality for graded spaces, we first address the formulation of a dual representation.
Given a representation $\rho$ on a vector space $V$, its dual representation $\rho^*$ on the dual space $V^*$ is defined as:
\begin{equation}
    \rho^* \colon g \triangleright \bra{v} \coloneqq \bra{v} \rho(g^{-1}).
\end{equation}
This definition is motivated by the prerequisite of being a linear representation, as well as ensuring the invariance of the duality pairing:
\begin{align}
    g \triangleright \ev{V}(v, w) = \ev{V}(g \triangleright v, g \triangleright w) &= \bra{v} \rho(g^{-1}) \rho(g) \ket{w} = \ev{V}(v, w)\\
    g \triangleright \diagram{equivariance}{9} &= \diagram{equivariance}{7} = \diagram{equivariance}{8} = \diagram{equivariance}{9}\ .
\end{align}

Expressed with respect to a basis, the dual representation takes the form 
\begin{equation}
    \rho^*(g) = \rho(g^{-1})^T = (\rho(g)^{-1})^T
\end{equation}
where the extra transpose is indeed necessary in order to satisfy the defining property of a representation, \ie\ $\rho^*(g_1) \rho^*(g_2) = \rho^*(g_1 \circ g_2)$. For a unitary representation, this further results in $\rho^*(g) = \overline{\rho(g)}$, also known as the conjugate representation. For an irrep $\rho^{(a)}$, the dual or conjugate representation $\rho^{(a)*}$ is again irreducible, and we use the notation $\bar{a}$ for the irrep label to which it is equivalent: $\rho^{(a)*} \eqsim \rho^{(\bar{a})}$. In principle, a nontrivial isomorphism (similarity transform) might be needed to turn this equivalence into an equality, and we return to this in the next section. For the case of abelian groups, where all irreps are one-dimensional, we do not need to consider this complication. 

More generally, the dual space of a graded space containing several irreps is naturally graded, with this grading naturally reflected the canonical dual basis $\{\gradedbra{i}{a}\}$ and the pairing being diagonal in the charges:
\begin{equation}
    \begin{aligned}
    \eta \colon& V^* \otimes V \rightarrow \complexs\\
               &\gradedbra{i}{a} \gradedket{j}{b} \mapsto \delta_{ij} \delta_{ab}
    \end{aligned}
\end{equation}
For the structural properties of the tensor data, determined by the fusion of the different charges in the domain and codomain, there is distinction between a charge block $a$ in a dual space (or thus, a tensor index associated with $\gradedbra{i}{a}$), or a charge $\bar{a}$ in a normal space (or thus, a tensor index associated with $\gradedket{i}{\bar{a}}$). As such, in the labeling of the tensor components, we will always use the sector as if it were a normal space. Whether a certain index is associated with a normal space (down arrow) or dual space (up arrow) is of course important for other tensor operations and is stored in the structural information associated with a tensor map. 

We can now consider line bending operations, starting from the elementary right band
\begin{equation}
\label{eq:abeliantranspose}
\begin{aligned}
    \diagram{abelian}{1} &\mapsto \diagram{abelian}{3} \coloneqq \diagram{abelian}{2} \\
    \gradedtensor{t}{i_1\ldots i_{N_1}; j_1\ldots j_{N_2}}{a_1 \ldots a_{N_1}; b_1\ldots b_{N_2}}
                         &\mapsto \gradedtensor{s}{k_1\ldots k_{N_1-1}; l_1\ldots l_{N_2+1}}{a_1 \ldots a_{N_1 - 1}; b_1 \ldots b_{N_2} a_{N_1}^*}
\end{aligned}
\end{equation}
The interaction with our matrix representation poses a complexity when a line is bent. Components of the block $\gradedtensor{t}{}{c}$ with coupled charge $c = \bigotimes a_1 \ldots a_{N_1} = \bigotimes b_1 \ldots b_{N_2}$ will be mapped to components of the block $\gradedtensor{s}{}{\tilde{c}}$ with coupled charge $\tilde{c} = \bigotimes a_1 \ldots a_{N_1 - 1} = \bigotimes b_1 \ldots b_{N_2} a_{N_1}^*$. As $c$ and $\tilde{c}$ are typically distinct, different components of the same input block are mapped to different output blocks, and vice versa, components from different input blocks are required for every output block. Indeed, it will be a recurring property of the class of tensor operations that we designated as `index manipulations' that they rearrange components within the matrix blocks. In contrast, tensor map compositions (contractions), adjoints and decompositions discussed in the following sections will preserve the matrix block structure.

To illustrate this more explicitly, we return to the example from \eqref{eq:abeliangroupexample}. Bending the second space from the domain to the right alters the matrix representation as follows:
\begin{equation}
    \tilde{t} =
    \begin{pmatrix}
        \gradedtensor{t}{}{++;++} & \cdot \\
        \gradedtensor{t}{}{--;++} & \cdot \\
        \gradedtensor{t}{}{-+;+-} & \cdot \\
        \gradedtensor{t}{}{+-;+-} & \cdot \\
        \cdot & \gradedtensor{t}{}{-+;-+} \\
        \cdot & \gradedtensor{t}{}{+-;-+} \\
        \cdot & \gradedtensor{t}{}{++;--} \\
        \cdot & \gradedtensor{t}{}{--;--}
    \end{pmatrix}
    \mapsto
    \begin{pmatrix}
        \gradedtensor{\tilde{t}}{}{+++;+} & \cdot \\
        \gradedtensor{\tilde{t}}{}{--+;+} & \cdot \\
        \gradedtensor{\tilde{t}}{}{-+-;+} & \cdot \\
        \gradedtensor{\tilde{t}}{}{+--;+} & \cdot \\
        \cdot & \gradedtensor{\tilde{t}}{}{-++;-} \\
        \cdot & \gradedtensor{\tilde{t}}{}{+-+;-} \\
        \cdot & \gradedtensor{\tilde{t}}{}{++-;-} \\
        \cdot & \gradedtensor{\tilde{t}}{}{---;-} 
    \end{pmatrix}
\end{equation}

These bending operations can again be chained to create the tensor transpose operation, which cyclically permutes the indices of any tensor map.
We propose an alteration to \algref{alg:transpose} for tensors with abelian symmetries, outlined in \algref{alg:abeliantranspose}.

This algorithm is a specific implementation of \eqref{eq:bend_coeffs}, where all components of that equation can be filled in.
By exploiting the knowledge of the structure of the different maps, it is possible to trade some simplicity for increased efficiency.
This is done by increasing the required bookkeeping, as shown in \algref{alg:abeliantranspose}, at a reduced cost, as the elements that are zero due to symmetry constraints are automatically discarded.

\begin{algorithm}[ht]
\caption{Abelian Tensor Transpose}\label{alg:abeliantranspose}
\SetKwInOut{Input}{Inputs}\SetKwInOut{Output}{Output}
\SetKwFunction{transpose}{transpose}
\DontPrintSemicolon

\Input{A tensor map $A$ and two tuples of integers $(p, q)$}
\Output{A transposed tensor map $C$}
\BlankLine
\For{$(a_1 \ldots; b_1 \ldots) \leftarrow$ fusion channels of $A$}{
    $x$ $\leftarrow$ reduced array elements of $A$ associated with $(a_1 \ldots; b_1\ldots)$\;
    $y$ $\leftarrow$ \transpose{$x$, $(p\ldots, q\ldots)$}\;
    \BlankLine
    \tcp{Here we determine the cyclic shift of the irreps}
    \tcp{Note how the duality of the irrep labels changes}
    \For{$i$ $\leftarrow$ $1$ to $|p|$}{
        $e_i$ $\leftarrow$ $p_i$th element of $(a_1\ldots; b_1^*\ldots)$\;
    }
    \For{$j$ $\leftarrow$ $1$ to $|q|$}{
        $f_j$ $\leftarrow$ $q_j$th element of $(a_1^*\ldots; b_1\ldots)$\;
    }
    \BlankLine
    reduced array elements of $C$ associated with $(e_1\ldots; f_1\ldots)$ $\leftarrow$ y\;
}
\end{algorithm}

\subsubsection{Permutations}

The integration of permutations with abelian symmetries involves minimal alterations.
The swapping map $\tau$ extends in a straightforward way to accommodate the graded structure:
\begin{equation}
    \begin{aligned}
        \tau \colon V \otimes W \rightarrow W \otimes V \\
        \gradedtensor{\tau}{ij;kl}{ab;ef} = \delta_{il}\delta_{af}\delta_{jk}\delta_{be}
    \end{aligned}
\end{equation}

Given this map, the algorithm for permutations in tensor maps with abelian symmetries is akin to the algorithm for tensor transposes.
This is outlined in \algref{alg:abelianpermute}.
Again, unlike for tensor transposes, this algorithm is not restricted to cyclic permutations, providing greater flexibility.
This algorithm similarly is a specific implementation of \eqref{eq:permute_coeffs}, which optimally makes use of the known structure of the swapping maps.

\begin{algorithm}[ht]
\caption{Abelian Tensor Permutation}\label{alg:abelianpermute}
\SetKwInOut{Input}{Inputs}\SetKwInOut{Output}{Output}
\SetKwFunction{permute}{permute}
\DontPrintSemicolon

\Input{A tensor map $A$ and two tuples of integers $(p, q)$}
\Output{A permuted tensor map $C$}
\BlankLine
\For{$(a_1\ldots; b_1\ldots) \leftarrow$ fusion channels of $A$}{
    $x$ $\leftarrow$ reduced array elements of $A$ associated with $(a_1\ldots; b_1\ldots)$\;
    $y$ $\leftarrow$ \permute{$x$, $(p\ldots, q\ldots)$}\;
    \BlankLine
    \tcp{Here we determine the permutation of the irreps}
    \tcp{Note how the duality of the irrep labels changes}
    \For{$i$ $\leftarrow$ $1$ to $|p_1|$}{
        $e_i$ $\leftarrow$ $p_i$th element of $(a_1\ldots; b_1^*\ldots)$\;
    }
    \For{$j$ $\leftarrow$ $1$ to $|q|$}{
        $f_j$ $\leftarrow$ $q_j$th element of $(a_1^*\ldots; b_1\ldots)$\;
    }
    \BlankLine
    reduced array elements of $C$ associated with $(e_1\ldots; f_1\ldots)$ $\leftarrow$ y\;
}
\end{algorithm}

\subsubsection{Traces}

Before discussing general partial traces, we first review an example where a trace is performed over the right-most indices of a tensor map $A \colon W_1 \otimes \cdots \otimes W_N \otimes V \rightarrow V$.
According to the definition of the trace, we compute this operation through application of the duality pairing maps:
\begin{equation}
\label{eq:abelian_trace}
\begin{aligned}
    \diagram{abelian}{4} &\mapsto \diagram{abelian}{6} \coloneqq \diagram{abelian}{5} \\
    \gradedtensor{A}{i_1; j_1\ldots j_N j_{N+1}}{a_1; b_1\ldots b_N b_{N+1}} 
               &\mapsto \gradedtensor{C}{; j_1\ldots j_N}{;b_1\ldots b_N}
               \coloneqq \gradedtensor{A}{i_1; j_1\ldots j_N j_{N+1}}{a_1; b_1\ldots b_N b_{N+1}} \delta_{a_1 b_{N+1}} \delta_{i_1 j_{N+1}}
\end{aligned}
\end{equation}

This operation extends similarly to the general case by first transposing permuting the traced indices to be adjacent.
In particular, this can be summarized as a loop over all possible fusion channels, summing over the ``diagonal'' traced indices and charges.
This is detailed in \algref{alg:abelian_trace}.

\begin{algorithm}[ht]
\caption{Abelian Tensor Trace}\label{alg:abelian_trace}
\SetKwInOut{Input}{Inputs}\SetKwInOut{Output}{Output}
\SetKwFunction{trace}{trace}
\DontPrintSemicolon

\Input{A tensor map $A \colon W_1 \otimes \cdots \otimes W_{N_2} \rightarrow V_1 \otimes \cdots \otimes V_{N_1}$, two tuples of integers for tracing $(p^{\tau}, q^{\tau})$, and two tuples of integers for permuting $(p^{\pi}, q^{\pi})$}
\Output{A traced tensor map $C$}
\BlankLine
\For{$i \leftarrow 1$ to $|p^\pi|$}{
    $V^\prime_i \leftarrow$ $p^\pi_i$th vector space of $(V_1 \ldots, W_1^* \ldots)$\;
}
\For{$j \leftarrow 1$ to $|q^\pi|$}{
    $W^\prime_j \leftarrow$ $q^\pi_j$th vector space of $(V_1^* \ldots, W_1 \ldots)$\;
}
\BlankLine
Assign $C \colon W^\prime_1 \otimes \cdots \otimes W^\prime_{|q^\pi|} \rightarrow V^\prime_1 \otimes \cdots \otimes V^\prime_{|p^\pi|}$\;
\BlankLine
\tcp{Adapt permutation tuples to account for traced indices}
$\tilde{p}^\pi \leftarrow$ $p^\pi$, shifted with indices from $p^\tau, q^\tau$ removed\;
$\tilde{q}^\pi \leftarrow$ $q^\pi$, shifted with indices from $p^\tau, q^\tau$ removed\;
\BlankLine
\For{$(a_1 \ldots; b_1\ldots) \leftarrow$ fusion channels of $A$}{
    \For{$i$ $\leftarrow$ $1$ to $|p^\tau|$}{
        $d_i$ $\leftarrow$ $(p^\tau_i){}^\text{th}$ element of $(a_1\ldots; b_1^*\ldots)$\;
    }
    \For{$j$ $\leftarrow$ $1$ to $|q^\tau|$}{
        $e_j$ $\leftarrow$ $(q^\tau_j)^\text{th}$ element of $(a_1^*\ldots; b_1\ldots)$\;
    }
    \If{$c_i == d_i$}{
        $a$ $\leftarrow$ reduced array elements of $A$ associated with $(a_1\ldots; b_1\ldots)$\;
        \For{$i$ $\leftarrow$ $1$ to $|p^\pi|$}{
            $f_i$ $\leftarrow$ $(p^\pi_i)^\text{th}$ element of $(a_1\ldots; b_1^*\ldots)$\;
        }
        \For{$j$ $\leftarrow$ $1$ to $|q^\pi|$}{
            $g_j$ $\leftarrow$ $(p^\pi_j)^\text{th}$ element of $(a_1^* \ldots; b_1 \ldots)$\;
        }
        $c$ $\leftarrow$ \permute{\trace{$a$, $(p^\tau \ldots, q^\tau \ldots$)}, $(\tilde{p}^\pi \ldots, \tilde{q}^\pi \ldots)$}\;
        reduced array elements of $C$ associated with $(f_1\ldots; g_1\ldots)$ $+=$ $c$\;
    }
}
\end{algorithm}

\subsection{Contractions}

In order to extend the contraction of tensors to the cases involving abelian symmetries, we can simply focus on the efficient implementation of tensor map composition.
The outer product is a special case of this operation and the general pairwise contraction can be reduced to tensor map composition.
This then allows us to handle all possible contractions.
Specifically, we previously found that this operation can be implemented efficiently by considering the matrix representations of the tensor maps involved.
Here, we improve the efficiency over \eqref{eq:tm_composition} by working directly with the block-diagonal structure of these matrices.
In particular, they can be multiplied block-by-block, which yields for each coupled sector $c$:
\begin{equation}
\label{eq:abelian_compostion}
    \gradedtensor{C}{I;J}{c} = \gradedtensor{A}{I;K}{c} \gradedtensor{B}{K;J}{c}
\end{equation}
One subtle observation here: it is possible for the blocks of $A$ and $B$ to have dimension $0$ for some of the blocks. This happens when a charge $c$ is present in the coupled charges of the codomain of $A$ and the domain of $B$, but not in the domain of $A$ (which equals the codomain of $B$). In this case, $K$ varies over an empty range. The resulting block of $C$ then needs to be filled with $0$ entries.

Again, this operation can be generalized by incorporating pre- and post-processing permutation steps using the algorithms described above.
We summarize the algorithm for a generic abelian pairwise tensor contraction in \algref{alg:abelian_contraction}.

\begin{algorithm}[ht]
    \caption{Abelian tensor contraction}\label{alg:abelian_contraction}
\SetKwInOut{Input}{Inputs}\SetKwInOut{Output}{Output}
\SetKwFunction{permute}{permute}
\DontPrintSemicolon

\Input{A tensor map $A$ with permutation indices $(p^A, q^A)$, a tensor map $B$ with permutation indices $(p^B, q^B)$ and permutation indices for their resulting contraction $(p^{AB}, q^{AB})$}
\Output{The contracted tensor map $C$}
\BlankLine

$\tilde{A}$ $\leftarrow$ \permute{$A$, $(p^A, q^A)$}\;
$\tilde{B}$ $\leftarrow$ \permute{$B$, $(p^B, q^A)$}\;
\BlankLine
Assign $\tilde{C} \colon \text{domain}(\tilde{B}) \rightarrow \text{codomain}(\tilde{A})$\;
\BlankLine
\For{$c$ $\leftarrow$ coupled charges of $C$}{
    $a$ $\leftarrow$ block in matrix representation of $\tilde{A}$ associated with $c$\;
    $b$ $\leftarrow$ block in matrix representation of $\tilde{B}$ associated with $c$\;
    \If{$\mathrm{dim}(a) \neq 0$}{
        block in matrix representation of $\tilde{C}$ associated with $c$ $\leftarrow$ $a \cdot b$ \;
    }
}
\BlankLine
$C$ $\leftarrow$ \permute{$\tilde{C}$, $(p^{AB}, q^{AB})$}
\end{algorithm}

\subsubsection{Adjoint}

The definition of the adjoint for tensor maps with abelian symmetries is rather straightforward.
In particular, the canonical isomorphisms defined in \eqref{eq:abeliangrouping} are permutation matrices, which are unitary by construction.
As a result, the implementation of the adjoint simply boils down to taking the Hermitian conjugate of the block-diagonal matrix representation, which is implemented efficiently block-by-block.

\subsection{Decompositions}

In order to extend the tensor decompositions to symmetric tensors, we can again make use of the matrix representation of the tensor maps.
Because the grouping and splitting isomorphisms that relate the tensor map with its block diagonal matrix representation are unitary, we can guarantee that the orthogonality properties of the decompositions are preserved.
Because of this, the decompositions can be carried out block-by-block, which is again a significant advantage in terms of efficiency.
Without reiterating the details of each decompositions, we can summarize the general approach in \algref{alg:abelian_decomposition}.

\begin{algorithm}[ht]
    \caption{Abelian tensor decomposition}\label{alg:abelian_decomposition}
    \SetKwInOut{Input}{Inputs}\SetKwInOut{Output}{Output}
    \SetKwFunction{decompose}{decompose}
    \SetKwFunction{permute}{permute}
    \DontPrintSemicolon

    \Input{A tensor map $A$ and a tuple of integers $(p, q)$}
    \Output{The decomposed tensor maps $F_1, \ldots$}
    \BlankLine
    $\tilde{A} \leftarrow$ \permute{$A$, $(p, q)$}\;
    \For{$c$ $\leftarrow$ coupled charges of $\tilde{A}$}{
        $a$ $\leftarrow$ block in matrix representation of $\tilde{A}$ associated with $c$\;
        $f_1, \ldots$ $\leftarrow$ \decompose{$a$}\;
        blocks in matrix representations of $F_1\ldots$ associated with $c$ $\leftarrow$ $f_1,\ldots$\;
    } 
\end{algorithm}

There is one subtlety associated with obtaining eigenvectors from the eigenvalue decomposition, singular vectors from the singular value decomposition, or similar constructions for other decompositions.
To illustrate this, we consider the matrix representation of the eigenvectors of a tensor map $A \colon V \rightarrow V$.
By construction, this matrix is block-diagonal, with each block associated with a coupled charge $c$.
The eigenvectors therefore only have support on a single block, and can be assigned to the corresponding coupled charge.
However, unless the charge is trivial, these eigenvectors cannot be constructed as $v \colon \complexs \rightarrow V$, as doing so would not respect the graded structure.
Instead, we can remedy this by generalizing the eigenvectors to the maps $v:\gradedspace{V}{c} \to V$, with a nontrivial domain that exactly corresponds to the space associated with irrep $c$ (and that is therefore still one-dimensional in the abelian case).
Often, these auxiliary spaces will be represented with squiggly lines.
This leads to the following eigenvalue equation:
\begin{equation}
    A \circ \gradedtensor{v}{}{c} = \lambda_c \gradedtensor{v}{}{c} 
        \iff \diagram{abelian}{7} = \lambda_c \diagram{abelian}{8}
\end{equation}

\section{Non-abelian Symmetries}
\label{sec:non-abelian}

Building on our understanding of abelian symmetries, this section delves into the additional complexities associated with non-abelian symmetries.
Unlike abelian groups, non-abelian groups can have higher-dimensional irreducible representations, introducing intricacies in the representation and manipulation of symmetric tensors.
Crucial for handling these complexities is the use of fusion trees, which describe the symmetric internal structure of a tensor.

We begin by introducing fusion diagrams and fusion trees, illustrating their role in representing and implementing non-abelian symmetries.
These tools not only provide a mathematical framework for describing the fusion of representations but also allow for a visual way of understanding them.
Importantly, they admit efficient algorithms to handle their manipulations.

Subsequently, we explore how these structures influence the tensor operations discussed in Section \ref{sec:fundamentals}.
By comparing these advanced scenarios with the simpler context of abelian symmetries in Section \ref{sec:abeliansymmetries}, we highlight both the challenges as well as the enhanced efficiency of our framework when extended to accommodate generic symmetries.
This section aims not only to elaborate on the theoretical underpinnings of non-abelian symmetries but also to justify the algorithmic choices made within \TensorKit, demonstrating the practical impact of these sophisticated symmetry considerations on computational methods in tensor network theory.

\subsection{What Is A Fusion Tree?}
\label{subsec:whatisfusiontree}

Transitioning from abelian to non-abelian groups introduces complexities due to the multi-dimensional irreps associated with non-abelian symmetries.
For abelian symmetries, the internal structure of the invariant subspaces can be straightforwardly managed since they are one-dimensional.
In contrast, non-abelian groups require a more nuanced approach to handle the internal degrees of freedom of the invariant subspaces.

We recall from \eqref{eq:gradedbasis} that a basis for a graded vector space $V$ can be described by $\gradedket[m]{i}{c}$, where $c$ denotes the charge, $i$ is the \emph{outer label} indexing which copy of $\gradedspace{V}{c}$ the basis vector inhabits, and $m$ is the \emph{inner label} indicating internal degrees of freedom within $\gradedspace{V}{c}$.
A symmetric operator $A \colon V \rightarrow V$ expressed in this basis achieves a block-diagonal form due to Schur's lemma:
\begin{equation}
    A = \bigoplus_c \gradedtensor{A}{}{c} \otimes \id{\gradedspace{V}{c}} \equiv \gradedtensor{A}{i;k}{c} \delta_{mn} \gradedket[m]{i}{c}\gradedbra[n]{j}{c}
\end{equation}
Although this argument holds for non-abelian groups, additional complexities arise when considering tensor products of vector spaces.
For instance, a tensor map $t \colon W_1 \otimes W_2 \rightarrow V$ requires the decomposition of the tensor product representation on $W_1 \otimes W_2$ into its irreducible components to effectively apply Schur's lemma.

To that end, we can first consider the space of equivariant maps $\gradedspace{V}{a} \otimes \gradedspace{V}{b} \to \gradedspace{V}{c}$ for the fusion of two irrep spaces.
The fusion rules dictate that there are $\NSymbol{c}{ab}$ independent maps between these spaces, which we denote as $\XMatrix{ab}{c;\mu}$ with $1 \leq \mu \leq \NSymbol{c}{ab}$.
These maps are often referred to as \emph{fusion tensors}, and their inverses $\XMatrix{c;\mu}{ab}$ are known as \emph{splitting tensors}.
Graphically, they are depicted as follows:
\begin{equation}
    \XMatrix{ab}{c;\mu} = \diagram{wignereckart}{3},\quad
    \XMatrix{c;\mu}{ab} = \diagram{wignereckart}{12}
\end{equation}
With respect to the basis $\{\gradedket{m}{a}\}$ of $\gradedspace{V}{a}$ and similar for $\gradedspace{V}{b}$, we can expand the splitting tensor as
\begin{equation}
\gradedket{m_1}{a} \otimes \gradedket{m_2}{b} \mapsto \sum_n \left(\XMatrix{c;\mu}{ab}\right)_{n; m_1 m_2} \gradedket{n}{c}    
\end{equation}
The components $\left(\XMatrix{c;\mu}{ab}\right)_{n; m_1 m_2}$ are better known in the literature on representation theory as Clebsh-Gordan coefficients, in particular for the case of $\SU{2}$ in the context of angular momentum recoupling in quantum physics. 

These elementary splitting tensors can now be used to construct a grouping isomorphism from the tensor product of general graded spaces $W_1 = \bigoplus_a \complexs^{\NSymbol{a}{W_1}} \otimes \gradedspace{V}{a}$ and $W_2 = \bigoplus_b \complexs^{\NSymbol{b}{W_2}} \otimes \gradedspace{V}{b}$ to a single space $W \eqsim W_1 \otimes W_2 = \bigoplus_c \complexs^{\NSymbol{c}{W}} \otimes \gradedspace{V}{c}$, given by
\begin{equation}
\label{eq:groupingnonabelian}
    \gradedket[m_1]{i_1}{a} \gradedket[m_2]{i_2}{b} 
    \mapsto \sum_c \sum_{\mu=1}^{\NSymbol{c}{ab}}
        G_{j; i_1 i_2}^{c \mu; a b}
        \left(\XMatrix{c;\mu}{ab}\right)_{n; m_1 m_2} 
        \gradedket[n]{j}{c}
\end{equation}
where $G$ is a trivial grouping map with the structure of a permutation matrix, similar to the abelian grouping map in \eqref{eq:abeliangrouping}. In particular, this map embeds the indices $i_1=1,\ldots,\NSymbol{a}{W_1}$, $i_2=1,\ldots,\NSymbol{a}{W_2}$ and the different fusion channels $\mu=1,\ldots,\NSymbol{c}{ab}$ into the range of $j = 1, \ldots \NSymbol{c}{W}$, with thus $\NSymbol{c}{W} = \sum_{a,b} \NSymbol{a}{W_1} \NSymbol{b}{W_2} \NSymbol{c}{ab}$.

It will be an important property of our framework that we do not need to know the specific basis representation of the fusion and splitting tensors, \ie\ we will not be using Clebsch-Gordan coefficients explicitly. Rather, all tensor operations will be performed by a sequence of elementary manipulations that employ the structural properties of these fusion and splitting tensors.
The first of these properties is orthogonality and completeness, which can be expressed as
\begin{equation}
\label{eq:fusiontensororthogonality}
    \diagram{fusiontrees}{1} = \delta_{cc^\prime} \delta_{\mu \mu^\prime} \diagram{fusiontrees}{2}, \qquad\qquad
    \diagram{fusiontrees}{4} = \sum_{c \mu} \diagram{fusiontrees}{3}\ .
\end{equation}
These properties can be used to show that the grouping isomorphism in \eqref{eq:groupingnonabelian} is unitary.

Using this grouping isomorphism, a tensor map $t \colon W_1 \otimes W_2 \rightarrow V$ can be projected onto the basis of $W$, which exposes the representation of $t$ as a block diagonal matrix in line with Schur's lemma: 
\begin{align}
\label{eq:wignereckartdecomposition}
    \gradedket[m_1]{i_1}{a} \gradedket[m_2]{i_2}{b} 
    &\mapsto \sum_c \sum_{\mu=1}^{\NSymbol{c}{ab}}
        G_{j; i_1 i_2}^{c \mu; a b}
        \left(\XMatrix{c;\mu}{ab}\right)_{n; m_1 m_2} 
        \gradedket[n]{j}{c}\\
    \gradedket[n]{j}{c}
        &\mapsto \gradedtensor{\tilde{t}}{k; j}{c}
        \gradedket[n]{k}{c}
\end{align}
Defining the \emph{reduced tensor coefficients} $\gradedtensor{\tilde{t}}{k,i_1i_2}{c,\mu;ab} = \sum_{j} \gradedtensor{\tilde{t}}{k; j}{c}G_{j; i_1 i_2}^{c \mu; a b}$, we obtain the well-known Wigner-Eckart theorem, which fuels the computational handling of non-abelian symmetric tensors in tensor networks:
\begin{theorem}[Wigner-Eckart theorem]
    A tensor map $t \colon W_1 \otimes W_2 \rightarrow V$ invariant under the action of a group $\group{G}$ can be decomposed into two parts:
    the fusion tensor elements, which are independent of the outer degeneracy labels of the irreps, and the reduced tensor elements, which are independent of the inner labels of the irreps.
    \begin{align}
        t &\equiv \bigoplus_{abc} \left(\bigoplus_\mu^{\NSymbol{c}{ab}} \XMatrix{c;\mu}{ab} \otimes \gradedtensor{\tilde{t}}{}{c,\mu;ab}\right) \\
        \diagram{wignereckart}{1} &= \bigoplus_{abc} \left(\bigoplus_\mu^{\NSymbol{c}{ab}} \diagram{wignereckart}{12} \otimes \diagram{wignereckart}{2} \right)
    \end{align}
\end{theorem}


As we extend our discussion to tensor product spaces involving more than two components, additional complexities arise.
Consider, for example, a tensor map $t \colon V \rightarrow W_1 \otimes W_2 \otimes W_3$.
The decomposition of the tensor product of irreps, $a \otimes b \otimes c \rightarrow \bigoplus \NSymbol{d}{abc} d$, can proceed through multiple paths, either sequentially combining $(a \otimes b) \otimes c$ or $a \otimes (b \otimes c)$.
The corresponding splitting tensors for these sequences are given by:
\begin{align}
\label{eq:canonicalsplitting}
    \XSymbol{abc}{d}{\mu,e,\nu} &= 
        \left( \XSymbol{ab}{e}{\mu} \otimes \id{c} \right) \circ \XSymbol{ec}{d}{\nu} \\
\label{eq:noncanonicalsplitting}
    \tXSymbol{abc}{d}{\kappa,f,\lambda} &=
    \left( \id{a} \otimes \XSymbol{bc}{f}{\kappa} \right)
    \circ \XSymbol{af}{d}{\lambda}
\end{align}

This leads to multiple, albeit equivalent, decompositions for such tensor maps, each path yielding a distinct but valid set of reduced tensor coefficients.
These pathways and their implications can be clearly illustrated through the following diagrammatic representations:
\begin{equation}
\label{eq:fusionorder}
    \diagram{wignereckart}{8} = \bigoplus \diagram{wignereckart}{4} \otimes \diagram{wignereckart}{6} = \bigoplus \diagram{wignereckart}{5} \otimes \diagram{wignereckart}{7}
\end{equation}

These different fusion pathways present two primary challenges in our notation.
Firstly, the introduction of additional labels to denote intermediate fusion states becomes essential, to uniquely identify the structure.
Secondly, the fusion order affects the expansion coefficients, necessitating some bookkeeping to keep our framework consistent.

To address these complexities, we introduce the concept of a \emph{fusion tree}.
A fusion tree is a full binary tree whose edges are labeled by charges and whose vertices are labeled by integers, providing a structured representation of the fusion sequence.
At every vertex, the fusion rules must be satisfied in the form of the integer satisfying $\mu \in 1,\ldots,\NSymbol{c}{ab}$, with $\NSymbol{c}{ab} = 0$ if charge $c$ does not appear in the fusion product of $a$ and $b$ (making the vertex inadmissable).
Similarly, a \emph{splitting tree} involves the same kind of structure to represent the splitting sequence.
From the orthonormality and completeness of the binary fusion and splitting tensors, it follows that the complete set of admissible fusion trees associated with a particular fusion sequence forms an orthonormal basis.
However, fusion trees associated with different fusion sequences are not necessarily orthogonal, as we will discuss in subsequent sections.
Hence, to standardize the grouping and splitting isomorphisms, we adopt a canonical basis through a fixed sequence where fusion and splitting proceeds from left to right.
The orthonormality of the splitting trees can then be illustrated as
\begin{equation}
\label{eq:fusiontreeorthogonality}
    \diagram{fusiontrees}{5} = \delta_{dd^\prime} \delta_{ee^\prime} \delta_{\mu\mu^\prime} \delta_{\nu\nu^\prime} \diagram{fusiontrees}{6}, \qquad
    \diagram{fusiontrees}{8} = \sum_{de\mu\nu} \diagram{fusiontrees}{7}
\end{equation}

In this framework, the symmetry constraints dictate that reduced tensor coefficients can be associated to every admissible pair of canonical fusion and splitting trees.
This leads to the canonical expansion of a tensor map $t \colon V_1 \otimes V_2 \otimes V_3 \otimes V_4 \rightarrow W_1 \otimes W_2 \otimes W_3 \otimes W_4$:
\begin{equation}
\label{eq:canonicalfusiontree}
    \diagram{wignereckart}{9} = \bigoplus_{\text{fusion channels}} \diagram{wignereckart}{10} \otimes \diagram{wignereckart}{11}
\end{equation}

For further reference, we also introduce a short-hand notation to denote the reduced tensor components.
By assuming a fixed order of fusion channels, we enumerate all fusion (splitting) trees as $f_J$ ($s_I$), where $J$ and $I$ encase all labels for the trees, including outer uncoupled charges, the coupled charge on the central line connecting the fusion and splitting tree, intermediate charges on the inner lines of the trees, and multiplicity labels on the vertices.
Using this notation, the full tensor can be succinctly expressed as
\begin{equation}
    t = \bigoplus_{(s_I, f_J) \in \text{fusion channels}} \left(s_I, f_J\right) \otimes \gradedtensor{t}{i_1\ldots ; j_1\ldots}{s_I; f_J}
\end{equation}

Finally, we can embed the different reduced tensor coefficients \gradedtensor{t}{i_1\ldots ; j_1\ldots}{s_I; f_J} in a block diagonal matrix representation, with blocks labeled by the central coupled charge, as
\begin{equation}
\gradedtensor{t}{I,J}{c} = \sum_{s_I, f_J} G^{c; s_I}_{I;i_1,\ldots,i_{N_1}} \gradedtensor{\tilde{t}}{i_1\ldots i_{N_1}; j_1\ldots j_{N_2}}{s_I; f_J} G^{c; f_J}_{J;j_1,\ldots, j_{N_2}}
\end{equation}
where the sum ranges over all splitting and fusion trees for which the coupled charge equals $c$, and $G^{c;s_I}_{I; i_1\ldots i_{N}}$ is a generalization of the trivial grouping map that simply embeds the different compatible splitting trees $s_I$ with coupled charge $c$ and the associated outer indices into a single linear index $I$.

In comparison to the abelian case, the block diagonal matrix that appears after the total grouping isomorphisms have been applied, now takes the form $\bigoplus_c \gradedtensor{\tilde{t}}{}{c} \otimes \id{\gradedspace{V}{c}}$, \ie\ every block $\gradedtensor{\tilde{t}}{}{c}$ appears $d_c$ times.
This however does not affect that we can directly act on these blocks when composing and decomposing tensor maps, or taking their adjoint.
As in the abelian case, the index manipulations are the operations that reorganize the structure of reduced tensor components in the different blocks, as we now explore in the next section.

\subsection{Fusion Tree Manipulations}
As we extend the discussions from Section \ref{sec:indexmanipulations} to include non-abelian symmetries, we first outline some general techniques and strategies that are applicable across a range of index manipulations.

These processes typically involve several distinct steps: initially, we begin with a tensor represented in its canonical basis.
We then apply a series of manipulations, as defined in \ref{sec:indexmanipulations}, which are executed separately on both the fusion-splitting trees and the reduced tensor coefficients.
Since the reduced tensor coefficients lack further structure, they can be manipulated in accordance with the methods detailed in \ref{sec:indexmanipulations}.
However, the manipulations on the fusion-splitting trees lead to non-canonical diagrams.
The goal then becomes to re-express these tensors in their canonical basis, as summarized in \algref{alg:nonabelian-transform}.
The process can be diagrammatically represented as follows, where the hatched area indicates an arbitrary fusion diagram, which is not necessarily of the canonical tree form.
\begin{equation}
    \fusiontreemanipulations{8} \otimes \fusiontreemanipulations{10} 
        \mapsto \fusiontreemanipulations{9} \otimes \fusiontreemanipulations{11} 
        = \fusiontreemanipulations{8} \otimes \fusiontreemanipulations{12}
\end{equation}

\begin{algorithm}[ht]
\caption{Non-abelian Tensor Transformation}\label{alg:nonabelian-transform}
\SetKwInOut{Input}{Inputs}\SetKwInOut{Output}{Output}
\SetKwFunction{transform}{transform}
\SetKwFunction{permute}{permute}
\DontPrintSemicolon

\Input{A tensor map $A$, two tuples of integers $(p, q)$ and a tree transformation \transform}
\Output{A transformed tensor map $C$}
\BlankLine
\For{$(s_I, f_J) \leftarrow$ fusion-splitting tree pairs of $A$}{
    $x$ $\leftarrow$ reduced array elements of $A$ associated with $(s_I, f_J)$\;
    \tcp{Loop over output trees that have non-zero contribution $c$}
    \For{$(s_K, f_L, c) \leftarrow $ \transform{$f$}}{
        $y$ $\leftarrow$ reduced array elements of $C$ associated with $(s_K, f_L)$\;
        $y += c \cdot $ \permute{$x$, $(p\ldots, q\ldots)$}\;
    }
}
\end{algorithm}

This last projection step to reinstate the canonical basis, or in other words the determination of \texttt{transform}, can be executed through multiple methods.
A straightforward approach leverages the fact that fusion trees make up an orthogonal basis.
Therefore, by calculating their overlaps we can project the manipulated fusion trees back onto the canonical basis.
This process is the generalization of the projection steps of Section \ref{sec:indexmanipulations}, illustrated by:
\begin{equation}
    \fusiontreemanipulations{13} \otimes \fusiontreemanipulations{12} = \fusiontreemanipulations{14} \otimes \fusiontreemanipulations{11}
\end{equation}

However, this method scales unfavorably with an increase in the number of tensor indices, or in the presence of symmetry groups with large irreducible representations, in which cases computing these overlaps becomes prohibitively expensive.
An alternative, more scalable method that involves a series of localized diagrammatic manipulations is outlined in the remainder of this section.
This approach not only manages the scalability issues but also seamlessly integrates more generalized symmetries, as this no longer requires explicit representations of the fusion tensors.
This point will be further discussed in Section \ref{sec:category}.

These manipulations only depend on the structural components of tensors, which are determined by the underlying symmetry.
This makes these operations ideal for memoized strategies, which are crucial for performance optimization.
In \TensorKit, substantial resources are devoted to ensuring that this data is cached efficiently and accessed in a thread-safe manner.

\subsubsection{Elementary Fusion Tree Manipulations}
\label{sec:fusiontreemanipulations}



As outlined, we start by discussion a number of elementary operations on fusion trees and pairs of fusion and splitting threes, that can then be composed in order to describe the effect of the larger index manipulation operations such as a general transposition or permutation.

\paragraph{$F$-moves}{
    One of the most used manipulations allows the transition between the various splitting sequences that can be used to define higher-rank splitting tensors, specifically between \eqref{eq:canonicalsplitting} and \eqref{eq:noncanonicalsplitting}.
    This recoupling of splitting tensors is commonly referred to as an $F$-move, governed by an isomorphism in terms of $F$-symbols, which are defined as:
    \begin{equation}
        \fusiontreemanipulations{15} =
        \sum_{\kappa f \lambda} \FSymbol{abc}{d}{\nu,e,\mu}{\lambda,f,\kappa}
            \fusiontreemanipulations{16}
    \end{equation}
    The inverse of this operation also exists, and due to the unitarity of the splitting trees the $F$-symbol itself is unitary:
    \begin{equation}
        \sum_{\kappa f \lambda} \FSymbol{abc}{d}{\nu,e,\mu}{\lambda,f,\kappa} \FSymbol*{abc}{d}{\nu^\prime,e^\prime,\mu^\prime}{\lambda,f,\kappa} = \delta_{\nu\nu^\prime}\delta_{\mu\mu^\prime}\delta_{ee^\prime}
    \end{equation}
    
    Similarly, we can define a recoupling of the fusion trees as follows:
    \begin{equation}
        \fusiontreemanipulations{17} =
        \sum_{\lambda f \kappa} \FSymbol{d}{abc}{\mu,e,\nu}{\kappa,f,\lambda}
            \fusiontreemanipulations{18}
    \end{equation}
    Due to the adjoint relationship, these $F$-moves are interrelated, leading to the following identity:
    \begin{equation}
        \FSymbol{d}{abc}{\mu,e,\nu}{\kappa,f,\lambda} = \FSymbol*{abc}{d}{\nu,e,\mu}{\lambda,f,\kappa}
    \end{equation}
    This relation between splitting and fusion tree manipulations is generically applicable and allows us to primarily focus on the splitting trees, with the understanding that manipulations on the fusion trees follow by applying the adjoint map.
    
    In practical applications to tensor maps, this operation often needs to be performed across all fusion trees in the available fusion channels.
    Each fusion tree's recoupling is represented as a linear combination of canonical fusion trees, which can be effectively visualized as a (sparse) matrix where each column details the linear combinations needed for a single fusion tree's recoupling.
    This matrix interpretation proves valuable as it translates the composition of subsequent diagrammatic manipulations into matrix multiplications, enhancing both clarity and computational efficiency.
}

\paragraph{Tree insertion}{
    Building upon the elementary $F$-moves, fusion tree insertion involves attaching one fusion tree to an uncoupled leg of another and then projecting the result back onto the canonical fusion tree basis.
    This process is typically initiated at the tip of the fusion tree to be inserted, where it can be effectively integrated through an $F$-move.
    The process is depicted diagrammatically as follows:
    \begin{equation}
        \fusiontreemanipulations{19} = \sum_{\kappa f \lambda} \FSymbol*{abc}{d}{\nu,e,\mu}{\lambda,f,\kappa} \fusiontreemanipulations{20}
    \end{equation}
    
    This operation reduces the number of uncoupled legs of the inserted fusion tree by one at each step.
    If the inserted fusion tree has $N$ uncoupled legs, recursive application of this procedure manipulates the structure until the canonical basis is achieved after $N-1$ steps.
    In terms of computational implementation, this series of $F$-moves can be represented using matrix operations, as previously discussed.
    The entire insertion process can thus be encapsulated as a sequence of matrix multiplications, effectively streamlining the insertion procedure:
    \begin{equation}
        \label{eq:fusiontreeinsertion}
        \fusiontreemanipulations{21} 
        = \sum \underbrace{\FSymbol*{\cdots}{\cdot}{\cdot,\cdot,\cdot}{\cdot,\cdot,\cdot} \cdots \FSymbol*{\cdots}{\cdot}{\cdot,\cdot,\cdot}{\cdot,\cdot,\cdot}}_{N-1 \text{ factors}}
            \fusiontreemanipulations{22}
    \end{equation}
}

\subsubsection{Duality Fusion Tree Manipulations}
As discussed in Subsection~\ref{subsec:abeliantransposition}, the dual representation $\rho^{(a)*}$ associated with in irrep $\rho^{(a)}$ is itself irredicuble and thus isomorphic to an irrep with the label $\bar{a}$.
In the abelian case, irreps $a$ and $\bar{a}$ had the property that they fuse to the trivial irrep.
In the non-abelian case, this generalizes to $\NSymbol{a \bar{a}}{I} = 1$, whereas $\NSymbol{a b}{I} = 0$ for all $b \neq \overline{a}$.
Furthermore, in the non-abelian case, we need to consider the isomorphisms $Z_a:(\gradedspace{V}{a})^* \to \gradedspace{V}{\bar{a}}$ relating $\rho^{(a)*}$ to $\rho^{(\bar{a})}$ via
\begin{equation}
\label{eq:definitionZ}
    Z_a \circ \rho^{(a)*}(g) = \rho^{(\bar{a})}(g) \circ Z_a, \forall g \in G\,
\end{equation}
henceforth depicted as
\begin{equation}
    Z_a = \fusiontreemanipulations{23}, \quad Z_a^\dagger = \fusiontreemanipulations{24}\ .
\end{equation}
The unitary of the representations implies that also $Z_a$ is unitary
\begin{equation}
\label{eq:Zmorphismunitary}
    \fusiontreemanipulations{25} = \fusiontreemanipulations{28}, \quad \fusiontreemanipulations{26} = \fusiontreemanipulations{27}\ .
\end{equation}
These isomorphism also allow us to establish a relationship between the (co)evaluation maps and the fusion and splitting tensors of $a$ and $\bar{a}$ to the identity, namely
\begin{equation}
    X^I_{a \bar{a}} = \frac{1}{\sqrt{d_a}} \left( \mathbb{1}_a \otimes Z_a \right) \circ \eta_a = \frac{1}{\sqrt{d_a}} \left(Z_a^T \otimes \mathbb{1}_{\bar{a}}\right) \circ \tilde{\eta}_{\bar{a}}
\end{equation}
\begin{equation}
    \fusiontreemanipulations{29} = \frac{1}{\sqrt{d_a}} \fusiontreemanipulations{30} = \frac{1}{\sqrt{d_a}} \diagram{fusiontreemanipulations2}{1}\ .
\end{equation}
As used before, $d_a$ represents the dimension of the irrep and can be obtained by tracing the identity map on $\gradedspace{V}{a}$, often depicted diagrammatically as \emph{popping bubbles}:
\begin{equation}
    d_a = \fusiontreemanipulations{49}\ .
\end{equation}
Here, we also notice the appearance of $Z_{a}^T: (\gradedspace{V}{\bar{a}})^* \to \gradedspace{V}{a}$, which has the same domain and codomain as $Z_{\bar{a}}$, and can indeed also be used to relate $\rho^{(a)*}$ to $\rho^{(\bar{a})}$, as follows trivially from transposing the defining relation in \eqref{eq:definitionZ}.
Since the isomorphism relating both irreps should be unique up to an overall constant, or actually a phase in the case of unitary isomorphisms, we obtain
\begin{equation}
Z_a^T = \chi_a Z_{\bar{a}}
\end{equation}
where $\chi_a$ is known as the \emph{Frobenius-Schur phase}\footnote{Whenever $a = \bar{a}$, this coincides with the usual notion of the Frobenius-Schur indicator, which takes the values $\{1, 0, -1\}$ and can distinguish between real, complex and quaternionic representations.}.
If $a \neq \bar{a}$, we can define $Z_{\bar{a}}$ such that $\chi_a=1$, though we do not require this in our framework.
For $a = \bar{a}$, the Frobenius-Schur phase must equal $\chi_a = \pm 1$ and is topologically protected.
It designates whether the representation $\rho^{(a)}$ is real ($\chi_a = +1$) or quaternionic ($\chi_a=-1$).

Using the $Z$ isomorphisms, fusion and splitting diagrams can be adapted to accommodate various arrow configurations, as for example 
\begin{equation}
    \fusiontreemanipulations{31} \coloneqq \fusiontreemanipulations{32}
\end{equation}

Henceforth, we will extend our definition of the canonical basis of fusion-splitting tree pairs.
Internal arrows will always be oriented downwards, so that only the standard fusion and splitting tensors appear at every vertex.
However, on the final uncoupled legs of the fusion or splitting trees, isomorphisms $Z$ or $Z^\dagger$ will be added, respectively, when they are associated with dual spaces appearing in the domain or codomain.

\paragraph{Bending along the right}
To bend lines towards the right in a fusion tree, the manipulation can be focused on a single vertex at a time.
This simplification involves the duality pairing combined with a splitting tensor:
\begin{equation}
\label{eq:ftbendright}
\begin{aligned}
    \fusiontreemanipulations{33} 
        &= \sqrt{d_b} \fusiontreemanipulations{34}
        = \sqrt{d_b} \fusiontreemanipulations{35} \\
        &= \sqrt{d_b} \sum_{\nu d \mu^\prime} \FSymbol*{a}{ab\bar{b}}{\nu,c^\prime,\mu^\prime} {1,I,1}\fusiontreemanipulations{36} 
        = \sqrt{d_b} \sum_\nu \FSymbol*{a}{ab\bar{b}}{\nu,c,\mu}{1,I,1} \fusiontreemanipulations{37} \\
        &= \sqrt{\frac{d_c}{d_a}} \sum_\nu \BSymbol{ab}{c}{\mu}{\nu} \fusiontreemanipulations{37}\ .
\end{aligned}
\end{equation}
The $B$-symbol defined here as
\begin{equation}
    \BSymbol{ab}{c}{\mu}{\nu} = \sqrt{\frac{d_a d_b}{d_c}} \FSymbol*{a}{ab\bar{b}}{\nu,c,\mu}{1,I,1} = \sqrt{\frac{d_a d_b}{d_c}} \FSymbol{ab\bar{b}}{a}{\mu,c,\nu}{1,I,1} 
\end{equation}
represents the coefficients of this transformation, with the normalization to ensure unitarity
\begin{equation}
    \sum_\nu \BSymbol{ab}{c}{\mu}{\nu} \cdot \BSymbol*{ab}{c}{\lambda}{\nu} = \delta_{\mu\lambda}\ .
\end{equation}

However, special caution is required when a line already includes a $Z$-morphism.
Making the morphism explicit allows the remainder of the diagram to be transformed using \eqref{eq:ftbendright}.
While the line bending turns this splitting tensor into a normal fusion tensors with all arrows downwards, the initial $Z$-morphism is not triviall cancelled. Instead, the resulting diagram contains the combination $Z_{\bar{b}} \circ (Z_{b}^\dagger)^T$. However, using the relation $Z^T_{b} = \chi_b Z_{\bar{b}}$, we obtain $Z_{\bar{b}} \circ (Z_{b}^\dagger)^T = Z_{\bar{b}} \circ (\chi_b Z_{\bar{b}})^\dagger = \overline{\chi_b} \id{\bar{b}}$, as illustrated in the following diagram:
\begin{equation}
\begin{aligned}
    \diagram{fusiontreemanipulations2}{2} &= \diagram{fusiontreemanipulations2}{3} \\
        &= \sqrt{\frac{d_c}{d_a}} \sum_\nu \BSymbol{a\bar{b}}{c}{\mu}{\nu} \diagram{fusiontreemanipulations2}{4} \\
        &= \overline{\chi_b} \sqrt{\frac{d_c}{d_a}} \sum_\nu \BSymbol{a\bar{b}}{c}{\mu}{\nu} \diagram{fusiontreemanipulations2}{5}
\end{aligned}
\end{equation}

\paragraph{Folding along the left}{
Bending a line towards the left involves multiple vertices of the fusion tree, introducing a sequence of diagrammatic manipulations.
This process, referred to as \emph{folding}, initiates by inserting a resolution of the identity at the junction where the bending line intersects with the line connecting the splitting and fusion trees.
Subsequently, the top and bottom segments of the diagram are independently reverted to the canonical basis, utilizing the insertion techniques specified in \eqref{eq:fusiontreeinsertion}.
\begin{equation}
    \fusiontreemanipulations{41} = \fusiontreemanipulations{42}
    \xrightarrow[]{\text{insertion}} \fusiontreemanipulations{43}
    \xrightarrow[\text{insertion}]{}  \fusiontreemanipulations{44}
\end{equation}
}

The completion of this process involves eliminating the tadpole diagram, yielding a diagram that conforms to the canonical structure:
\begin{equation}
    \label{eq:tadpole}
    \fusiontreemanipulations{45}
    = \fusiontreemanipulations{46}
    = \sqrt{d_a} \fusiontreemanipulations{47}
    = \sqrt{d_a} \delta_{cI}\; \fusiontreemanipulations{48} 
\end{equation}

\subsubsection{Braiding Fusion Tree Manipulations}
\label{sec:nonabelianbraiding}

In order to further permute the indices of fusion and splitting trees, we additionally have to incorporate the swapping map $\tau$.
Using this map, we can swap neighbouring indices of a fusion tree, for which we distinguish two cases:

\paragraph{$R$-move}{
The $R$-move is defined by swapping the indices of a single splitting tensor:
\begin{equation}
\label{eq:Rmove}
    \fusiontreemanipulations{1}
        = \sum_\nu \RSymbol{ba}{c}{\mu}{\nu} \fusiontreemanipulations{2}
\end{equation}
where the $R$-symbol represents these braiding coefficients, which again form a unitary matrix.
This operation can be used to swap the left-most indices of a canonical splitting tree.
}

\paragraph{Braiding move}{
Expanding on the $R$-move, the braiding move enables arbitrary permutations of adjacent indices by employing a combination of $R$- and $F$-moves\footnote{There is an alternative sequence involving two $F$-moves and an $R$-move that achieves the same result.
As $F$-moves are typically computationally more demanding, the ``$RFR$'' sequence is however favored over the ``$FRF$'' sequence.}:
\begin{align}
\label{eq:RFRmove}
    \fusiontreemanipulations{3} &= \fusiontreemanipulations{4}
    = \sum_\rho \RSymbol{ec}{d}{\mu}{\rho} \fusiontreemanipulations{5} \\
                                &= \sum_{\rho} \sum_{\sigma f \kappa} \RSymbol{ec}{d}{\mu}{\rho} \FSymbol*{cab}{d}{\kappa,f,\sigma}{\nu,e,\rho}  \fusiontreemanipulations{6} \\
                                &=  \sum_\rho \sum_{\sigma f \kappa} \sum_\lambda\RSymbol{ec}{d}{\mu}{\rho} \FSymbol*{cab}{d}{\kappa,f,\sigma}{\nu,e,\rho} \RSymbol*{ac}{f}{\lambda}{\kappa} \fusiontreemanipulations{7}
\end{align}
Each of these steps involves specific transformations that reconfigure the fusion tree according to manipulations we defined previously.
}

To implement arbitrary permutations for fusion-splitting tree pairs efficiently, the process starts by converting all indices to a form amenable to splitting tree manipulations.
This is achieved by bending all indices to the splitting tree.
The permutation is then decomposed into a series of adjacent swaps, each resolved using the specified $R$- and $F$-moves.

\subsection{Index Manipulations}

Equipped with the mechanisms for manipulating fusion trees, we now extend these techniques to handle the index manipulations required for tensor maps with non-abelian symmetries.
This section outlines necessary adaptations to the algorithms presented in Section \ref{sec:abeliansymmetries}, focusing specifically on modifications to the storage scheme and methods for transposing, permuting and tracing indices.

For clarity, we will illustrate these techniques using the example of a $\SU{2}$-symmetric tensor map $t \colon V \otimes V \rightarrow V \otimes V$, where $V = \gradedspace{V}{0} \oplus \gradedspace{V}{\frac{1}{2}}$.
As discussed in Subsection~\ref{subsec:whatisfusiontree}, the reduced tensor elements of a symmetric tensor are associated with the valid canonical fusion-splitting tree pairs.
For this example, there are five distinct splitting trees (where the arrows are dropped for convenience):
\begin{equation}
\label{eq:fusionchannelssu2}
\begin{aligned}
    (1) = \diagram{su2fusionchannels}{1}&
    (2) = \diagram{su2fusionchannels}{4}&
    (3) = \diagram{su2fusionchannels}{2} \\
    (4) = \diagram{su2fusionchannels}{3}&
    (5) = \diagram{su2fusionchannels}{5}&
\end{aligned}
\end{equation}
Because the domain equals the codomain, valid fusion trees are the mirrored counterparts to those splitting trees, and valid pairs consist of all combinations with the same coupled charge.
In particular, the tensor map $t$ contains 9 reduced tensor elements $t^{(f,s)}$, namely for $(f,s)\in \{ (1,1), (1,2), (2,1), (2,2), (3,3), (3,4), (4,3), (4,4), (5,5)\}$ using the labelling from \eqref{eq:fusionchannelssu2}.

 \subsubsection{Grouping And Splitting Indices}
As furthermore discussed in Subsection~\ref{subsec:whatisfusiontree}, the valid fusion trees of a tensor product space can be used to define a grouping isomorphism to a single equivalent space, namely by collecting all fusion trees with a fixed coupled sector $c$, and combining them with a trivial grouping of the different outer indices associated with trees into a linear index.
This isomorphism can be used to write down the block-diagonal form of the tensor map $t$:
\begin{equation}
    \label{eq:nonabelianblockdiagonal}
    G^\dagger t G = \bigoplus_c \gradedtensor{t}{I; J}{c} \otimes \id{\gradedspace{V}{c}}
\end{equation}



We can illustrate this through the example of $t \colon V \otimes V \rightarrow V \otimes V$.
The 9 different reduced tensor elements are then organized into the following block-diagonal matrix form, where degeneracy indices have been omitted:
\begin{equation}
\begin{aligned}
    t &= \begin{pmatrix}
            t^{\left(1,1\right)} & t^{\left(1,2\right)} \\
            t^{\left(2,1\right)} & t^{\left(2,2\right)}
        \end{pmatrix} \otimes \gradedtensor{\mathbb{1}}{}{0}\\
    &\oplus \begin{pmatrix}
            t^{\left(3,3\right)} & t^{\left(3,4\right)} \\
            t^{\left(4,3\right)} & t^{\left(4,4\right)}
        \end{pmatrix} \otimes \gradedtensor{\mathbb{1}}{}{\frac{1}{2}}\\
    &\oplus \begin{pmatrix}
            t^{\left(5,5\right)}
        \end{pmatrix} \otimes \gradedtensor{\mathbb{1}}{}{1}
\end{aligned}
\end{equation}
This matrix form also highlights the computational advantages of non-abelian symmetries.
Without these symmetry constraints, specifying a tensor map representation would require $(\dim{V})^2 \times (\dim{V})^2 = 9 \times 9 = 81$ coefficients, in stark contrast to the mere nine degrees of freedom required here.

\subsubsection{Transpositions}

Transpositions involve permuting the indices of the reduced tensor coefficients and demand a recoupling of the fusion diagrams, effectively resulting in a re-summation procedure.

For illustration, we consider cycling the indices of a tensor map $t$ clockwise.
This operation transforms $t$ to a tensor map $s \colon V^* \otimes V \rightarrow V \otimes V^*$, yielding a tensor map with the same structural composition as \eqref{eq:fusionchannelssu2} but with altered arrow orientations.
Employing the techniques from Section \ref{sec:fusiontreemanipulations}, the transformation matrix that restores canonical basis to the nine fusion channels can be computed, and if we represent these fusion channels according to their linear order in a column major matrix, this leads to:
\begin{equation}
    \fusiontreemanipulations{50} = \begin{pmatrix}
        1 & \cdot & \cdot & \cdot & \cdot & \cdot & \cdot & \cdot & \cdot \\
        \cdot & \cdot & \cdot & \cdot & \cdot & \cdot & \cdot & \frac{1}{\sqrt{2}} & \cdot \\
        \cdot & \cdot & \cdot & \cdot & -\frac{1}{\sqrt{2}} & \cdot & \cdot & \cdot & \cdot \\
        \cdot & \cdot & \cdot & \frac{-1}{2} & \cdot & \cdot & \cdot & \cdot & -\frac{1}{2} \\
        \cdot & \sqrt{2} & \cdot & \cdot & \cdot & \cdot & \cdot & \cdot & \cdot \\
        \cdot & \cdot & \cdot & \cdot & \cdot & \cdot & 1 & \cdot & \cdot \\
        \cdot & \cdot & \cdot & \cdot & \cdot & 1 & \cdot & \cdot & \cdot \\
        \cdot & \cdot & -\sqrt{2} & \cdot & \cdot & \cdot & \cdot & \cdot & \cdot \\
        \cdot  & \cdot & \cdot & -\frac{3}{2} & \cdot & \cdot & \cdot & \cdot & \frac{1}{2} \\
    \end{pmatrix} \cdot
    \fusiontreemanipulations{51}.
\end{equation}

Consequently, the transposed tensor map $s$ is computed to be:
\begin{equation}
\begin{aligned}
    s &= \begin{pmatrix}
            t^{\left(1,1\right)} & -\sqrt{2} t^{\left(3,3\right)} \\
            \sqrt{2} t^{\left(4,4\right)} & -\frac{1}{2} t^{\left(2,2\right)} -\frac{3}{2} t^{\left(5,5\right)} 
        \end{pmatrix} \otimes \gradedtensor{\mathbb{1}}{}{0} \\
    &\oplus \begin{pmatrix}
            -\frac{1}{\sqrt{2}} t^{\left(2,1\right)} & t^{\left(4,3\right)} \\
            t^{\left(3,4\right)} & \frac{1}{\sqrt{2}} t^{\left(1,2\right)}
        \end{pmatrix} \otimes \gradedtensor{\mathbb{1}}{}{\frac{1}{2}} \\
    &\oplus \begin{pmatrix}
            -\frac{1}{2} t^{\left(2,2\right)} + \frac{1}{2} t^{\left(5,5\right)} 
        \end{pmatrix} \otimes \gradedtensor{\mathbb{1}}{}{1}.
\end{aligned}
\end{equation}
This example illustrated that, compared to the abelian case, transpositions do not simply amount to simply reshuffling the reduced tensor elements, but generalize to taking linear combinations.
Other transpositions, as well as the other index manipulations discussed below, follow the same approach of computing a transition matrix for the fusion diagrams and subsequently expressing the new reduced tensor coefficients as linear combinations of the original ones.

In this example, the reduced tensor components are scalars, but in general they will be multidimensional arrays with dimensions given by the degeneracies of the uncoupled charges of the fusion - splitting tree pair in their respective spaces.
In that case, the tensor components themselves also need to be permuted, exactly as in the abelian case.

\subsubsection{Permutations}

Permuting tensor maps requires a similar approach to transpositions, where the fusion diagrams are now braided and recoupled to compute the transition matrix.
This matrix then fuels the re-summation procedure to the permuted reduced tensor coefficients.
To demonstrate, we consider swapping the indices of both the domain and codomain of the tensor map $t$, resulting in a new tensor map $s \colon V \otimes V \rightarrow V \otimes V$.

The transition matrix for this permutation is calculated using the braiding techniques discussed in Section \ref{sec:nonabelianbraiding}.
The resulting matrix is computed to be:
\begin{equation}
    \fusiontreemanipulations{52} = \begin{pmatrix}
        1 & \cdot & \cdot & \cdot & \cdot & \cdot & \cdot & \cdot & \cdot \\
        \cdot & -1 & \cdot & \cdot & \cdot & \cdot & \cdot & \cdot & \cdot \\
        \cdot & \cdot & -1 & \cdot & \cdot & \cdot & \cdot & \cdot & \cdot \\
        \cdot & \cdot & \cdot & 1 & \cdot & \cdot & \cdot & \cdot & \cdot \\
        \cdot & \cdot & \cdot & \cdot & \cdot & \cdot & \cdot & 1 & \cdot \\
        \cdot & \cdot & \cdot & \cdot & \cdot & \cdot & 1 & \cdot & \cdot \\
        \cdot & \cdot & \cdot & \cdot & \cdot & 1 & \cdot & \cdot & \cdot \\
        \cdot & \cdot & \cdot & \cdot & 1 & \cdot & \cdot & \cdot & \cdot \\
        \cdot  & \cdot & \cdot & \cdot & \cdot & \cdot & \cdot & \cdot & 1 \\
    \end{pmatrix} \cdot
    \fusiontreemanipulations{53}
\end{equation}

Therefore, the permuted tensor map $s$ becomes:
\begin{equation}
\begin{aligned}
    s &= \begin{pmatrix}
            t^{\left(1,1\right)} & -t^{\left(1,2\right)} \\
            -t^{\left(2,1\right)} & t^{\left(2,2\right)}
        \end{pmatrix} \otimes \gradedtensor{\mathbb{1}}{}{0}\\
    &\oplus \begin{pmatrix}
            t^{\left(4,4\right)} & t^{\left(4,3\right)} \\
            t^{\left(3,4\right)} & t^{\left(3,3\right)}
        \end{pmatrix} \otimes \gradedtensor{\mathbb{1}}{}{\frac{1}{2}}\\
    &\oplus \begin{pmatrix}
            t^{\left(5,5\right)}
        \end{pmatrix} \otimes \gradedtensor{\mathbb{1}}{}{1}
\end{aligned}
\end{equation}

This method again illustrates how tensor map permutation can be achieved in general, by computing the transition matrix for the fusion diagrams and subsequently expressing the new reduced tensor coefficients as linear combinations of the original ones.

\subsubsection{Traces}

For (partial) traces, the process remains similar.
This operation involves connecting one or more pairs of the fusion tree legs.
The resulting diagram can be manipulated back to the canonical basis by first making sure the paired legs are adjacent and then recoupling the result to a tadpole diagram.
This tadpole can be neutralised by virtue of \eqref{eq:tadpole}, resulting in a canonical fusion tree.

To illustrate, we obtain the following transformation matrix for a partial trace of a tensor map $t \colon V \otimes V \rightarrow V \otimes V$:
\begin{equation}
    \fusiontreemanipulations{54} = \begin{pmatrix}
        1 & \cdot \\
        \cdot & \cdot \\
        \cdot & \cdot \\
        \cdot & 0.5 \\
        2 & \cdot \\
        \cdot & \cdot \\
        \cdot & \cdot \\
        \cdot & 1 \\
        \cdot & 1.5 \\
    \end{pmatrix} \cdot
    \fusiontreemanipulations{55}
\end{equation}
As a result, the traced tensor map $\tilde{t} \colon V \rightarrow V$ is computed to be:
\begin{equation}
    \tilde{t} = \left( \left(\gradedtensor{t}{}{1,1} + 2 \gradedtensor{t}{}{4,4}\right) \otimes \gradedtensor{\mathbb{1}}{}{0} \right)
    \oplus \left( \left(0.5 \gradedtensor{t}{}{2,2} + \gradedtensor{t}{}{3,3} + 1.5 \gradedtensor{t}{}{5,5} \right) \otimes \gradedtensor{\mathbb{1}}{}{\frac{1}{2}} \right)
\end{equation}

\subsection{Contractions and compositions, adjoints, and decompositions}

The grouping isomorphism that transform a tensor map into a block diagonal matrix representation are no longer simple permutation matrices, as they are constructed using the fusion and splitting tensors.
However, they are still unitary.
As a consequence, there is no need to adapt the techniques from Section \ref{sec:abeliansymmetries}, and the necessary computations can be performed directly on the blocks $\gradedtensor{t}{}{c}$

For contractions, after the necessary transpositions or permutations, they amount to compositions of the maps involved, where the unitary grouping isomorphism in the domain of the leftmost map cancels with the adjoint thereof appearing in the codomain of the rightmost tensor map.
Similarly, applying the adjoint to the tensor map naturally interchanges and transforms the grouping maps in the domain and codomain, leaving an adjoint on the block diagonal matrix representation of the tensor map.
Finally, the same holds true for tensor decompositions involving unitary or isometric maps, where furthermore new indices always appear as a single space without tensor product structure, which do not require additional grouping isomorphisms. 

Furthermore, there is an improved computational benefit over the abelian case, due to the presence of identity morphisms on the internal labels associated with the irrep spaces $\gradedspace{V}{c}$.
Without imposing the symmetry, these would result in the block with coupled charge $c$ appearing $d_c$ times.
Here, we only need to apply the computation to each block $\gradedtensor{t}{}{c}$ once, instead of $d_c$ times.



\section{Categorical Symmetries}
\label{sec:category}

This section provides an overview of tensor categories, with the aim of establishing a framework for understanding tensor maps.
The central question we address is: "What constitutes a consistent framework for tensor maps?"
Alternatively, because of the heavy use of the graphical calculus, we could also ask: "What makes up consistent rules for manipulating diagrams?"
The succinct answer is that tensor maps are morphisms in a unitary fusion category, which may optionally be braided.
Our discussion is not intended to serve as an exhaustive introduction to category theory.
Instead, it aims to connect categorical concepts with their practical applications in tensor maps.

While this introduction aims to acquaint readers with the language of category theory and provide insights into how these theoretical constructs relate to practical applications in tensor maps, it is not intended to be mathematically rigorous or exhaustive.
Instead, we selectively explore topics and theorems that support a coherent analytical framework and offer references for those seeking a deeper exploration \cite{beer2018categoriesanyonstravelogue,etingofTensorCategories2015, turaevMonoidalCategoriesTopological2017, mugerStructureModularCategories2003, kasselQuantumGroups1995, selingerSurveyGraphicalLanguages2011, nLab_2008}.
In fact, as will become clear, we have already encountered all the relevant mathematical concepts and structures in the discussion of the (irreducible) representations of groups, and only a handful of further generalizations are required.

We begin by exploring the qualifiers used in describing categories such as \emph{unitary}, \emph{fusion}, and \emph{braided}, clarifying their significance for tensor maps.
This discussion will elucidate the foundational aspects that make tensor maps consistent and functional within this categorical framework.

Following this, the complexities and nuances that arise when generalizing beyond group representations are examined, focusing on their implementation within \TensorKit.
These subtleties highlight the broader applicability and flexibility of the categorical framework, demonstrating its capacity to handle a diverse range of symmetries and representations.

To make these theoretical insights more tangible, we illustrate through examples how this formalism encompasses group representations already familiar to us.
Additionally, we explore how the framework extends to systems with fermionic characteristics and incorporates theories of anyonic statistics.
These examples not only broaden the theoretical foundation but also enhance the practical utility of tensor maps in diverse quantum systems.

\subsection{What Is A Fusion Category?}

The connection between category theory and tensor maps is not as remote as it might initially appear.
The basic definition of a category $\category{C}$ involves
\begin{itemize}
    \item a collection of objects $V, W, \ldots \in \Ob(\category{C})$;
    \item a set of morphisms between every pair of objects $V$ and $W$, denoted as $\Hom(V, W)$;
    \item an associative composition law for morphisms $f \colon V \rightarrow W$ and $g \colon W \rightarrow Z$, resulting in $g \circ f \colon V \rightarrow Z$;
    \item an identity morphism $\id{V}: V\to V$ for every object $V$, such that $\id{W} \circ f = f = f \circ \id{V}$ for all $f:V \to W$.
\end{itemize}
This is analogous to the action of tensor maps (morphisms) on vector spaces (objects).
Tensor maps, however, possess more structure, allowing for numerous additional operations defined by certain constraints.
Our focus here is to identify which of these structures align with this approach, inspired by the structure of fusion trees.

For vector spaces, all $d$-dimensional vector spaces are isomorphic, so often it can be convenient to consider \emph{the} $d$-dimensional vector space as a single entity, instead of all the isomorphic objects separately.
A similar statement can be made about the tensor maps, as we typically consider all $(d_1 \times d_2)$-dimensional morphism spaces as a single entity.
As a result, the required structures and constraints can often be expressed \emph{up to isomorphism}, while still retaining the essential properties.
Formalizing these notions gives rise to the concept of a \emph{fusion category}.

A fusion category is characterized by being rigid, semi-simple, linear ($\category{Vect}$-enriched), and monoidal with a simple unit.
It encompasses only finitely many isomorphism classes of simple objects, which aligns with the practical limitations of finite memory in computers\footnote{In \TensorKit, this condition is relaxed slightly, since we also allow categories that are not fusion but still semi-simple and have an infinite set of simple objects, but where each object is finitely generated, \ie~ locally finite categories.
This allows us to also include the representation theory of the classical semi-simple Lie groups.}.

Unpacking this definition, we start with linearity.
To preserve the framework of linear algebra, morphisms must form vector spaces over some number field, which we typically take to be the field of complex numbers $\complexs$.
This requirement ensures that tensor maps allow for addition and scalar multiplication in a consistent manner.

Furthermore, in order to retain Schur's lemma, we also require semi-simplicity.
This requirement allows the decomposition of objects into a direct sum of simple objects, which can be defined as objects for which $\End(a) \simeq \field{C}$.
Then, Schur's lemma dictates that morphisms between simple objects are either isomorphisms or zero morphisms.
In particular, for an object $V = W_1 \oplus W_2$, this requires the existence of inclusion maps $i_\alpha$ and projection maps $q_\alpha$ that provide a natural generalization for writing a tensor map as a a direct sum of fusion channels:
\begin{align}
    q_\alpha \circ i_\beta &= \delta_{\alpha\beta}\id{W_\alpha} \\
    \sum_\alpha i_\alpha \circ q_\alpha &= \id{V}
\end{align}

The monoidal structure refers to the capability of taking tensor products.
This implies the existence of a binary operation, denoted $\otimes$, which acts on objects to produce new objects and on morphisms to produce new morphisms.
A unit object $I \in \Ob(C)$ is also required for this operation, such that $I \otimes V \simeq V \simeq V \otimes I$ for any object $V$.
For technical reasons, we restrict the discussion to the cases where the unit object is simple.\footnote{This restriction distinguishes fusion categories from multi-fusion categories, a more general concept that is also useful in phsyics and for which \TensorKit support is currently being developed \cite{borisdevos}.}
Furthermore, this operation must be associative, which requires the existence of an isomorphism $\alpha_{V_1,V_2,V_3} \colon (V_1 \otimes V_2) \otimes V_3 \rightarrow V_1 \otimes (V_2 \otimes V_3)$.

Rigidity encompasses the ability to define dual objects, as well as to provide an exact pairing between objects and their duals.
This is captured in additional maps, which represent the different ways of bending lines.
In the categorical literature, they are commonly referred to as the left (right) evaluation and co-evaluation maps, denoted with $\eta$ ($\tilde{\eta}$) and $\epsilon$ ($\tilde{\epsilon}$).

The existence of a swapping isomorphism $\tau \colon V \otimes W \rightarrow W \otimes V$ is captured by braided categories.
The term \emph{braided} also captures that this operation is not necessarily symmetric, \ie~ $\tau_{V\otimes W}^{-1}$ is not necessarily equal to $\tau_{W \otimes V}$, leading to additional complexities as over-crossings and under-crossings need to be distinguished.
We also note that this structure is optional, as there are applications that do not require the ability to swap indices, such as most algorithms in the context of the one-dimensional matrix product states.

A categorical structure that is relevant to physics, and in particular quantum physics, is the notion of an adjoint or dagger.
This structure defines an anti-linear map $\dagger$ on morphisms $f \colon V \rightarrow W$ by associating a morphism $f^\dagger \colon W \rightarrow V$ so that $(f \circ g)^{\dagger} = g^\dagger \circ f^\dagger$ and $(f^\dagger)^\dagger = f$.
It is furthermore assumed that $f^\dagger \circ f$ is nonzero for any nonzero morphism $f$, and can then define $\langle f, g\rangle = \Tr{f^\dagger \circ g}$ as inner product between morphisms $f, g \in \Hom(V, W)$.
Finally, we can define inner-product preserving maps (\emph{unitary} maps), which are isomorphisms $f \colon V \rightarrow W$ for which $f^{-1} = f^\dagger$.

To avoid categorical technicalities, we will always assume unitary fusion categories, which imposes several additional constraints on the other structures \cite{etingofFusionCategories2005}.
In particular, the inclusion maps $i_\alpha$ should be isometric such that the corresponding projections are $q_\alpha = i_\alpha^\dagger$, and $V = W_1 \oplus W_2$ is an orthogonal direct sum decomposition.
Furthermore, the associators $\alpha$ as well as the braiding maps $\tau$ are all required to be unitary, and $(f \otimes g)^\dagger = f^\dagger \otimes g^\dagger$.
Finally, the evaluation and co-evaluation maps $\ev{}$ and $\coev{}$ also have some form of unitarity requirement, in the sense that $\ev{a}^\dagger = \coev{a^*}$ and $\coev{a}^\dagger = \ev{a^*}$.


\subsection{Topological Data}

We now examine more closely the data required for completely fixing such a category $\category{C}$, along with the consistency equations that they must satisfy.
In particular, we could equally well take these data as the definition, allowing \TensorKit to support all possible implementations of these data that are consistent.

\subsubsection{Objects And Morphisms}

The first piece of data we require is the generalization of the vector spaces, which form the objects $\Ob(\category{C})$ of the category.
Because of the semi-simplicity, we know that any object $V$ is isomorphic to a direct sum of a finite number of simple objects.

As we have already encountered in the case of group representations, it is possible for different simple objects to be isomorphic, and any one of those equivalent choices can be used in the direct sum decomposition of a general object.
Hence, the first defining piece of data is the set $\Irr(\category{C})$ of all isomorphism classes of simple objects, where henceforth we substitute the classes by a single fixed representative.
Put differently, for every $a, b \in \Irr(\category{C})$, Schur's lemma yields $\dim(\Hom(a, b)) = \delta_{a,b}$.
In a fusion category, one of these objects should correspond to the monoidal unit $I$, which we typically take to be the first object when $\Irr(\category{C})$ is represented as a list or an iterator.

Within \TensorKit, we allow for $\Irr(\category{C})$ itself to be an infinite set.
General objects $V$ can then build as finite direct sums of objects in $\Irr(\category{C})$:
\begin{equation}
V = \bigoplus_{a \in \Irr(\category{C})} \bigoplus_{\mu=1}^{\NSymbol{V}{a}} a\ .
\end{equation}
The number of times a simple object $a$ appears is referred to the degeneracy $\NSymbol{V}{a} = \NSymbol{a}{V}$, which corresponds to $\dim(\Hom(V, a)) = \dim(\Hom(a, V))$.
Therefore, general objects are represented as a list of pairs $(a \in \Irr(\category{C}), \NSymbol{V}{a} \in \mathbb{N})$.
However, general objects in \TensorKit also carry a duality specifier, and we will separately store objects that correspond to tensor products of two or more of these direct sum objects.
Hence, different general objects can be constructed that may be isomorphic.

The morphism space $\Hom(W, V)$ is then given by
\begin{equation}
\label{eq:morphismspacereduced}
    \Hom(W,V) \simeq \bigoplus_{a \in \Irr(\category{C})} \Hom(W, a) \otimes \Hom(a, V) 
        \simeq \bigoplus_{a \in \Irr(\category{C})} \field{C}^{\NSymbol{V}{a} \times \NSymbol{W}{a}} \otimes \End(a).
\end{equation}
with $\End(a)$ the one-dimensional space generated by $\id{a}$.
Hence, the morphisms between $W$ and $V$ are equivalent to a set of matrices of size $\NSymbol{V}{a} \times \NSymbol{W}{a}$, one for each simple object $a \in \Irr(\category{C})$.
This parameterization constitutes the basis for our representation of general tensor maps in \TensorKit, in complete analogy to how they were defined before.

\subsubsection{$N$-Symbols}

For monoidal categories, we must have that $a \otimes b \in \Ob(\category{C})$, such that it is equal to the direct sum of simple objects as well.
This is captured in non-negative integers $\NSymbol{ab}{c}$ for each triple of simple objects, which denote the multiplicity of $c \in \Irr(\category{C})$ in $a \otimes b$.
In order for the tensor product to have a unit, be associative, and have a dual, these $N$-symbols must obey:
\begin{align}
    &\NSymbol{Ia}{b} = \NSymbol{aI}{b} = \delta_{ab} \\
    &\NSymbol{abc}{d} = \sum_{e \in \Irr(\category{C})}\NSymbol{ab}{e} \NSymbol{ec}{d} = \sum_{f\in \Irr(\category{C})} \NSymbol{af}{d} \NSymbol{bc}{f} \\
    &\forall a \in \Irr(\category{C}) \text{, there is a unique } \bar{a} \in \Irr(\category{C}) \text{, with } \NSymbol{ab}{I} = \NSymbol{ba}{I} = \delta_{b\bar{a}}
\end{align}

Similar to the dimension of a vector space, we can associate a notion of dimension $d_a$ to each $a \in \Irr(\category{C})$, called the \emph{Frobenius-Perron dimension} or \emph{quantum dimension}.
For unitary fusion categories, this is fully fixed by requiring $d_a > 0$, along with the condition that the dimensions of the simple objects form a one-dimensional representation of the fusion rules:
\begin{equation}
\label{eq:fusiondimrep}
    d_a d_b = \sum_{c} \NSymbol{ab}{c} d_c.
\end{equation}
For a proof of this statement, we refer to \cite[Prop.~8.23]{etingofFusionCategories2005}.

Whereas in the case of (irreducible) representations of groups, the quantum dimensions are integers and coincide with the dimensions of the representation spaces, in a general fusion category, quantum dimensions can be arbitrary real numbers and there are not actual vector spaces that are intrinsically associated with the irreducible objects of a fusion category $\category{C}$.

\subsubsection{Fusion Spaces}

The morphisms $a \otimes b \rightarrow c$ then form a $\NSymbol{ab}{c}$-dimensional $\field{C}$-vector space for each triple of simple objects.
The specific projection maps $q_c = \XMatrix{c;\mu}{ab} \colon a \otimes b \rightarrow c$ make up the fusion tensors with $1 \leq \mu \leq \NSymbol{c}{ab}$ and provide a basis for the space $\Hom(a \otimes b, c)$.
Similarly, the inclusion maps $i_c = \XSymbol{ab}{c}{\mu} \colon c \rightarrow a \otimes b$ make up the splitting tensors, providing a basis for $\Hom(c, a \otimes b)$.
The splitting tensors can be chosen in relation to the fusion tensors so that they satisfy
\begin{equation}
    \XMatrix{c;\mu}{ab} \circ \XSymbol{ab}{d}{\nu} = \delta_d^c \delta_\nu^\mu \id{c}\ .
\end{equation}
and
\begin{equation}
    \sum_{c,\mu} \XSymbol{ab}{c}{\mu}\circ \XMatrix{c;\mu}{ab} = \id{a \otimes b}\ .
\end{equation}

Unitarity of the fusion category further implies that this choice of splitting tensors can be obtained as adjoints of the corresponding fusion tensors:
\begin{equation}
    \XSymbol{ab}{c}{\mu} = \XMatrix{c;\mu}{ab}^\dagger\ .
\end{equation}

Note that exact expressions for these maps are not strictly required for working with tensor maps, so we can leave these as abstract objects.
In other words, we only state that a basis for these spaces exists.
We do not need to specify the exact form of these basis elements.
The only requirement is that these abstract objects satisfy the properties outlined throughout the remainder of this section.

\subsubsection{$F$-Symbols}

The associators $\alpha_{abc} \colon (a \otimes b) \otimes c \rightarrow a \otimes (b \otimes c)$ can be used to relate the different choices of basis, corresponding to different orders of combining fusion tensors, where the matrix representation of this basis change is known as the $F$-symbol $\FSymbol{abc}{d}{}{}$.
In a unitary category, this matrix is unitary and can be related to the fusion and splitting tensors as follows:
\begin{equation}
    \label{eq:associator}
    \alpha_{abc} \circ \left(\XSymbol{ab}{e}{\mu} \otimes \id{c}\right) \circ \XSymbol{ec}{d}{\nu} 
    = \sum_{\kappa f \lambda}\FSymbol{abc}{d}{\mu,e,\nu}{\lambda,f,\kappa}  \circ \left(\id{a} \otimes \XSymbol{bc}{f}{\lambda}\right) \circ \XSymbol{af}{d}{\kappa}
\end{equation}

In order for these associators to be consistent, we further require that any two ways of mapping between different fusion orders must agree, as expressed by the triangle and pentagon equations.
The \emph{triangle equation} is obtained by expressing the two possible ways to map from $((a \otimes I) \otimes b)$ to $(a \otimes (I \otimes b))$:
\begin{align}
    \diagram{categories}{2} 
    \xrightarrow{\FSymbol{aIb}{c}{1,a,\mu}{1,b,\nu}} &\diagram{categories}{3} 
        \xrightarrow{I \otimes b = b} \diagram{categories}{1} \\
    \diagram{categories}{2}
        &\xrightarrow{a \otimes I = a} \diagram{categories}{1}
\end{align}
This leads to the following constraints on the $F$-symbols:
\begin{align}
    \FSymbol{Iab}{c}{1,a,\mu}{\nu,c,1} &= \delta_\mu^\nu~,
        &\FSymbol{aIb}{c}{1,a,\mu}{1,b,\nu} &= \delta_\mu^\nu~,
        &\FSymbol{abI}{c}{1,c,\mu}{1,b,\nu} &= \delta_\mu^\nu
\end{align}

The \emph{pentagon equation} expresses that the two possible ways to map between $(((a \otimes b) \otimes c) \otimes d)$ and $(a \otimes (b \otimes (c \otimes d)))$ must agree.
\begin{align}
    \diagram{categories}{4}
    \xrightarrow{\FSymbol{fcd}{e}{\kappa,g,\mu}{\nu,h,\gamma}} &\diagram{categories}{5}
    \xrightarrow{\FSymbol{abh}{e}{\lambda,f,\gamma}{\rho,i,\sigma}} \diagram{categories}{6} \\
    \diagram{categories}{4}
    \xrightarrow{\FSymbol{abc}{g}{\lambda,f,\kappa}{\alpha,j,\beta}} \diagram{categories}{7}
    &\xrightarrow{\FSymbol{ajd}{e}{\beta,g,\mu}{\tau,i,\sigma}} \diagram{categories}{8}
    \xrightarrow{\FSymbol{bcd}{i}{\alpha,j,\tau}{\nu,h,\rho}} \diagram{categories}{6}
\end{align}
As a result, the pentagon equation for the $F$-symbols reads:
\begin{equation}
    \label{eq:pentagon}
    \sum_{\gamma} \FSymbol{fcd}{e}{\kappa,g,\mu}{\nu,h,\gamma} \cdot \FSymbol{abh}{e}{\lambda,f,\gamma}{\rho,i,\sigma}
    = \sum_{\beta j \alpha} \sum_{\tau} \FSymbol{abc}{g}{\kappa,f,\lambda}{\alpha,j,\beta} \cdot \FSymbol{ajd}{e}{\beta,g,\mu}{\tau,i,\sigma} \cdot \FSymbol{bcd}{i}{\alpha,j,\tau}{\nu,h,\rho}
\end{equation}

A result known as the coherence theorem for monoidal categories, proven by MacLane \cite{MacLane:1963unh}, further states that the triangle and pentagon equations are sufficient to ensure the consistency of any other diagrammatic manipulation involving recouplings and insertions of the unit.
This theorem thus serves as a powerful underpinning that validates our usage of the diagrammatic notation for tensor maps, as well as the consistency of the manipulations defined in \ref{sec:fusiontreemanipulations}.


\subsubsection{Duality}

The duality of objects, or the ability to bend lines, is captured by the existence of two families of morphisms, the evaluation and coevaluation maps $\ev{}$ and $\coev{}$.
\begin{align}
    \ev{a} \colon a^* \otimes a \rightarrow I,\qquad & \coev{a} \colon I \rightarrow a \otimes a^* \\
    \ev{a^*} \colon a \otimes a^* \rightarrow I, \qquad& \coev{a^*} \colon I \rightarrow a^* \otimes a
\end{align}
These maps are required to satisfy the following \emph{snake equations}:
\begin{align}
    \label{eq:snake}
     \left( \id{a} \otimes \ev{a} \right) \circ \left( \coev{a} \otimes \id{a} \right)
        &= \id{a} 
        = \left( \ev{a^*} \otimes \id{a} \right) \circ \left( \id{a} \otimes \coev{a^*} \right),&
    \diagram{categories}{14} 
        = \diagram{categories}{12} 
        = \diagram{categories}{9}\\
    \left( \ev{a^*} \otimes \id{a^*} \right) \circ \left( \id{a^*} \otimes \coev{a^*} \right)
        &= \id{a^*} 
        = \left( \id{a^*} \otimes \ev{a} \right) \circ \left( \coev{a} \otimes \id{a^*} \right),&
    \diagram{categories}{13} 
        = \diagram{categories}{11} 
        = \diagram{categories}{10}
\end{align}

These maps provide an exact pairing between objects and their duals, enabling the transitions $\Hom(W, V) \simeq \Hom(I, V \otimes W^*) \simeq \Hom(V^*, W^*)$.
When considering the dual of simple objects, the fixed list of representatives $\Irr(\category{C})$ causes a complication, in that now the dual $a^*$ may not be contained in that list directly. This necessitates introducing the isomorphism $Z_a: a^* \to \bar{a}$. The coevaluation map $\eta_a \colon I \rightarrow a \otimes a^*$ can then be combined with $Z_a$ to yield a morphism in $\Hom(I, a \otimes \bar{a})$, containing the single linearly independent element $\XSymbol{a\bar{a}}{I}{1}$. Hence, we can relate both quantities and make the specific choice
\begin{align}
    \XSymbol{a\bar{a}}{I}{1} = \frac{1}{\sqrt{d_a}} (\id{a} \otimes Z_a) \circ \coev{a} = \frac{1}{\sqrt{d_a}} (Z^T_{a} \otimes \id{\bar{a}}) \circ \coev{\bar{a}}\ ,
\end{align}
where we have also chosen the normalization so $\tr(\id{a}) = \eta_a \circ \tilde{\epsilon}_a = d_a$\footnote{A different convention that often occurs is the isotopic normalisation, which would make these numbers $1$ and make diagrams invariant under the actions of (co)evaluation maps.}.

Recall that we also need to be slightly careful with relating $Z^T_{a}$ and $Z_{\bar{a}}$, which need only be isomorphic, and can thus differ up to a phase, \ie~ $Z^T_a = \chi_a Z_{\bar{a}}$.
Though not required, this phase can be absorbed in the definition of $Z_{\bar{a}}$ whenever $a \neq \bar{a}$.
In contrast, when $a = \bar{a}$, we find $Z_a = (Z^T_a)^T = \chi_a Z^T_{\bar{a}} = \chi_a \chi_a Z_a$, such that $\chi_a = \pm 1$ but cannot be eliminated.

All of this data is however already encoded in the $F$-symbols and therefore does not constitute additional data of the fusion category.
In particular, we note the following diagrammatic relations:
\begin{equation}
    \frac{\chi_a}{d_a} \diagram{categories}{12} = \frac{1}{d_a} \diagram{categories}{15} = \frac{1}{d_a} \diagram{categories}{16} = \diagram{categories}{17} = \FSymbol{a\bar{a}a}{a}{I}{I} \diagram{categories}{12}
\end{equation}
such that 
\begin{align}
    \frac{1}{d_a} &= \left| \FSymbol{a\bar{a}a}{a}{1,I,1}{1,I,1} \right|, &
    \chi_a &= \mathrm{sign}\left( \FSymbol{a\bar{a}a}{a}{1,I,1}{1,I,1} \right)
\end{align}

Additionally, the following diagrams elegantly prove that our normalization choice coincides with the dimensions of the simple objects, \ie\ $d_a$ coincides with the definition of \eqref{eq:fusiondimrep}:
\begin{equation}
    d_a d_b = \diagram{categories}{31} 
        = \sum_{\nu c \mu} \diagram{categories}{32} 
        = \sum_{\nu c \mu} \diagram{categories}{33}
        = \sum_{c,\mu} \diagram{categories}{34}
        = \sum_{c} \NSymbol{ab}{c} d_c 
\end{equation}

As a final remark on duality, we note that a priori it might not be clear that this concept is left-right symmetric, \ie~ that $a^* = {}^*a$.
However, in a unitary fusion category, there is a canonical choice of the evaluation and coevaluation maps that satisfies the snake equations and ensures that this is the case.
This is a direct corollary of Proposition 4.8.1 in \cite{etingofTensorCategories2015}.
In particular, this also implies that left and right traces coincide, such that $\tr_l(\id{a}) = d_a = \tr_r(\id{a})$, or in other words that left and right dimensions coincide.
As such, we will assume these properties throughout \TensorKit, and will not distinguish between left and right duals, traces or dimensions.

\subsubsection{$R$-Symbols}

Finally, we study the possible braiding structure of a unitary fusion category.
The existence of a braiding isomorphism $\tau_{V,W} \colon V \otimes W \rightarrow W \otimes V$ requires at the very least that $\NSymbol{ab}{c} = \NSymbol{ba}{c}$.
In a unitary category, this isomorphism is unitary and can be expressed in the splitting basis as:
\begin{equation}
    \label{eq:braiding}
    \tau_{a,b} \circ \XSymbol{ab}{c}{\mu} = \sum_\nu \RSymbol{ab}{c}{\mu}{\nu} \XSymbol{ba}{c}{\nu}
\end{equation}

In order for this braiding to be consistent, the following two relations must hold:
Firstly, the braiding must be consistent with the tensor product (formally referred to as \emph{naturality}), which is expressed as
\begin{equation}
    \diagram{categories}{18} = \diagram{categories}{19}.
\end{equation}

Secondly, there is a coherence condition called the \emph{hexagon equation}, which expresses the consistency of the braiding with the associators.
\begin{align}
    \diagram{categories}{20}
    &\xrightarrow{\RSymbol{cd}{e}{\nu}{\rho}} \diagram{categories}{24}
    \xrightarrow{\FSymbol*{dab}{e}{\sigma,g,\gamma}{\mu,c,\rho}} \diagram{categories}{25}
    \xrightarrow{\RSymbol*{ad}{g}{\alpha}{\sigma}} \diagram{categories}{23} \\
    \diagram{categories}{20}
    &\xrightarrow{\FSymbol{abd}{e}{\mu,c,\nu}{\kappa,f,\lambda}} \diagram{categories}{21}
    \xrightarrow{\RSymbol{bd}{f}{\kappa}{\theta}} \diagram{categories}{22}
    \xrightarrow{\FSymbol*{adb}{e}{\alpha,g,\beta}{\theta,f,\lambda}} \diagram{categories}{23}
\end{align}

In a braided unitary fusion category, we can also define the so-called \emph{twist} $\theta_a \colon a \rightarrow a$, which corresponds to a \emph{self-braiding} of the object.
For simple objects, this map is an element of $\End(a) \simeq \field{C}$, and as a result, consists of a scalar times the identity.
In particular, unitarity implies that it is a phase:
\begin{equation}
    \diagram{categories}{26} = \sum_{b\mu} \diagram{categories}{27} = \sum_{b\mu\nu} \RSymbol{aa}{b}{\mu}{\nu} \diagram{categories}{28} = \sum_{b\mu} \frac{d_b}{d_a} \RSymbol{aa}{b}{\mu}{\mu} \diagram{categories}{29}
\end{equation}
\begin{equation}
    \theta_a = \sum_{b\mu} \frac{d_b}{d_a} \RSymbol{aa}{b}{\mu}{\mu} \id{a}
\end{equation}
Whenever this twist is non-trivial, we therefore have to be careful with our diagrammatic manipulations, as we are no longer allowed to twist or untwist lines without consequences.

\subsection{Index and tensor manipulations}
The topological data of a (braided) fusion category specified in the previous paragraph coincides with that of the representations of non-abelian groups (see also Subsection~\ref{subsec:examples}).
We can thus provide the most general definition of a tensor map in \TensorKit as being a morphism in such a category.
In particular, the properties of these categories are precisely what is required to ensure that the operations we defined in \ref{sec:indexmanipulations} remain consistent.

The main difference to the case of group-based symmetries is that, in the general case, the fusion and splitting tensors are abstract objects that merely define a basis for the $\Hom$-spaces.
However, all that is needed is the transformation properties of these objects under elementary manipulations, which is exactly what the topological data provides.
The techniques and algorithms for manipulating tensor maps discussed in the previous section on non-abelian symetries can thus be directly replicated for categorical symmetries, because we have been careful not to rely on an explicit representation of the fusion and splitting tensors in any step.
In particular, the abstract fusion tensors also define (abstract) unitary grouping isomorphisms that map the domain and codomain of any tensor map to an isomorphic object that is a direct sum of simple objects in explicit form, and with respect to which the tensor map acquires the block diagonal form dictated by \eqref{eq:morphismspacereduced}.

There is one remaining complication when specifying more complicated operations such as a full tensor network contraction, resulting from the fact that representations of groups have trivial braiding properties resulting from the ordinary interchange of the representation spaces in the tensor product.
This subtlety is addressed in the next subsection.

\subsection{Planar Operations}
\label{sec:planaroperations}

Conventionally, a tensor network is specified by a graph, where the only information that is provided is the connectivity of the edges and the tensors that are associated with the vertices.
However, in a (braided) fusion category, we only know how to make sense of planar diagrams, where the specific planar projection of the graph bears significance.
In particular, without additional information, we can obtain multiple different diagrams from the same graph, which affects the outcome.

In order to avoid this ambiguity without the need to provide more information about the geometry of the network, we can impose that the graph must be planar.
Specifically, whenever a graph is planar, the planar projection of the graph is unique, which allows us to unambiguously define the tensor network \cite{whitneyCongruentGraphsConnectivity1932}.
Furthermore, this restriction is necessary whenever we want to work with a non-braided category, since in these cases crossing lines are not defined.

Of course, not all tensor networks are planar, and we might want to work with graphs that have no planar projection.
For these cases, \TensorKit provides the ability to insert symbolic \emph{braiding tensors} at the crossing points, which in turn restore the planarity of the graph.
Here, it is important to note that in the most generic case these braiding tensors are not symmetric, so therefore we must make a distinction between over-crossings and under-crossings.

Finally, we note that we can drop all these considerations and return to the usual case of graphs whenever the following two conditions are met:
We require the braiding to be symmetric, which allows us to ignore the distinction between over- and under-crossings.
Additionally, we also require the twist to be trivial, which does not necessarily follow from the symmetry of the braiding.
In this case, we can work with any planar projection of the graph without altering the outcome, restoring the usual graphical calculus for tensor networks.
Both of these options are supported by \TensorKit.

\subsection{Examples}
\label{subsec:examples}
In this final section, we discuss some specific examples.
Aside from $\category{Vect}$ and $\category{SVect}$, the most important example for dealing with tensors with (conventional) symmetries, \eg\ in many-body physics, is $\Rep{G}$, the category of representations of a group.
We also include the categories $\category{Fib}$ and $\category{Ising}$ as prominent cases for anyonic theories.
Many additional categories exist, and as long as the topological data is known, \TensorKit easily allows for the inclusion of custom categories.
Notably, there is an exhaustive list of all multiplicity-free categories up to six simple objects \cite{bridgemanGitHubJCBridgemanSmallRankUnitaryFusionData}, which is supported through \library{CategoryData.jl} \cite{githubGitHubLkdvosCategoryDatajl}.

\subsubsection{The Category $\Rep{G}$}

Of course, as we intended this framework as a generalization of group-based symmetries, these should also be described by a fusion category.
The category $\Rep{G}$ describes representations of a group $G$, which we assume to be either a finite group or a compact Lie group for simplicity.
We will also restrict our discussion to finite-dimensional representations, which can then be chosen to be unitary.
In particular, the ability to wrap this data in a unitary braided fusion category underpins the consistency of all manipulations defined before.

The objects of this category consist of the representations of the group, and the morphisms consist of the $G$-invariant linear maps on the carrier spaces.
The monoidal structure is found by way of the tensor product representation, with the trivial representation acting as the unit.
The dual of a representation $U$ is the contragradient representation $U^* = (U^{-1})^T$, which coincides with the conjugate representation $\bar{U}$ due to unitarity.
Each representation can be decomposed into irreducible representations, which make up $\Irr(\Rep{G})$.
The fusion rules are determined by this direct product decomposition and the dimensions coincide with the usual vector space dimension of the carrier space.

As an explicit example we consider $\group{G} = \SU{2}$, where the irreducible representations are labeled by half-integers $j = 0, \frac{1}{2}, 1, \ldots$
The fusion rules follow from the Clebsch-Gordan decomposition:
\begin{equation}
    \label{eq:SU2NSymbol}
    \NSymbol{j_1 j_2}{j_3} = \begin{cases}
        1 & \qquad \left|j_1 - j_2\right| \leq j_3 \leq j_1 + j_2 \\
        0 & \qquad \text{else}
    \end{cases}
\end{equation}
From this also follows that all irreps are self-dual, and their dimensions are $d_j = 2j + 1$.
We also note that, as $\NSymbol{ab}{c} \leq 1$, we can drop the multiplicity labels in the other symbols.

In this case, the $F$-symbol can be explicitly computed.It relates to the $6j$-symbol via
\begin{equation}
    \label{eq:SU2FSymbol}
    \FSymbol{j_1j_2j_3}{j_4}{j_5}{j_6} 
        = \sqrt{d_{j_5} d_{j_6}}
        (-1)^{j_1 + j_2 + j_3 + j_4}
        \begin{Bmatrix}
            j_1 & j_2 & j_5 \\
            j_3 & j_4 & j_6
        \end{Bmatrix}.
\end{equation}
The $R$-symbol is given by 
\begin{equation}
    \label{eq:SU2RSymbol}
    \RSymbol{j_1j_2}{j_3}{}{} = (-1)^{j_1 + j_2 + j_3},
\end{equation}
whereby the resulting twist is trivial, $\theta_j = 1$.
Finally, the Frobenius-Schur phase is $\chi_j = (-1)^{2j}$.

In general, the topological data for $\Rep{G}$ can be obtained by inspecting the representation theory of the group.
Data for the representations of a vast amount of finite groups is available in the literature, for example through the \library{GAP} library \cite{GAP4}.
More specifically, given the fusion rules for the representations along with the Clebsch-Gordan coefficients, one can compute the $F$-symbols and $R$-symbols, and from there the twist and Frobenius-Schur phases.
As the Clebsch-Gordan coefficients form an explicit representation of the splitting and fusion tensors, this is achieved by projecting \eqref{eq:associator} onto the basis of fusion tensors:
\begin{equation}
    \diagram{categories}{35} = \FSymbol{abc}{d}{\nu,e,\mu}{\rho,f,\sigma} \diagram{categories}{36}
\end{equation}

Similarly, the $R$-symbols can be computed by projecting \eqref{eq:braiding} onto the basis of splitting tensors:
\begin{equation}
    \diagram{categories}{37} = \RSymbol{ab}{c}{\mu}{\nu} \diagram{categories}{38}
\end{equation}

\subsubsection{Fermions and fusion categories}

Fermionic symmetries can be accommodated within the framework of fusion categories through the category $\category{SVect}$ of super vector spaces.
This category has two simple objects: the monoidal unit $I = \field{C}^{1|0}$ and the simple odd object $J = \field{C}^{0|1}$.
A general object in this category is $V = \field{C}^{n_0|n_1} = I^{\oplus n_0} \oplus J^{\oplus n_1}$.
The fusion rules are defined by $J \otimes J = I$ and the trivial relations expressing that $I$ is the monoidal unit.
The non-zero values of $\NSymbol{ab}{c}$ are thus given by $\NSymbol{JJ}{I} = \NSymbol{IJ}{J} = \NSymbol{JI}{J} = \NSymbol{II}{I} = 1$.

In $\category{SVect}$, there are no fusion multiplicities, and the associativity is expressed by a trivial $F$-symbol:
\begin{equation}
    \label{eq:trivialF}
    \FSymbol{abc}{d}{e}{f} = \NSymbol{ab}{e} \NSymbol{ec}{d} \NSymbol{bc}{f} \NSymbol{af}{d}
\end{equation}
Duality is straightforward, given by $I^* = I$ and $J^* = J$, with dimensions $d_I = d_J = 1$ and Frobenius-Schur phases $\chi_I = \chi_J = 1$.
The category is braided, with $\RSymbol{JJ}{I}{}{} = -1$ and $\RSymbol{IJ}{J}{}{} = \RSymbol{JI}{J}{}{} = \RSymbol{II}{I}{}{} = 1$.
The resulting twists are given by $\theta_I = 1$ and $\theta_J = -1$.

This provides an explicit example of a category with a symmetric braiding but a non-trivial twist.
This typically necessitates planar diagrammatic rules as defined in \ref{sec:planaroperations}.
However, a slight modification of the framework can restore the ability to assign unique results to tensor network graphs.
This is achieved by ``pushing the twist into the right-evaluation map'', which redefines the natural pairing between vectors and co-vectors by absorbing the parity factor:
\begin{align}
    \tilde{\ev{V^*}} \colon V \otimes V^* &\rightarrow \complexs\\
    \tilde{\ev{V^*}} &= \ev{V^*} \circ \left( \theta_V \otimes \id{V^*} \right) \\
    \ket{v} \otimes \bra{w} &\mapsto (-1)^{|v||w|} \delta_{vw}
\end{align}
With this redefinition, the twist becomes trivial.

While this alteration allows the use of graphs instead of planar diagrams for specifying tensor networks, it comes with the drawback of incompatibility with the unitary structure.
Specifically, the resulting dimensions are no longer positive, as $d_I = 1$ and $d_J = -1$.
This shows up when trying to compute inner products as tensor contractions, where additional parity factors must be inserted to cancel the twist that is automatically inserted by the contraction.
For a comprehensive exposition of this framework and its applications in tensor networks, we refer to \cite{mortier2025a}.

\subsubsection{Deligne Product Categories}

Given two or more fusion categories $\category{C}_1, \category{C}_2, \ldots, \category{C}_n$, there is a framework to combine these categories into a single fusion category, the Deligne product category $\category{C}_1 \boxtimes \category{C}_2 \boxtimes \ldots \boxtimes \category{C}_n$.
The objects in this combined category are given by the Cartesian product of the objects of the individual categories, and the morphisms are given by the tensor product of the morphisms of the individual categories.

Using this definition, the topological data of the Deligne product category can be described in terms of the topological data of the individual categories.
Importantly, if the input categories are unitary fusion categories, the resulting Deligne product category is also a unitary fusion category.
For simplicity, we will define the Deligne product for two input categories $\category{C}_1$ and $\category{C}_2$, but this can be extended sequentially to more categories.
This construction is associative and commutative, and changing the order of the extension process yields equivalent categories.

The simple objects in the Deligne product category are given by the Cartesian product of objects, $\Irr(\category{C}_1 \boxtimes \category{C}_2) = \Irr(\category{C}_1) \times \Irr(\category{C_2})$.
The fusion rules are derived from the tensor product of the fusion rules of the individual categories:
\begin{equation}
    \NSymbol{a_1 \boxtimes a_2 ~ b_1 \boxtimes b_2}{c_1 \boxtimes c_2} = \NSymbol{a_1b_1}{c_1} \cdot \NSymbol{a_2b_2}{c_2}.
\end{equation}
As a consequence, the unit object in the Deligne product category is the product of the unit objects from each category, $I_1 \boxtimes I_2$, and the dual object is the product of the dual objects from each category, $\bar{a}_1 \boxtimes \bar{a}_2$.
The dimensions of the objects are the product of the dimensions of the individual categories, \ie\ $d_{a_1 \boxtimes a_2} = d_{a_1} d_{a_2}$.

The associators in the Deligne product category are given by the tensor product of the associators from the individual categories.
This implies that the $F$-symbols are the appropriate Kronecker product of the $F$-symbols from the individual categories:
\begin{equation}
    \FSymbol{a_1 \boxtimes a_2 ~ b_1 \boxtimes b_2 ~ c_1 \boxtimes c_2}{\,\,d_1 \boxtimes d_2}{\mu_1\mu_2,e_1 \boxtimes e_2,\nu_1\nu_2}{\kappa_1 \kappa_2, f_1 \boxtimes f_2, \lambda_1 \lambda_2} = 
        \FSymbol{a_1 b_1 c_1}{d_1}{\mu_1,e_1,\nu_1}{\kappa_1,f_1,\lambda_1} \cdot \FSymbol{a_2 b_2 c_2}{d_2}{\mu_2,e_2,\nu_2}{\kappa_2,f_2,\lambda_2}
\end{equation}

Similarly, if all input categories are braided, the $R$-symbols are given by the Kronecker product of the $R$-symbols from the individual categories:
\begin{equation}
    \RSymbol{a_1 \boxtimes a_2 ~ b_1 \boxtimes b_2}{c_1 \boxtimes c_2}{\mu_1 \mu_2}{\nu_1 \nu_2}
        = \RSymbol{a_1 b_1}{c_1}{\mu_1}{\nu_1} \cdot \RSymbol{a_2 b_2}{c_2}{\mu_2}{\nu_2}.
\end{equation}

This construction provides a generic way to extend the representations of direct product groups to categories.
Notably, we have the equivalence $\category{Rep}^{\group{G_1} \times \group{G_2}} \simeq \category{Rep}^{\group{G_1}} \boxtimes \category{Rep}^{\group{G_2}}$.

\subsubsection{Anyons And Fusion Categories}

Many (unitary) fusion categories with non-trivial braiding rules arise in the context of representations of quantum groups and quasi-triangular Hopf algebras.
The resulting simple objects in these categories appear as anyons or topological sectors in the study of topological phases in (quasi)-two-dimensional quantum matter or as conformal towers in the study of conformal field theories.
The details thereof would lead us too far.
Instead, we provide two well-known examples.

The Fibonacci category $\category{Fib}$ has two simple objects, $I$ and $\tau$, with $\tau \otimes \tau = I \oplus \tau$ as the only non-trivial fusion rule.
This leads to $\bar{\tau} = \tau$ and $d_I = 1$ and $d_\tau = \phi$, with $\phi = \frac{1 + \sqrt{5}}{2}$ being the golden ratio.
The standard choice for the $F$-symbol is
\begin{align}
    \FSymbol{\tau\tau\tau}{\tau}{}{} &= \begin{pmatrix}\phi^{-1} & \phi^{-1/2} \\ \phi^{-1/2} & -\phi^{-1} \end{pmatrix}
\end{align}
where the matrix should be read with respect to the order of the basis $\{I, \tau\}$.
For all other index combinations, the $F$-symbol has the trivial value, similar to \eqref{eq:trivialF}.
Finally, there is a braiding for which the non-trivial elements are given by $\RSymbol{\tau \tau}{I}{}{} = e^{+4 \pi i/ 5}$ and $\RSymbol{\tau \tau}{\tau}{}{} = e^{-3\pi i/5}$.

The category $\category{Ising}$ has three simple objects, denoted as $I$, $\sigma$ and $\psi$, with the non-trivial fusion rules given by
\begin{align}
    \sigma \otimes \sigma &= I \oplus \psi,
    & \sigma \otimes \psi &= \sigma,
    &\psi \otimes \psi &= I.
\end{align}
All objects are thus self-dual, and we have $d_I = d_\psi = 1$ and $d_\sigma = \sqrt{2}$.
The only non-trivial components of the $F$-symbol are given by
\begin{align}
    \FSymbol{\sigma \sigma \sigma}{\sigma}{}{} 
        &= \frac{1}{\sqrt{2}} \begin{pmatrix} 
        1 & 0 & 1\\ 
        0 & 0 & 0\\ 
        1 & 0 & -1 
    \end{pmatrix},& \FSymbol{\sigma \psi \sigma}{\psi}{\sigma}{\sigma} &= \FSymbol{\psi \sigma \psi}{\sigma}{\sigma}{\sigma} = -1
\end{align}
where the matrix has to be read with respect to the basis ordered as $\{I, \sigma, \psi\}$.
All other cases reduce to the trivial \eqref{eq:trivialF}.
The non-trivial elements of the $R$-symbol are given by
\begin{align}
    \RSymbol{\sigma\sigma}{I}{}{} &= e^{-\pi i/8}, &\RSymbol{\sigma\sigma}{\psi}{}{} &= e^{+3\pi i/8}, &\RSymbol{\psi\psi}{I}{}{} &= -1, &\RSymbol{\sigma\psi}{\sigma}{}{} &= \RSymbol{\psi\sigma}{\sigma}{}{} = -i.
\end{align}

\section{Benchmarks}

Finally, in order to highlight the capabilities of the framework and demonstrate the benefits of correctly imposing the symmetry, we also provide some benchmark data in the context of one-dimensional quantum physics.
Symmetries show up in a variety of models, and while here we focus on the performance gains, there are other valuable insights that follow from having access to symmetry-resolved data.
For simplicity, we focus our attention to some prominent contractions in DMRG, one of the most common tensor network algorithms, with a variety of symmetries implemented.

Matrix product states (MPS) and the DMRG algorithm offer a highly successful way to obtain ground states of (local) Hamiltonians.
Here we mimic some of the key parts of that algorithm for our benchmarks.
In an attempt to eliminate the noise on our benchmark results and the dependency on various hyper-parameters and implementation details within the DMRG algorithm, we refrain from benchmarking a full run of the algorithm.
Instead, we first focus our attention towards the contractions that appear as eigenvalue equations in the single- and two-site version of DMRG, which tend to dominate the total runtime of the algorithm for sufficiently large dimensions of the various tensors.
Additionally, we note that for the two-site version the eigenvalue solver is followed by a (truncated) singular value decomposition, which we therefore also include in our benchmark cases.
The workloads are depicted schematically in \algref{alg:benchmark1} and \algref{alg:benchmark2}.

The results presented below are all run on Intel\textcopyright Xeon\textcopyright Gold 6244 CPU in a single-threaded environment to make comparisons fair between the different symmetries and models.
The benchmark code and precise results are found in the attached GitHub repository \cite{tensorkitbenchmarks}.

\begin{algorithm}[ht]
\caption{Single-site Benchmark}\label{alg:benchmark1}
\SetKwInOut{Input}{Inputs}\SetKwInOut{Output}{Output}
\SetKwFunction{H}{transpose}
\DontPrintSemicolon

\Input{$\diagram{mps}{7}$, $\diagram{mps}{12}$, $\diagram{mps}{13}$ and $\diagram{mps}{14}$}\;
\Output{$\diagram{mps}{7}$}
\BlankLine
\BlankLine
\For{$i \leftarrow$ repetitions}{
    $\diagram{mps}{7} \leftarrow \diagram{mps}{4}$\;
}
\end{algorithm}


\begin{algorithm}[ht]
\caption{Two-site Benchmark}\label{alg:benchmark2}
\SetKwInOut{Input}{Inputs}\SetKwInOut{Output}{Output}
\SetKwFunction{H}{transpose}
\DontPrintSemicolon

\Input{$\diagram{mps}{7}$, $\diagram{mps}{12}$, $\diagram{mps}{13}$ and $\diagram{mps}{14}$}
\Output{$\diagram{mps}{7}$}
\BlankLine
$\diagram{mps}{8} \leftarrow \text{contract}\left(\diagram{mps}{7}, ~\diagram{mps}{7}\right)$\;
\BlankLine
\For{$i \leftarrow$ repetitions}{
    $\diagram{mps}{8} \leftarrow \diagram{mps}{9}$\;
}
\tcp{Use truncated SVD to obtain single-site tensor again}
$U, \Sigma, V^\dagger \leftarrow \text{SVD}\left(\diagram{mps}{8}\right)$\;
$\diagram{mps}{7} = U$
\end{algorithm}

We perform the benchmarks with tensors with random entries, but to ensure that these examples are in fact representative, we will use \MPSKit to obtain the different spaces $P$, $V$ and $W$ that are involved in these benchmarks, as follows:

\begin{align*}
    \diagram{mps}{7} \colon V \otimes P \leftarrow V \qquad
    \diagram{mps}{8} &\colon V \otimes P \leftarrow V \otimes P^* \qquad
    \diagram{mps}{12} \colon W \otimes P \leftarrow P \otimes W \\
    \diagram{mps}{13} \colon V \leftarrow V \otimes W &\qquad
    \diagram{mps}{14} \colon W \otimes V^* \leftarrow V^*
\end{align*}

We then focus on prominent toy models with a variety of relevant symmetries: the (isotropic) spin $1$ Heisenberg model, a modified \SU{3} Heisenberg model in the adjoint representation and the Hubbard model at half-filling.

\subsection{Heisenberg model}

The isotropic Heisenberg model is characterized by the Hamiltonian \ref{eq:heisenberg}, where $\vec{S} \equiv (S_x,~S_y,~S_z)$ are the spin $1$ versions of the Pauli matrices:
\begin{equation}
    \label{eq:heisenberg}
    H_{\text{Heisenberg}} = J \sum_{\langle i,j\rangle} \vec{S}_i \cdot \vec{S}_j\ .
\end{equation}
Importantly, this operator is invariant under global \SU{2} rotations, and as a result we may construct the tensor map $\vec{S} \cdot \vec{S} \colon P \otimes P \leftarrow P \otimes P$ as a $(2,2)$-tensor map, with $P = \gradedspace{P}{1}$ containing one triplet.
We can construct the relevant tensors for our benchmarks with $W = 2 \cdot \gradedspace{W}{0} \oplus \gradedspace{W}{1}$, and $V = \bigoplus_i d_i \cdot \gradedspace{V}{i + \frac{1}{2}}$.
The exact values of $d_i$ are obtained by studying the ground state of this Hamiltonian on an infinite chain using \MPSKit, and can be found in the benchmark repository.

\begin{figure}
    \centering
    \includegraphics[width=0.8\linewidth]{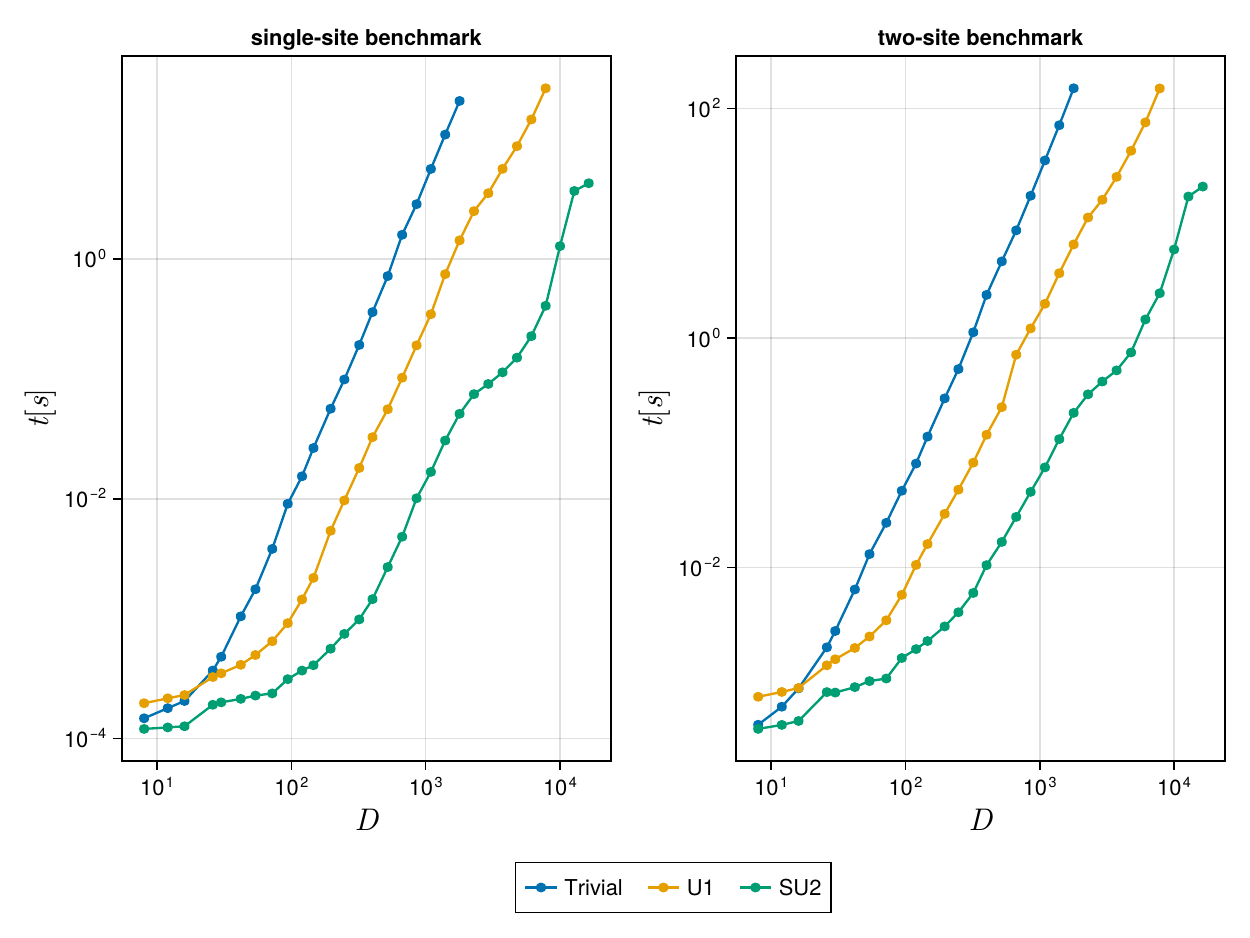}
    \caption{Benchmark results for the application of \algref{alg:benchmark1} and \algref{alg:benchmark2}, applied to typical space sizes for the spin $1$ \SU{2} Heisenberg model, for various imposed symmetries.}
    \label{fig:bench_heisenberg}
\end{figure}

For comparison, we will also consider the same Hamiltonian while imposing only subgroups of \SU{2}, notably \U{1} and $\Z{1}$, corresponding to the trivial case with no symmetry constraints.
This allows us to compare respectively non-abelian, abelian, and trivial symmetry implementations of our benchmarks, in function of the dimension $D = \dim(V)$, as shown in Figure \ref{fig:bench_heisenberg}.
From these results, we note that for all imposed symmetries, the benchmarks show approximately the same scaling of $\mathcal{O}(D^3)$, which is the expected theoretical cost of these contractions.
Nevertheless, we find that imposing more symmetry leads to a decrease in runtime at similar dimensions of orders of magnitude.
Also interesting to note is that when imposing only \U{1} compared to the trivial symmetry, we have a crossover point around $\dim(D) = 25$ where the overhead of the symmetry bookkeeping is outweighed by the computational gains.
When using \SU{2}-symmetric tensors however, the computational benefit immediately dominates the added overhead, and we end up with roughly a two orders of magnitude speed-up compared to non-symmetric implementations.

\subsection{\SU{3}-symmetric Heisenberg model}

We then consider a modified version of the Heisenberg model with \SU{3} symmetry using the Hamiltonian in \eqref{eq:su3heisenberg}, where we replaced the usual Pauli-matrices by the generators $T^a$ of \SU{3} in the $\mathbf{8}$-dimensional adjoint representation.
\begin{equation}
    \label{eq:su3heisenberg}
    \tilde{H}_{\text{Heisenberg}} = J \sum_{\langle i,j \rangle} \vec{T}_i \cdot \vec{T}_j\ .
\end{equation}

Here, we construct the tensor map $\vec{T} \cdot \vec{T} \colon P_{\SU{3}} \otimes P_{\SU{3}} \leftarrow P_{\SU{3}} \otimes P_{\SU{3}}$ with $P_{\SU{3}} = \gradedspace{P}{\mathbf{8}}$.
We construct the relevant tensors for our benchmark, now with $W = 2 \cdot \gradedspace{W}{\mathbf{1}} \oplus \gradedspace{W}{\mathbf{8}}$, and $V = \bigoplus_i d_i \gradedspace{V}{\mathbf{i}}$.
Again, we obtain the relevant values of $d_i$ by studying the ground state of this Hamiltonian on an infinite chain, and plot the results in terms of the total dimension $D = \dim(V)$.

\begin{figure}
    \centering
    \includegraphics[width=0.8\linewidth]{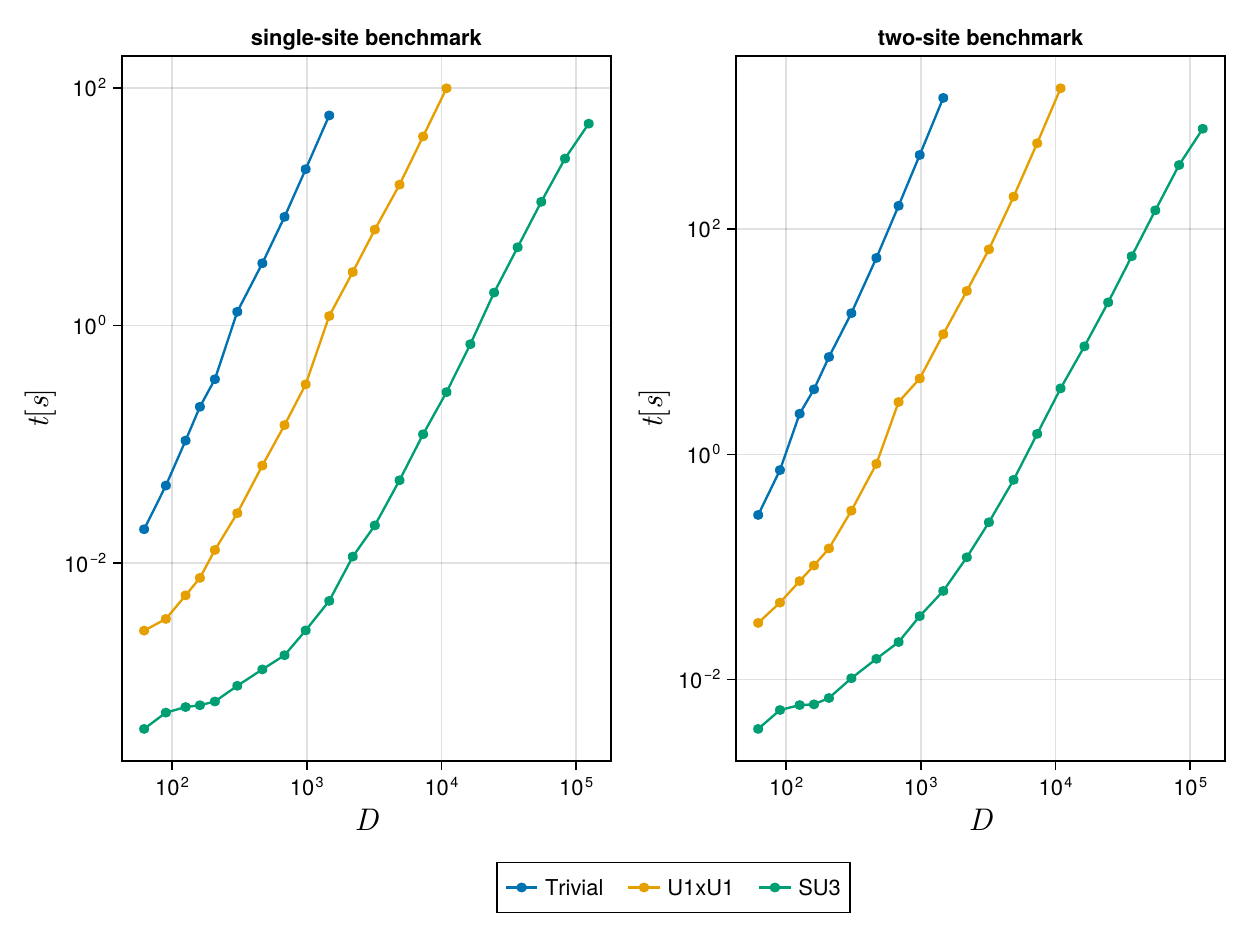}
    \caption{Benchmark results for the application of \algref{alg:benchmark1} and \algref{alg:benchmark2}, applied to typical space sizes for the modified \SU{3} Heisenberg model in the adjoint representation, for various imposed symmetries.}
    \label{fig:bench_heisenberg_su3}
\end{figure}

We compare the results by contrasting imposing only subgroups of \SU{3}, in particular $\U{1} \times \U{1}$ and the trivial $\Z{1}$.
The different runtimes are depicted in Figure \ref{fig:bench_heisenberg_su3}.
As expected, this again does not alter the asymptotic scaling of $\mathcal{O}(D^3)$, but we find an even more exaggerated improvement by implementing larger symmetry groups.
In particular, the \SU{3}-symmetric tensors in this case yield an approximate four orders of magnitude reduction in runtime.

\subsection{Hubbard model}

Finally, we may consider the Hubbard model at half filling by looking at the Hamiltonian in \eqref{eq:hubbard}, where $c_\sigma^\dagger$ ($c_\sigma$) are the fermionic creation (annihilation) operators and $n_\sigma = c_\sigma c_\sigma^\dagger$ is the number operator. 
We immediately present a form that hints at the various symmetries that are present in this model.
\begin{equation}
\label{eq:hubbard}
\begin{aligned}
    H_{\text{Hubbard}} = &-t \sum_{\langle i,j \rangle} \sum_\sigma \left(c_{i,\sigma}^\dagger c_{j,\sigma} + \text{h.c.}\right) \\
    &+ U \sum_i \left(n_{i,\uparrow} - \frac{1}{2}\right)\left(n_{i,\downarrow} - \frac{1}{2}\right)
    - \mu \sum_i \left(n_{i,\uparrow} + n_{i,\downarrow}\right)
\end{aligned}
\end{equation}
In particular, we have a fermionic parity symmetry $f\Z{2}$, particle number conservation $\U{1}$ and spin rotation $\SU{2}$, which at half filling may be promoted to a combined $f\Z{2} \times \SO{4} / \Z{2}$ symmetry by making use of the particle-hole symmetry \cite{yang1990}.
Here, we will implement this through the appropriate mapping to $f\Z{2} \times \SU{2} \times \SU{2}$, where the allowed combination of \SU{2} irreps have integer values for the sum of their spins.

\begin{figure}
    \centering
    \includegraphics[width=0.8\linewidth]{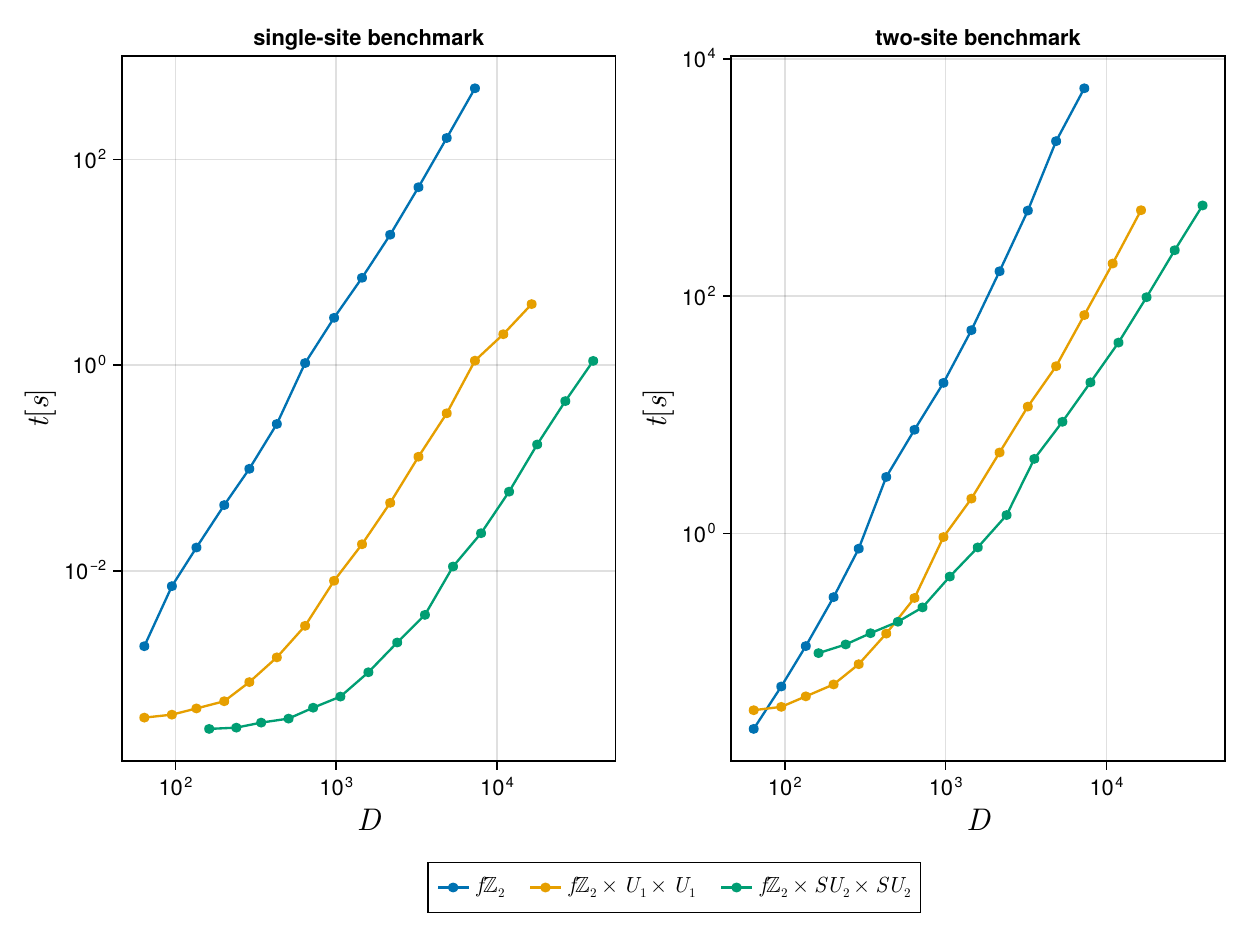}
    \caption{Benchmark results for the application of \algref{alg:benchmark1} and \algref{alg:benchmark2}, applied to typical space sizes for the Hubbard model at half-filling, for various imposed symmetries.}
    \label{fig:bench_hubbard}
\end{figure}

We can again evaluate the performance by considering the same Hamiltonian and only imposing subgroups of $f\Z{2} \times \SU{2} \times \SU{2}$.
For conciseness, we choose $f\Z{2} \times \U{1} \times \SU{2}$, $f\Z{2} \times \U{1} \times \U{1}$ and $f\Z{2}$, where the second factor denotes particle number and the final factor denotes spin symmetry.
Note that other combinations are also valid and compatible with our framework, but we cannot \emph{break} the fermionic parity symmetry without altering the results.
We present our results in Figure \ref{fig:bench_hubbard}.

Again we conclude that the symmetry strongly improves the efficiency of the algorithm, now with an effective speed-up of slightly over two orders of magnitude over the case where only fermionic parity is imposed.
Nevertheless, we now do find that there is a larger region where the overhead of the more complex symmetry group $f\Z{2} \times \SU{2} \times \SU{2}$ is not countered by the reduced cost.
In general, the exact details of this effect will be dependent on the exact sizes of the different spaces, the chosen multi-threading settings, etc.
However, we do consistently find that it is possible to get to the scaling regime where we have the expected cubic behavior and large performance benefits.

\section{Conclusion and Outlook}
Tensor networks have emerged as a powerful and flexible tool across many areas of computational science, from condensed matter physics and quantum information to high-energy physics, mathematics, machine learning and data compression.
Underlying most tensor network algorithms is a common set of basic operations--index manipulations, tensor contractions and decomposition--which are then composed in various ways.
In this work, we have analyzed these core operations in varying levels of generality and abstraction, identifying the structures that enable efficient implementation, particularly in the presence of symmetries as they are commonly encountered in the context of many-body physics.

\TensorKit is designed to capture these patterns and optimizations in a way that is both high-performing and user-friendly.
The examples in this paper demonstrate how the framework enables the development of scalable tensor network algorithms with minimal overhead, while still achieving maximal performance.
Crucially, \TensorKit handles much of the complexity internally, allowing users to benefit from sophisticated symmetry-aware and memory-efficient algorithms without requiring a deep understanding of the underlying representation theory or data layout schemes.

Looking forward, while the core functionality of \TensorKit has reached a level of maturity and stability, several promising directions for future development remain.
On the theoretical front, there is room to go beyond the current focus on regular fusion categories by exploring multi-fusion or even more general categorical settings, as briefly discussed in Section \ref{sec:category}.
On the practical side, extending support for additional symmetry types, e.g.\ point-group symmetries used in quantum chemistry applications, and categories is an ongoing goal.
Developing numerical tools to automatically compute or approximate the necessary topological data would open the door to an even broader class of applications.

From a performance perspective, ongoing experiments with the interplay of the internal structure of the tensor maps and multi-threading, multi-core parallelism, and hardware acceleration (e.g., GPU support) hold the potential to further improve runtime and scalability.

Overall, we hope that \TensorKit can serve both as a robust foundation for current tensor network applications and as a flexible platform for exploring new theoretical and computational frontiers.
As the field continues to evolve, so too will the need for (open-source!) software that combines the necessary mathematical structures with a high-performance implementation.
This is precisely where \TensorKit aims to thrive.
As tensor networks grow in scope, we believe \TensorKit is positioned to support both established practices and the exploration of new theoretical directions.

\section*{Acknowledgements}

The design and development of the \TensorKit package have benefited from countless discussions with many people, including most current and former members of the Quantum Group at Ghent University.
Being an open-source software project developed over the course of many years, we also thank all past, current and future contributors, including the bug reports and feature requests that have shaped this package.
In particular, we like to thank Maarten Van Damme, who initiated the \MPSKit package on top of \TensorKit early-on, and has as such had a strong influence on the development and design decisions of the \TensorKit package.

With regards to this manuscript, we would especially like to acknowledge Jacob C. Bridgeman for many helpful discussions and providing the \LaTeX\ package to typeset our heavy notation, as well as Lander Burgelman for a careful review.

\paragraph{Funding information}
The \TensorKit development has been supported directly by funding from the Research Foundation Flanders (FWO) [Grant No. GOE1520N], as well as indirectly through the European Union’s Horizon 2020 program [Grant Agreement No.~715861 (ERQUAF) and No.~101125822 (GAMATEN)].
Sustainable long-term funding for further development and maintenance would be very welcome.

The Flatiron Institute is a division of the Simons Foundation (LD).


\bibliography{TensorKit2.bib}

\end{document}